\documentclass[11pt]{article}

\usepackage{epsfig}
\usepackage{rotating}

\begin{document}         

\newcommand{\Rparity}{$R$-parity}
\newcommand{\Rp}  {$R_{p}$}
\newcommand{\lb}  {$\lambda$}
\newcommand{\lbp} {$\lambda^{'}$}
\newcommand{\lbpp} {$\lambda^{''}$}

\newcommand{\ee}{{\mathrm e}^+ {\mathrm e}^-}
\newcommand{\sq}{\tilde{\mathrm q}}
\newcommand{\seff}{\tilde{\mathrm f}}
\newcommand{\sele}{\tilde{\mathrm e}}
\newcommand{\ellp}{\ell^+}
\newcommand{\ellm}{\ell^-}
\newcommand{\sell}{\tilde{\ell}}
\newcommand{\squark}{\tilde{q}}
\newcommand{\snu}{\tilde{\nu}}
\newcommand{\smu}{\tilde{\mu}}
\newcommand{\stau}{\tilde{\tau}}
\newcommand{\chp}{\tilde{\chi}^+_1}
\newcommand{\chip}{\tilde{\chi}^+_1}
\newcommand{\chim}{\tilde{\chi}^-_1}
\newcommand{\chpm}{\tilde{\chi}^\pm_1}
\newcommand{\chipm}{\tilde{\chi}^\pm_1}
\newcommand{\nt}{\tilde{\chi}^0}
\newcommand{\qq}{{\mathrm q}\bar{\mathrm q}}
\newcommand{\sleppair}{\sell^+ \sell^-}
\newcommand{\nunu}{\nu \bar{\nu}}
\newcommand{\mumu}{\mu^+ \mu^-}
\newcommand{\tautau}{\tau^+ \tau^-}
\newcommand{\ellell}{\ell^+ \ell^-}
\newcommand{\nulqq}{\nu \ell {\mathrm q} \bar{\mathrm q}'}
\newcommand{\MZ}{M_{\mathrm Z}}
\newcommand{\Wstar}{W$^{(*)}$}

\newcommand {\stopm}         {\tilde{\mathrm{t}}_{1}}
\newcommand {\stopn}         {\tilde{\mathrm{t}}}
\newcommand {\stops}         {\tilde{\mathrm{t}}_{2}}
\newcommand {\stopbar}       {\bar{\tilde{\mathrm{t}}}_{1}}
\newcommand {\stopx}         {\tilde{\mathrm{t}}}
\newcommand {\sneutrino}     {\tilde{\nu}}
\newcommand {\slepton}       {\tilde{\ell}}
\newcommand {\stopl}         {\tilde{\mathrm{t}}_{\mathrm L}}
\newcommand {\stopr}         {\tilde{\mathrm{t}}_{\mathrm R}}
\newcommand {\stoppair}      {\tilde{\mathrm{t}}_{1}
\bar{\tilde{\mathrm{t}}}_{1}}
\newcommand {\gluino}        {\tilde{\mathrm g}}

\newcommand {\chin}          {\tilde{\chi }^{0}_{1}}
\newcommand {\neutralino}    {\tilde{\chi }^{0}_{1}}
\newcommand {\neutrala}      {\tilde{\chi }^{0}_{2}}
\newcommand {\neutralb}      {\tilde{\chi }^{0}_{3}}
\newcommand {\neutralc}      {\tilde{\chi }^{0}_{4}}
\newcommand {\bino}          {\tilde{\mathrm B}^{0}}
\newcommand {\wino}          {\tilde{\mathrm W}^{0}}
\newcommand {\higginoa}      {\tilde{\rm H_{1}}^{0}}
\newcommand {\higginob}      {\tilde{\mathrm H_{1}}^{0}}
\newcommand {\chargino}      {\tilde{\chi }^{\pm}_{1}}
\newcommand {\charginop}     {\tilde{\chi }^{+}_{1}}
\newcommand {\KK}            {{\mathrm K}^{0}-\bar{\mathrm K}^{0}}
\newcommand {\ff}            {{\mathrm f} \bar{\mathrm f}}
\newcommand {\bstopm} {\mbox{$\boldmath {\tilde{\mathrm{t}}_{1}} $}}
\newcommand {\Mt}            {M_{\mathrm t}}
\newcommand {\mscalar}       {m_{0}}
\newcommand {\Mgaugino}      {M_{1/2}}
\newcommand {\rs}            {\sqrt{s}}
\newcommand {\WW}            {{\mathrm W}^+{\mathrm W}^-}
\newcommand {\MGUT}          {M_{\mathrm {GUT}}}
\newcommand {\Zboson}        {{\mathrm Z}^{0}}
\newcommand {\Wpm}           {{\mathrm W}^{\pm}}
\newcommand {\allqq}         {\sum_{q \neq t} q \bar{q}}
\newcommand {\mixang}        {\theta _{\mathrm {mix}}}
\newcommand {\thacop}        {\theta _{\mathrm {Acop}}}
\newcommand {\cosjet}        {\cos\thejet}
\newcommand {\costhr}        {\cos\thethr}
\newcommand {\djoin}         {d_{\mathrm{join}}}
\newcommand {\mstop}         {m_{\stopm}}
\newcommand {\msell}         {m_{\sell}}
\newcommand {\mchi}          {m_{\neutralino}}
\newcommand {\pp}{p \bar{p}}

\newcommand{\epair}{\mbox{${\mathrm e}^+{\mathrm e}^-$}}
\newcommand{\mupair}{\mbox{$\mu^+\mu^-$}}
\newcommand{\taupair}{\mbox{$\tau^+\tau^-$}}
\newcommand{\qpair}{\mbox{${\mathrm q}\overline{\mathrm q}$}}
\newcommand{\eeee}{\mbox{\epair\epair}}
\newcommand{\eemumu}{\mbox{\epair\mupair}}
\newcommand{\eetautau}{\mbox{\epair\taupair}}
\newcommand{\eeqq}{\mbox{\epair\qpair}}
\newcommand{\fs}{ final states}
\newcommand{\epairf}{\mbox{\epair\fs}}
\newcommand{\mupairf}{\mbox{\mupair\fs}}
\newcommand{\taupairf}{\mbox{\taupair\fs}}
\newcommand{\qpairf}{\mbox{\qpair\fs}}
\newcommand{\eeeef}{\mbox{\eeee\fs}}
\newcommand{\eemumuf}{\mbox{\eemumu\fs}}
\newcommand{\eetautauf}{\mbox{\eetautau\fs}}
\newcommand{\eeqqf}{\mbox{\eeqq\fs}}
\newcommand{\ffff}{four fermion final states}
\newcommand{\llnunu}{\mbox{\lpair\nul\nubar}}
\newcommand{\lnuqq}{\mbox{\lept\nubar\qpair}}
\newcommand{\zee}{\mbox{Zee}}
\newcommand{\zzg}{\mbox{ZZ/Z$\gamma$}}
\newcommand{\wenu}{\mbox{We$\nu$}}

\newcommand{\el}{\mbox{${\mathrm e}^-$}}
\newcommand{\selem}{\mbox{$\tilde{\mathrm e}^-$}}
\newcommand{\smum}{\mbox{$\tilde\mu^-$}}
\newcommand{\staum}{\mbox{$\tilde\tau^-$}}
\newcommand{\slept}{\mbox{$\tilde{\ell}^\pm$}}
\newcommand{\sleptm}{\mbox{$\tilde{\ell}^-$}}
\newcommand{\lept}{\mbox{$\ell^-$}}
\newcommand{\Hl}{\mbox{$\mathrm{L}^\pm$}}
\newcommand{\Hm}{\mbox{$\mathrm{L}^-$}}
\newcommand{\Hnu}{\mbox{$\nu_{\mathrm{L}}$}}
\newcommand{\nul}{\mbox{$\nu_\ell$}}
\newcommand{\nubar}{\mbox{$\overline{\nu}_\ell$}}
\newcommand{\spair}{\mbox{$\tilde{\ell}^+\tilde{\ell}^-$}}
\newcommand{\lpair}{\mbox{$\ell^+\ell^-$}}
\newcommand{\staupair}{\mbox{$\tilde{\tau}^+\tilde{\tau}^-$}}
\newcommand{\smupair}{\mbox{$\tilde{\mu}^+\tilde{\mu}^-$}}
\newcommand{\selepair}{\mbox{$\tilde{\mathrm e}^+\tilde{\mathrm e}^-$}}
\newcommand{\ch}{\mbox{$\tilde{\chi}^\pm_1$}}
\newcommand{\chpair}{\mbox{$\tilde{\chi}^+_1\tilde{\chi}^-_1$}}
\newcommand{\chm}{\mbox{$\tilde{\chi}^-_1$}}
\newcommand{\chmp}{\mbox{$\tilde{\chi}^\pm_1$}}
\newcommand{\chz}{\mbox{$\tilde{\chi}^0_1$}}
\newcommand{\dch}{\mbox{\chm$\rightarrow$\chz\lept\nubar}}
\newcommand{\dslept}{\mbox{\sleptm$\rightarrow$\chz\lept}}
\newcommand{\dH}{\mbox{\Hm$\rightarrow$\lept\nubar\Hnu}}
\newcommand{\mch}{\mbox{$m_{\tilde{\chi}^\pm_1}$}}
\newcommand{\mslept}{\mbox{$m_{\tilde{\ell}}$}}
\newcommand{\mstau}{\mbox{$m_{\staum}$}}
\newcommand{\msmu}{\mbox{$m_{\smum}$}}
\newcommand{\msele}{\mbox{$m_{\selem}$}}
\newcommand{\mchz}{\mbox{$m_{\tilde{\chi}^0_1}$}}
\newcommand{\dm}{\mbox{$\Delta m$}}
\newcommand{\dmch}{\mbox{$\Delta m_{\ch-\chz}$}}
\newcommand{\dmslept}{\mbox{$\Delta m_{\slept-\chz}$}}
\newcommand{\dmhl}{\mbox{$\Delta m_{\Hl-\Hnu}$}}
\newcommand{\w}{\mbox{W$^\pm$}}

\newcommand{\acopc}{\mbox{$\phi^{\mathrm{acop}}$}}
\newcommand{\acolc}{\mbox{$\theta^{\mathrm{acol}}$}}
\newcommand{\acop}{\mbox{$\phi_{\mathrm{acop}}$}}
\newcommand{\acol}{\mbox{$\theta_{\mathrm{acol}}$}}
\newcommand{\pt}{\mbox{$p_{t}$}}
\newcommand{\pz}{\mbox{$p_{\mathrm{z}}^{\mathrm{miss}}$}}
\newcommand{\ptevt}{\mbox{$p_{t}^{\mathrm{miss}}$}}
\newcommand{\ptaxic}{\mbox{$a_{t}^{\mathrm{miss}}$}}
\newcommand{\stevt}{\mbox{$p_{t}^{\mathrm{miss}}$/\Ebeam}}
\newcommand{\staxic}{\mbox{$a_{t}^{\mathrm{miss}}$/\Ebeam}}
\newcommand{\dptaxic}{\mbox{missing $p_{t}$ wrt. event axis \ptaxic}}
\newcommand{\cosevt}{\mbox{$\mid\cos\theta_{\mathrm{p}}^{\mathrm{miss}}\mid$}}
\newcommand{\axicos}{\mbox{$\mid\cos\theta_{\mathrm{a}}^{\mathrm{miss}}\mid$}}
\newcommand{\pthet}{\mbox{$\theta_{\mathrm{p}}^{\mathrm{miss}}$}}
\newcommand{\athet}{\mbox{$\theta_{\mathrm{a}}^{\mathrm{miss}}$}}
\newcommand{\dcosevt}{\mbox{$\mid\cos\theta\mid$ of missing p$_{t}$}}
\newcommand{\daxicos}{\mbox{$\mid\cos\theta\mid$ of missing p$_{t}$ wrt. event
axis}}
\newcommand{\efdsw}{\mbox{$x_{\mathrm{FDSW}}$}}
\newcommand{\acopf}{\mbox{$\Delta\phi_{\mathrm{FDSW}}$}}
\newcommand{\acopm}{\mbox{$\Delta\phi_{\mathrm{MUON}}$}}
\newcommand{\acopt}{\mbox{$\Delta\phi_{\mathrm{trk}}$}}
\newcommand{\po}{\mbox{$E_{\mathrm{isol}}^\gamma$}}
\newcommand{\qprod}{\mbox{$q1$$*$$q2$}}
\newcommand{\lcode}{lepton identification code}
\newcommand{\nctro}{\mbox{$N_{\mathrm{trk}}^{\mathrm{out}}$}}
\newcommand{\necao}{\mbox{$N_{\mathrm{ecal}}^{\mathrm{out}}$}}
\newcommand{\mout}{\mbox{$m^{\mathrm{out}}$}}
\newcommand{\nctec}{\mbox{\nctro$+$\necao}}
\newcommand{\gfract}{\mbox{$F_{\mathrm{good}}$}}
\newcommand{\zz}       {\mbox{$|z_0|$}}
\newcommand{\dz}       {\mbox{$|d_0|$}}
\newcommand{\sint}      {\mbox{$\sin\theta$}}
\newcommand{\cost}      {\mbox{$\cos\theta$}}
\newcommand{\mcost}     {\mbox{$|\cos\theta|$}}
\newcommand{\dedx}     {\mbox{$dE/dx$}}
\newcommand{\wdedx}     {\mbox{$W_{dE/dx}$}}
\newcommand{\xe}     {\mbox{$x_E$}}

\newcommand{\ssix}     {\mbox{$\sqrt{s}$~=~161~GeV}}
\newcommand{\sthree}     {\mbox{$\sqrt{s}$~=~130--136~GeV}}
\newcommand{\mrecoil}     {\mbox{$m_{\mathrm{recoil}}$}}
\newcommand{\llmass}     {\mbox{$m_{ll}$}}
\newcommand{\sml}{\mbox{Standard Model \lpair$\nu\nu$ events}}
\newcommand{\sme}{\mbox{Standard Model events}}
\newcommand{\sig}{events containing a lepton pair plus missing transverse momentum}
\newcommand{\wpair}{\mbox{$W^+W^-$}}
\newcommand{\dW}{\mbox{W$^-\rightarrow$\lept\nubar}}
\newcommand{\dsele}{\mbox{\selem$\rightarrow$\chz e$^-$}}
\newcommand{\eeeell}{\mbox{\epair$\rightarrow$\epair\lpair}}
\newcommand{\eell}{\mbox{\epair\lpair}}
\newcommand{\llgam}{\mbox{$\ell\ell(\gamma)$}}
\newcommand{\nunugam}{\mbox{$\nu\bar{\nu}\gamma\gamma$}}
\newcommand{\acope}{\mbox{$\Delta\phi_{\mathrm{EE}}$}}
\newcommand{\nee}{\mbox{N$_{\mathrm{EE}}$}}
\newcommand{\eesum}{\mbox{$\Sigma_{\mathrm{EE}}$}}
\newcommand{\at}{\mbox{$a_{t}$}}
\newcommand{\spp}{\mbox{$p$/\Ebeam}}
\newcommand{\acoph}{\mbox{$\Delta\phi_{\mathrm{HCAL}}$}}

\newcommand{\roots}     {\sqrt{s}}
%
%
\newcommand{\thrust}    {T}
\newcommand{\nthrust}   {\hat{n}_{\mathrm{thrust}}}
\newcommand{\thethr}    {\theta_{\,\mathrm{thrust}}}
\newcommand{\phithr}    {\phi_{\mathrm{thrust}}}
\newcommand{\acosthr}   {|\cos\thethr|}
\newcommand{\thejet}    {\theta_{\,\mathrm{jet}}}
\newcommand{\acosjet}   {|\cos\thejet|}
\newcommand{\thmiss}    { \theta_{\mathrm{miss}} }
\newcommand{\cosmiss}   {| \cos \thmiss |}

\newcommand{\Evis}      {E_{\mathrm{vis}}}
\newcommand{\Rvis}      {E_{\mathrm{vis}}\,/\roots}
\newcommand{\Mvis}      {m_{\mathrm{vis}}}
\newcommand{\Rbal}      {R_{\mathrm{bal}}}

\newcommand{\Ecm}{\mbox{$E_{\mathrm{cm}}$}}
\newcommand{\Ebeam}{\mbox{$E_{\mathrm{beam}}$}}
\newcommand{\ipb}{\mbox{pb$^{-1}$}}
\newcommand{\wrt}{with respect to}
\newcommand{\sm}{Standard Model}
\newcommand{\smb}{Standard Model background}
\newcommand{\smp}{Standard Model processes}
\newcommand{\smc}{Standard Model Monte Carlo}
\newcommand{\mc}{Monte Carlo}
\newcommand{\btb}{back-to-back}
\newcommand{\tp}{two-photon}
\newcommand{\tpb}{two-photon background}
\newcommand{\tpp}{two-photon processes}
\newcommand{\lp}{lepton pairs}
\newcommand{\vto}{\mbox{$\tau$ veto}}
\newcommand{\gsim}{\;\raisebox{-0.9ex}
           {$\textstyle\stackrel{\textstyle >}{\sim}$}\;}
\newcommand{\lsim}{\;\raisebox{-0.9ex}{$\textstyle\stackrel{\textstyle<}
           {\sim}$}\;}
\newcommand{\degree}    {^\circ}

\newcommand{\phiacop}   {\phi_{\mathrm{acop}}}


%
%
\newcommand{\ZP}[3]    {Z. Phys. {\bf C#1} (#2) #3.}
\newcommand{\PL}[3]    {Phys. Lett. {\bf B#1} (#2) #3.}
\newcommand{\etal}     {{\it et al}.,\,\ }
\newcommand{\PhysLett}  {Phys.~Lett.}
\newcommand{\PRL} {Phys.~Rev.\ Lett.}
\newcommand{\PhysRep}   {Phys.~Rep.}
\newcommand{\PhysRev}   {Phys.~Rev.}
\newcommand{\NPhys}  {Nucl.~Phys.}
\newcommand{\NIM} {Nucl.~Instr.\ Meth.}
\newcommand{\CPC} {Comp.~Phys.\ Comm.}
\newcommand{\ZPhys}  {Z.~Phys.}
\newcommand{\IEEENS} {IEEE Trans.\ Nucl.~Sci.}
\newcommand{\EuroPhys}  {Euro.~Phys. \ Jour.}
%
%
\newcommand{\OPALColl}  {OPAL Collab.}
\newcommand{\JADEColl}  {JADE Collab.}
%
\newcommand{\onecol}[2] {\multicolumn{1}{#1}{#2}}
\newcommand{\ra}        {\rightarrow}   

\begin{center}
\Large 
EUROPEAN LABORATORY FOR PARTICLE PHYSICS
\end{center}

\begin{flushright}

\Large
CERN-EP/98-203 \\
11th December 1998

\end{flushright}

\vspace{0.8cm}

\begin{center}

{    \huge \bf \boldmath

Searches for R--Parity Violating Decays of Gauginos
at 183 GeV at LEP }

\normalsize

\vspace{0.5cm}

\LARGE

The OPAL Collaboration


\end{center}

\vspace{1.0cm}

\begin{abstract}
Searches for pair-produced charginos and neutralinos with \Rparity\ 
violating decays have been performed 
using a data sample corresponding to an integrated luminosity of
56~pb$^{-1}$ 
collected with the OPAL detector at LEP
at a centre-of-mass energy of $\sqrt{s}=$ 183 GeV.
An important consequence of \Rparity\ 
violation
is that 
the lightest supersymmetric particle becomes unstable. 
The searches have been performed under the assumptions 
that 
the
lightest supersymmetric particle promptly decays
and that only one 
\Rparity\ violating coupling is dominant 
for each of the decay modes considered. Such processes would 
yield multiple leptons, jets plus leptons, or multiple jets with or 
without significant missing energy in the final state.
No excess of such events above Standard Model backgrounds 
has been observed. 
Limits are presented on the production cross-sections of 
gauginos in \Rparity\ violating scenarios. 
Limits 
are also presented in the framework
of the Minimal Supersymmetric Standard Model.
\end{abstract}

\vspace{1.5cm}

\begin{center}

(Submitted to Euro. J. Phys. C.)



\end{center}



\vspace{2.5cm}


\newpage
\begin{center}{\Large        The OPAL Collaboration
}\end{center}\bigskip
\begin{center}{
G.\thinspace Abbiendi$^{  2}$,
K.\thinspace Ackerstaff$^{  8}$,
G.\thinspace Alexander$^{ 23}$,
J.\thinspace Allison$^{ 16}$,
N.\thinspace Altekamp$^{  5}$,
K.J.\thinspace Anderson$^{  9}$,
S.\thinspace Anderson$^{ 12}$,
S.\thinspace Arcelli$^{ 17}$,
S.\thinspace Asai$^{ 24}$,
S.F.\thinspace Ashby$^{  1}$
% ...
}\end{center}\bigskip
\bigskip
$^{  1}$School of Physics and Astronomy, University of Birmingham,
Birmingham B15 2TT, UK
\newline
$^{  2}$Dipartimento di Fisica dell' Universit\`a di Bologna and INFN,
I-40126 Bologna, Italy
\newline
$^{  3}$Physikalisches Institut, Universit\"at Bonn,
D-53115 Bonn, Germany
\newline
$^{  4}$Department of Physics, University of California,
Riverside CA 92521, USA
\newline
$^{  5}$Cavendish Laboratory, Cambridge CB3 0HE, UK
\newline
$^{  6}$Ottawa-Carleton Institute for Physics,
Department of Physics, Carleton University,
Ottawa, Ontario K1S 5B6, Canada
\newline
$^{  7}$Centre for Research in Particle Physics,
Carleton University, Ottawa, Ontario K1S 5B6, Canada
\newline
$^{  8}$CERN, European Organisation for Particle Physics,
CH-1211 Geneva 23, Switzerland
\newline
$^{  9}$Enrico Fermi Institute and Department of Physics,
University of Chicago, Chicago IL 60637, USA
\newline
$^{ 10}$Fakult\"at f\"ur Physik, Albert Ludwigs Universit\"at,
D-79104 Freiburg, Germany
\newline
$^{ 11}$Physikalisches Institut, Universit\"at
Heidelberg, D-69120 Heidelberg, Germany
\newline
$^{ 12}$Indiana University, Department of Physics,
Swain Hall West 117, Bloomington IN 47405, USA
\newline
$^{ 13}$Queen Mary and Westfield College, University of London,
London E1 4NS, UK
\newline
$^{ 14}$Technische Hochschule Aachen, III Physikalisches Institut,
Sommerfeldstrasse 26-28, D-52056 Aachen, Germany
\newline
$^{ 15}$University College London, London WC1E 6BT, UK
\newline
$^{ 16}$Department of Physics, Schuster Laboratory, The University,
Manchester M13 9PL, UK
\newline
$^{ 17}$Department of Physics, University of Maryland,
College Park, MD 20742, USA
\newline
$^{ 18}$Laboratoire de Physique Nucl\'eaire, Universit\'e de Montr\'eal,
Montr\'eal, Quebec H3C 3J7, Canada
\newline
$^{ 19}$University of Oregon, Department of Physics, Eugene
OR 97403, USA
\newline
$^{ 20}$CLRC Rutherford Appleton Laboratory, Chilton,
Didcot, Oxfordshire OX11 0QX, UK
\newline
$^{ 22}$Department of Physics, Technion-Israel Institute of
Technology, Haifa 32000, Israel
\newline
$^{ 23}$Department of Physics and Astronomy, Tel Aviv University,
Tel Aviv 69978, Israel
\newline
$^{ 24}$International Centre for Elementary Particle Physics and
Department of Physics, University of Tokyo, Tokyo 113-0033, and
Kobe University, Kobe 657-8501, Japan
\newline
$^{ 25}$Institute of Physical and Environmental Sciences,
Brunel University, Uxbridge, Middlesex UB8 3PH, UK
\newline
$^{ 26}$Particle Physics Department, Weizmann Institute of Science,
Rehovot 76100, Israel
\newline
$^{ 27}$Universit\"at Hamburg/DESY, II Institut f\"ur Experimental
Physik, Notkestrasse 85, D-22607 Hamburg, Germany
\newline
$^{ 28}$University of Victoria, Department of Physics, P O Box 3055,
Victoria BC V8W 3P6, Canada
\newline
$^{ 29}$University of British Columbia, Department of Physics,
Vancouver BC V6T 1Z1, Canada
\newline
$^{ 30}$University of Alberta,  Department of Physics,
Edmonton AB T6G 2J1, Canada
\newline
$^{ 31}$Duke University, Dept of Physics,
Durham, NC 27708-0305, USA
\newline
$^{ 32}$Research Institute for Particle and Nuclear Physics,
H-1525 Budapest, P O  Box 49, Hungary
\newline
$^{ 33}$Institute of Nuclear Research,
H-4001 Debrecen, P O  Box 51, Hungary
\newline
$^{ 34}$Ludwigs-Maximilians-Universit\"at M\"unchen,
Sektion Physik, Am Coulombwall 1, D-85748 Garching, Germany
\newline
\bigskip\newline
$^{  a}$ and at TRIUMF, Vancouver, Canada V6T 2A3
\newline
$^{  b}$ and Royal Society University Research Fellow
\newline
$^{  c}$ and Institute of Nuclear Research, Debrecen, Hungary
\newline
$^{  d}$ and Department of Experimental Physics, Lajos Kossuth
University, Debrecen, Hungary
\newline
$^{  e}$ on leave of absence from the University of Freiburg
\newline
\newpage
\section{Introduction}
In the general Lagrangian of the Minimal Supersymmetric extension of the
Standard Model (MSSM) \cite{ref:MSSM}, 
the terms violating lepton ($L$) and baryon ($B$) numbers
can be written 
as\footnote{There 
exists an additional \Rparity\ violating term: $\mu_i L_i H_u$, 
with $\mu_i$ a bilinear coupling and $H_u$ the up-type Higgs field.
This term is usually assumed to become zero by a rotation 
of the lepton field,
and is neglected in this paper.}~:

\[
{\cal L}_{RPV}  = 
\lambda_{ijk}   L_i L_j {\overline E}_k
+  \lambda^{'}_{ijk}  L_i Q_j {\overline D}_k
+  \lambda^{''}_{ijk} {\overline U}_i {\overline D}_j {\overline D}_k,
\]
where $i,j,k$ are the generation indices of the superfields 
$L, Q,E,D$ and $U$. $L$ and $Q$ are lepton and quark left-handed doublets,  
respectively. 
$\overline E$, $\overline D$ and $\overline U$ are right-handed 
singlet charge-conjugate superfields for the charged 
leptons and down- and up-type quarks, respectively. 
Yukawa couplings are denoted by
\lb\/, \lbp\/, and \lbpp\/.
The first term in ${\cal L}_{RPV}$ is 
anti-symmetric in $i$ and $j$, the third one 
anti-symmetric in $j$ and $k$, and $ i < j $ for \lb\ and
$ j < k $ for \lbpp. This makes a total of $ 9 + 27 + 9 = 45 $ 
parameters in addition
to those of the \Rparity\ conserving MSSM.

For a large range of values for \lb, \lbp, and \lbpp\,
these terms lead to effects like a short proton lifetime, in
contradiction with present experimental results. To avoid such effects, 
a new multiplicative quantum number, called $R$-parity, and defined as 
$ R_p = (-1)^{2S+3B+L} $ is introduced, where $S$ is the spin, and 
postulated to be conserved.
This is equivalent to setting all couplings \lb, \lbp, and \lbpp\ to zero.
\Rparity\ discriminates between ordinary and supersymmetric particles: 
\Rp\ = +1 for the \sm\ particles 
and \Rp\ = $-1$ for their supersymmetric partners.
\Rparity\ conservation implies that
supersymmetric particles are always pair-produced and 
always decay through
cascade decays to ordinary particles plus the 
lightest supersymmetric particle (LSP). 
The LSP has to be stable and is a cold dark matter candidate,
if neutral.

However there is no {\it a priori} law 
that requires the conservation of $R$-parity.
Strong experimental constraints only exist 
on the product of two \lb\--couplings\footnote{For a few individual 
couplings strong limits exist.}, and therefore
${\cal L}_{RPV} $ is not excluded by experimental results under the 
assumption that only one of the \lb\--couplings is significantly
different from zero. 
For example the non-observation of proton decay results in the
limits\footnote{All quoted limits are given 
for a sparticle mass of 100~GeV.}  
$\lambda_{11k}^{'} \cdot \lambda_{11k}^{''} \le 10^{-22} $
\cite{hinchliffe} for $k = 2,3$. A more general limit gives
$\lambda_{ijk}^{'} \cdot \lambda_{lmn}^{''} \le 10^{-10} $
\cite{smirnov}.
Limits on individual couplings are calculated e.g. from searches for
neutrinoless double beta decay or from tests of lepton universality
in pion or tau decays, and are of order $10^{-2}$ for most
couplings.
A complete listing of all existing limits is given in 
\cite{bhatta}.

The main consequence of \Rparity\ violation is an unstable LSP, 
yielding different experimental signatures compared to \Rparity\ conservation.
Also, in this case, the $\chin$ is not a cold dark matter candidate, 
and mass limits
for $\chin$ cannot be interpreted as such.
Results for decays via the coupling \lb\ have been presented by
ALEPH~\cite{ref:aleph}.
With \Rparity\ violation the production of single
sparticles becomes possible and limits from OPAL 
are given in \cite{opal_2fermion}.

The model used in this paper is a constrained Minimal Supersymmetric
Model (CMSSM)~\cite{Bartl,chargino-theory,Ambrosanio,Carena}. 
It has only five free 
parameters not counting the 
additional 45 Yukawa 
couplings \lb\/, \lbp\/, and \lbpp\/.
A common mass is assumed
for the gauginos, ($m_{1/2}$), and  
for the sfermions, ($m_0$), at the GUT
scale. The other free parameters are $\mu$,
the mixing parameter of the 
two Higgs field doublets, $\tan \beta$,
the ratio of the vacuum expectation values of the 
two Higgs doublets,
and $A$,  
a tri-linear coupling. 
By also assuming gauge unification at the GUT scale,
the masses at the electroweak scale 
of the U(1)$_{\mathrm Y}$ gaugino, ($M_1$), and of the 
soft SUSY breaking SU(2)$_{\mathrm L}$ gaugino, ($M_2$), 
are related by $M_1 = \frac{5}{3} \tan^2
\theta_{\mathrm W} M_2$ 
with $\theta_{\mathrm W}$ the weak mixing angle
and  $M_1 = 0.42  m_{1/2}$.

In this paper we present searches, assuming \Rparity\ violation, 
for the pair production of charginos ($\chipm$) and neutralinos ($\chin$)
with the OPAL detector at a centre-of-mass energy of 183~GeV
at the LEP e$^+$e$^-$ collider at CERN,
using an 
integrated luminosity of $\sim $ 56~pb$^{-1}$.
Decays via \lb\/, \lbp\/, and \lbpp\ couplings are searched for.
We further assume a prompt
decay of all SUSY particles, and design our searches to be sensitive 
only to particles decaying close to the interaction vertex.
This corresponds to a sensitivity for values of \lb, \lbp, and \lbpp\
greater than $ \sim 10^{-5}$.
The assumption of heavy sfermion masses is made in addition.

\section{Gaugino Production and Decay}

In electron-positron collisions, charginos and neutralinos can be 
pair-produced through $s$-channel processes involving $\Zboson$ 
or $\gamma$ exchange. They can also be produced through 
$t$-channel exchange of an electron-sneutrino, ($\snu_e$), or
a selectron, ($\sele$). 
The chargino (neutralino) pair production cross-section is 
reduced (enhanced) due to 
interference
between the $s$- and $t$-channels.

\subsection{Decay Modes}

\begin{figure}[htb]
\begin{center}
\epsfig{file=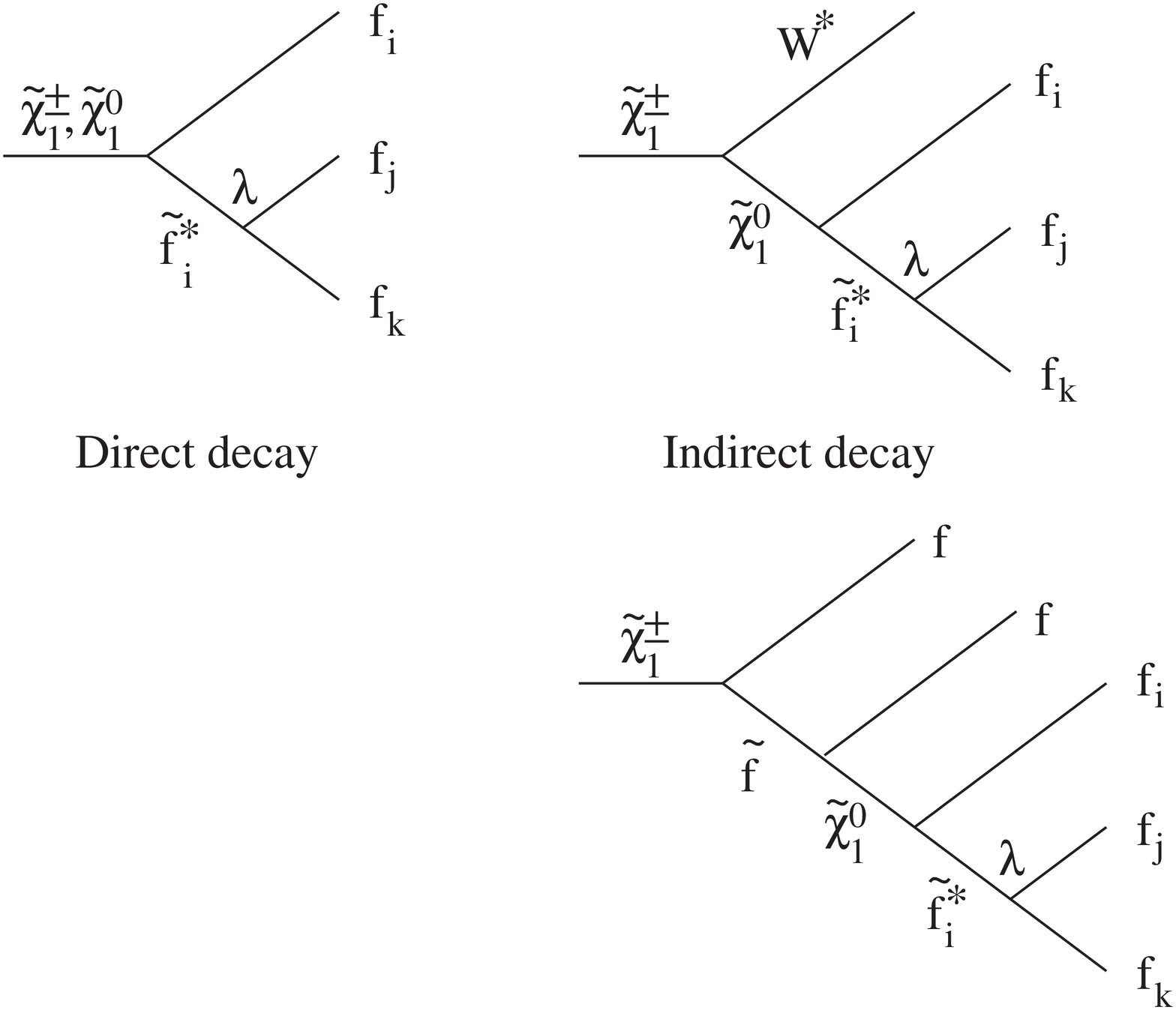,width=11.0cm}
\caption{\it Direct and indirect decays.}
\label{fig:decay}
\end{center}
\end{figure}

We distinguish direct and indirect decays, as shown 
in Figure~\ref{fig:decay}. 

In the direct mode, the gaugino decays into a fermion and 
a virtual sfermion which, in turn, decays
via the \Rparity\ violating Lagrangian. 
In the indirect mode,the \Rparity\ violating transition occurs
at a later stage in the decay sequence.
In this paper both the direct decays
of the $\chipm$ and $\chin$ 
and the indirect decays  
of the $\chipm$,
shown in Figure~\ref{fig:decay}, are considered. 
Throughout, we assume that 
only one of the couplings \lb, \lbp, or \lbpp\ is different from zero.

\subsubsection{Direct Decays}

The direct decay of a gaugino produces three fermions and the pair-production
of gauginos results in 6-fermion final states.
The type (lepton or quark) of the fermions is
determined by the couplings \lb, \lbp, and \lbpp\ while the flavour
is determined  by the indices of the couplings. 

For non-vanishing \lb, the decay of a $\chin $ via 
the \lb$_{ijk} L_i L_j {\overline E}_k$ operator
results in the following final states:

\begin{center}
$\nt_1 \rightarrow  \ell^-_i \nu_j \ell^+_k$,
\quad
$\nt_1 \rightarrow  \ell^+_i \overline{\nu}_j \ell^-_k$,
\quad
$\nt_1 \rightarrow   \nu_i \ell^-_j \ell^+_k$,
\quad
$\nt_1 \rightarrow   \overline{\nu}_i \ell^+_j \ell^-_k$
\end{center}

In each case,
one of the leptons is
a neutrino, and the
other two have opposite electric charge. 
The flavours of the leptons are not correlated. 
This results in final states with four charged leptons and 
missing energy.

For the decay of a $\chipm $ via the 
\lb$_{ijk} L_i L_j {\overline E}_k$ operator the final states are:
\begin{center}
$\chp \rightarrow \nu_i \nu_j \ell^+_k$,
\quad
$\chp \rightarrow \ell^+_i \ell^+_j \ell^-_k $,
\quad
$\chp \rightarrow \ell^+_i \overline{\nu}_j \nu_k $, 
\quad
$\chp \rightarrow   \overline{\nu}_i \ell^+_j \nu_k $
\end{center}

with either one charged lepton plus two neutrinos or three charged leptons.

The final states with one charged lepton
with index $i$ or $j$ are strongly suppressed because they involve
the decay of the $\chipm$ into a neutrino from the  left handed 
lepton doublet field and a slepton from the right handed singlet field.
These final states, which can occur via mixing of the left-- and right--handed
states into the mass eigenstates, 
are neglected within this paper.
Consequently, 
the final state consists 
of two leptons of the same flavour and missing energy, 
four charged leptons with missing energy 
or six charged leptons and no missing energy.

For non-vanishing \lbp, both the $\neutralino\ $
and the $\chargino\ $ 
decay into a lepton and two quarks through 
the $\lambda^{'}_{ijk} L_i Q_j {\overline D}_k$ operator. 
The possible decays are respectively:

\begin{center}
$\nt_1 \rightarrow  \ell^-_i u_j \overline{d}_k$,
\quad
$\nt_1 \rightarrow  \ell^+_i \overline{u}_j d_k $,
\quad
$\nt_1 \rightarrow  \nu_i d_j \overline{d}_k$,
\quad
$\nt_1 \rightarrow  \overline{\nu}_i \overline{d}_j d_k $
\end{center}

and

\begin{center}
$\chp \rightarrow  \nu_i u_j \overline{d}_k$,
\quad
$\chp \rightarrow  \overline{\nu}_i \overline{d}_j u_k $
\quad
$\chp \rightarrow  \ell^+_i \overline{d}_j d_k $,
\quad
$\chp \rightarrow  \ell^+_i \overline{u}_j u_k $ 
\end{center}

These decays result in 
final states with four jets and 
either missing energy, or one charged lepton with missing energy,
or two charged leptons.


For non-vanishing \lbpp, both the $\neutralino\ $
and the $\chargino\ $ 
decay into three quarks through 
the 
$  \lambda^{''}_{ijk} {\overline U}_i {\overline D}_j {\overline D}_k$
operator. 
The possible decays are:

\begin{center}
$\nt_1 \rightarrow  \overline{u}_i \overline{d}_j \overline{d}_k $,
\quad
$\nt_1 \rightarrow  u_i d_j d_k $
\end{center}

and

\begin{center}
$\chp \rightarrow  \overline{d}_i \overline{d}_j \overline{d}_k $,
\quad
$\chp \rightarrow  u_i u_j d_k $,
\quad
$\chp \rightarrow  u_i d_j u_k$
\quad
\end{center}

and the final states consist of six  jets in each case (with no 
missing energy).
Table~\ref{tab:direct_decays} 
lists the decay modes for the direct decays
of a chargino pair and a neutralino pair.

\subsubsection{Indirect Decays}

In the  indirect decay mode, the chargino decays via the \Rparity\ conserving
couplings to a neutralino  and \sm\ particles, and the neutralino 
decays via the \Rparity\ violating Lagrangian.

The $\chipm$ decays into five fermions
with three arising from the decay of the $\chin $
and two from the decay of a W$^{(*)}$ boson.
The decay products of the $\chin$ depend on the coupling 
\lb, \lbp, or \lbpp\ and are the same as in the direct decay of the
$\chin$.
The final states therefore consist of 10 fermions, varying between 
six leptons and missing energy and ten jets.

Besides decaying via $\chipm \ra \chin$ \Wstar, the chargino 
can also decay via $\chipm \ra \seff $f. In this case the subsequent 
decay of the sfermion $\seff \ra \chin \mathrm{f} \ra $ \mbox{(fff) f} leads
to chargino decays into five fermions, and the final state is  a subset 
of the final states from the indirect 
decays already considered.

Of the cascade decays involving a $\neutrala$ we only consider
the decay $\neutrala \ra \chin \gamma $.
Table~\ref{tab:indirect_decays} lists all final states for the 
indirect decay
of a chargino pair via a W$^{(*)}$ boson.
For decays via a sfermion, the final states are a subset of 
these states.



\subsubsection{Mixed Decays}

Whether a sparticle decays via the direct or the indirect mode depends
on the precise value of the 
MSSM parameters and the size of the \lb--coupling. When the 
decay width for direct and indirect decay modes are similar, 
the mixed mode occurs with one
sparticle decaying directly and the other indirectly.
We have not investigated these final states, except for decays via \lbpp;
but they are taken into account for all \lb-couplings when interpreting
the search results 
in Chapter~\ref{sec:interpretation}.


\begin{table}[tbp]
\begin{center}
\begin{tabular}{|r|ll|ll|}
\hline
& \multicolumn{2}{c|}{$ \chip \chim \rightarrow $} & 
\multicolumn{2}{c|}{$ \chin \chin \rightarrow $}   \\
\hline
\hline
\lb\ coupling & $ \ell_i \ell_j \ell_k $ & $ \ell_i \ell_j \ell_k $  &
                $ \nu \ellp \ellm  $ & $ \nu \ellp \ellm  $  \\
              & $ \ell_i \ell_j \ell_k $ & $ \nu \nu \ell_k   $ &
                                     &                        \\
              & $ \nu \nu \ell_k   $ & $ \nu \nu \ell_k   $ &
                                     &                        \\
\hline
\lbp\ coupling & $ \ell_i q q  $ & $ \ell_i q q  $ &
                 $ \ell_i q q  $ & $ \ell_i q q  $ \\
               & $ \ell_i q q  $ & $ \nu q q   $ & 
                $ \ell_i q q  $ & $ \nu q q   $ \\
               & $ \nu  q q  $ & $ \nu q q   $ &
                $ \nu  q q  $ & $ \nu q q   $ \\
\hline
\lbpp\ coupling & $ q q q  $ & $ q q q  $  & $ q q q  $ & $ q q q  $ \\
\hline
\end{tabular}
\end{center}
\caption{\it
Final states resulting from the 
direct decay modes of a $ \chipm $ and $\chin$ pair.
The final states consist of all 6 fermions listed in any one line.
Any fermion symbol ( $ \ell, \nu, q $ ) means particle or anti-particle
with arbitrary flavour, obeying the conditions from the Lagrangian. 
$ \ell_{i(j,k)}$ is a lepton with
flavour $ i (j,k) \quad (i=1,2,3)$.
The indices $i,j$, and $k$ correspond to the indices of the Yukawa couplings 
\lb\ and \lbp.} 
\label{tab:direct_decays}
\end{table}

\begin{table}[tbp]
\begin{center}
\begin{tabular}{|r|llll|}
\hline
$ \chip \chim \rightarrow $ & W$^{(*)}$ & W$^{(*)}$ & $\chin$ & $\chin$ \\
\hline
\hline
\lb\ coupling & $ \ellp \nu $ & $ \ellm \nu $ & $ \nu \ellp \ellm   $ & 
$ \nu \ellp \ellm  $ \\
& $ \ell \nu $ & $ q q $ & $ \nu \ellp \ellm   $ & $ \nu \ellp \ellm  $ \\
& $ q q $ & $ q q $ & $ \nu \ellp \ellm   $ & $ \nu \ellp \ellm  $ \\
\hline
\lbp\ coupling & $ \ellp \nu $ & $ \ellm \nu $ & $ \ell_i q q  $ & $ 
\ell_i q q $ \\
& $ \ellp \nu $ & $ \ellm \nu $ & $ \ell_i q q  $ & $ \nu_i q q $ \\
& $ \ellp \nu $ & $ \ellm \nu $ & $ \nu_i q q  $ & $ \nu_i q q $ \\
& $ \ell \nu $ & $ q q $ & $ \ell_i q q  $ & $ \ell_i q q $ \\
& $ \ell \nu $ & $ q q $ & $ \ell_i q q  $ & $ \nu_i q q $ \\
& $ \ell \nu $ & $ q q $ & $ \nu_i q q  $ & $ \nu_i q q $ \\
& $ q q $ & $ q q $ & $ \ell_i q q  $ & $ \ell_i q q $ \\
& $ q q $ & $ q q $ & $ \ell_i q q  $ & $ \nu_i q q $ \\
& $ q q $ & $ q q $ & $ \nu_i q q  $ & $ \nu_i q q $ \\
\hline
\lbpp\ coupling & $ \ellp \nu $ & $ \ellm \nu $ & $ q q q  $ & $ q q q $ \\
& $ \ell \nu $ & $ q q $ & $ q q q  $ & $ q q q $ \\
& $ q q $ & $ q q $ & $ q q q  $ & $ q q q $ \\
\hline
\end{tabular}
\end{center}
\caption{\it
Final states resulting from the 
indirect decay modes of a $ \chipm $ pair, including only decays 
directly to the $\chin $. Cascade decays  via other sparticles are not 
included.
The final states consist of all 10 fermions listed in any one line.
Any fermion symbol ( $ \ell, \nu, q $ ) means particle or anti-particle
with any flavour being allowed. $ \ell_i, \nu_i $ is a lepton with
flavour $ i (i=1,2,3)$,
where $i$ is determined by the first index in the \lbp\ coupling.
The final states listed here are for a decay via a W$^{(*)}$ boson.}
\label{tab:indirect_decays}
\end{table}

\subsection{Decay Widths}

The decay width of gauginos is governed by their field contents
and the size of 
the \lb\/-coupling\footnote{
        Within this section, the symbol \lb\ generically represents all 
        the \lb\, \lbp\ and \lbpp\ Yukawa couplings.}.
The full matrix elements needed to calculate the decay widths 
for $\chipm$ and $\chin$ are given in
\cite{dreiner_chipm} and \cite{dreiner_chin}, respectively.

For a pure photino-like $\chin$, the decay width for 
the \lb$_{ijk} L_i L_j {\overline E}_k$ operator
is given by~\cite{dawson}:

\[ \Gamma = \lambda^2 \frac{\alpha}{128 \pi^2}
\frac{(m_{\chin})^5}{(m_{\tilde{f}})^4},                \]

where $\alpha$ is the fine-structure constant and 
$m_{\tilde{f}}$ is the mass of the virtual sfermion in the decay.
For the \lbp$_{ijk} L_i Q_j {\overline D}_k$ operator
the decay width $\Gamma$ has to be multiplied by 
$ 3 \cdot e_q^2$, with $e_q$ the charge of the virtual squark. 

Assuming a decay length 
less than a few millimeters, and a
sfermion mass $m_{\tilde{f}}$ of 100~GeV,
gives  a sensitivity for \lb\ greater than   $ {\cal{O}} (10^{-5})$
for $\chin$ masses accessible at LEP2,
much lower than the strongest existing limits, of 
$\sim 0.0003$~\cite{Dimopoulus},
on any coupling \lb\ or \lbp\/.

For  very long lifetimes, 
the LSP decays outside the detector, and the event topology is 
exactly the same as in the \Rp\ conserving case.
This case is covered by 
gaugino searches assuming \Rparity\ conservation~\cite{OPALchipm}.

\section{Event Simulation}
\label{sec:MC}

Signal and background events have been generated, passed
through the full detector simulation~\cite{gopal} and the same analysis chain
as the real data. 

\subsection{Signal}

The simulation of the signal events has been done with the 
Monte Carlo program SUSYGEN~\cite{susygen}.
For the direct decays, events have been produced  for the mass values of
45, 70 and 90~GeV with a $m_0$ of 1~TeV and at a mass of 70~GeV for a 
$m_0$ of 48.4~GeV\footnote{This is the value of $m_0$ that gives the smallest 
expected number of events and is the smallest value still allowed 
from the limits on the sneutrino mass~\cite{pdg} and OPAL limits on the 
slepton masses~\cite{ref:slep172} in R-parity conserving decays.} 
for the $\chipm$.
For the $\chin$ in addition to the four mass points  mentioned above
also a mass value of 30~GeV has been generated, as 
there exists no direct mass limit from the LEP1 data. 
The mass values for the $\chipm$ have been chosen to cover the range between
the masses already excluded from the LEP1 data and the kinematic limit
of the 183~GeV centre-of-mass energy. 

For the indirect decay of the charginos, 
$\Delta m = m_{\chpm} - m_{\nt_1} = m_{\chpm}/2$
and $m_0$~=~1~TeV have been chosen for $\chipm$ mass values of
45, 70 and 90~GeV.
Additional events have been generated at $m_{\chargino}$~=~70~GeV 
and $\Delta m = m_{\chpm}/2$
with $m_0$~=~48.4~GeV,
leading to an enhanced t-channel contribution,
and at $m_{\chargino}$~=~90~GeV
for $\Delta m = $~5~GeV to account for 
changes in the event topologies from the model parameters.
The values of $\Delta m$ have been chosen to cover a large range 
for a limited number of Monte Carlo samples.
Differences in the efficiencies from these additional points are treated 
conservatively as inefficiencies that are applied to all mass values.

Events have been produced for each of the 
nine possible $\lambda_{ijk}$ couplings. 
Events 
have been produced for each 
lepton flavour, 
corresponding to the first index of \lbp.
The quark flavours corresponding to 
the second and third index of \lbp\ have been fixed to 
the first and second generation, with a few samples for
systematic checks also containing bottom quarks.
Events have been produced separately for the decay into either
charged or neutral leptons as well for the case in which 
one gaugino decays into a charged lepton and the other decays
into a neutral lepton.
For \lbpp\, events have been produced with the couplings 
$\lambda^{''}_{112}$
and
$\lambda^{''}_{223}$. Only for \lbpp\, the mixed final states with one 
$\chipm$ decaying directly and the other indirectly have also been produced.
All possible decays of the \Wstar\ have been considered.

For the decays via \lbpp\,
   quark triplets are not correctly handled by SUSYGEN/JETSET, as no gluon 
   radiation is developed.
   This problem could modify the jet structure of the events, and therefore 
   lead
   to a wrong estimate of the efficiency. 
   This effect has been studied using
  the pair-production process of squarks, where each squark decays as 
   $\squark \rightarrow q \chin$ and $\chin \rightarrow \nu qq$, leading to 
   final states containing 6 quarks, organised in three pairs (e.g. two pairs
   coming from $\chin$ decays and one pair from the two squark decays).
   Such processes have been generated for different squark and 
   neutralino masses,
   with the parton shower simulation switched on/off in JETSET.
   The variation of the 
   selection efficiency due to the presence/absence of gluon radiation 
has been estimated in this way to be 1.2\% for the analyses used in this paper,
and has been taken as a systematic error.

In part of the region of $\tan \beta < $ 2, 
for small $M_2 $
and
negative $\mu$, 
the branching ratio of $\chipm \ra W^{(*)} \neutrala$ becomes 
very large. We have generated events for the indirect decay of 
the $\chipm$ in this channel followed by 
$\neutrala \ra \chin\gamma$. This decay mode becomes dominant for 
parameter sets where  the $\chin$ is photino-like and
the $\neutrala$ higgsino-like. 

\subsection{Background}

The contribution to the background from two-fermion final states has been 
estimated using \mbox{BHWIDE}~\cite{ref:BHWIDE}
for the $\ee(\gamma)$ 
final states and KORALZ~\cite{ref:KORALZ} for the 
$\mumu(\gamma)$ and the $\tautau(\gamma)$ states. 
Multihadronic events, $\qq(\gamma)$,
have been simulated using PYTHIA~\cite{ref:JETSET1}. 

For the two-photon background, the PYTHIA~\cite{ref:JETSET1}, 
PHOJET~\cite{ref:PHOJET} and HERWIG~\cite{ref:herwig} Monte Carlo 
generators 
have been used for 
hadronic final states
and the Vermaseren~\cite{ref:VERMASEREN} generator 
for all $\ee \ell^+ \ell^-$ final states.
All other four-fermion final states have been 
simulated 
with grc4f~\cite{ref:grace4f}, which takes into 
account interferences between all four-fermion diagrams. 

As the cross-section for two-photon processes is very large, 
a minimum transverse momentum is already required at the generator
level to limit the sample size, 
leading to a deficit of Monte Carlo events compared to the data in 
early stages of many selections. After requiring a minimum
transverse momentum also in the data selection, generally a good agreement 
is obtained.

The produced number of events corresponds to at least 10 times the 
integrated luminosity of the data set, except for the 
 two-photon processes where it is at least twice as large.

For the small contributions to background final states with six or more 
primary fermions,
no Monte Carlo generator exists. 
These final states are therefore not included in the background Monte Carlo
 samples. 
Consequently the background could be slightly underestimated, which 
would lead to a conservative approach
when calculating upper bounds applying background subtraction.

\section{The OPAL Detector}

\label{sec:opaldet}

A complete description of the  OPAL detector can be found 
in Ref.~\cite{ref:OPAL-detector} and only a brief overview is given here.

The central detector consists of
a system of tracking chambers
providing charged particle tracking
over 96\% of the full solid 
angle\footnote
   {The OPAL coordinate system is defined so that the $z$ axis is in the
    direction of the electron beam, the $x$ axis is horizontal 
    and points towards the centre of the LEP ring, and  
    $\theta$ and $\phi$
    are the polar and azimuthal angles, defined relative to the
    $+z$- and $+x$-axes, respectively. The radial coordinate is denoted
    as $r$.}
inside a 0.435~T uniform magnetic field parallel to the beam axis. 
It is composed of a two-layer
silicon microstrip vertex detector, a high precision drift chamber,
a large volume jet chamber and a set of $z$ chambers measuring 
the track coordinates along the beam direction. 
A lead-glass electromagnetic (EM)
calorimeter located outside the magnet coil
covers the full azimuthal range with excellent hermeticity
in the polar angle range of $|\cos \theta |<0.82$ for the barrel
region  and $0.81<|\cos \theta |<0.984$ for the endcap region.
The magnet return yoke is instrumented for hadron calorimetry (HCAL)
and consists of barrel and endcap sections along with pole tip detectors that
together cover the region $|\cos \theta |<0.99$.
Four layers of muon chambers 
cover the outside of the hadron calorimeter. 
Electromagnetic calorimeters close to the beam axis 
complete the geometrical acceptance down to 24 mrad, except
for the regions where a tungsten shield is present to protect
the detectors from synchrotron radiation.
These include 
the forward detectors (FD) which are
lead-scintillator sandwich calorimeters and, at smaller angles,
silicon tungsten calorimeters (SW)~\cite{ref:SW}
located on both sides of the interaction point.
The gap between the endcap EM calorimeter and the FD
is instrumented with an additional lead-scintillator 
electromagnetic calorimeter,
called the gamma-catcher.

\section{Description of Analyses }

The final states resulting from the \Rparity\ violating decays
of gauginos are manifold.
The following sections describe 
the different analyses, denoted {\bf(A)} to {\bf(I)},  
for pure leptonic final
states, final states with jets plus leptons, 
(the case of  two taus plus at least four jets 
being handled separately), 
final states with four jets and missing energy,
final states with more than four jets and missing energy, and,
final states with 6 jets or more.
Table~\ref{tab:relation} lists the analyses and the corresponding 
final states.

\begin{table}[tbp]
\begin{center}
\begin{tabular}{|l|ll|}
\hline
Analysis & Production and decay sequence& \\
\hline
\hline
\bf{(A)} 
2 leptons + $E_{T miss}$ & 
$ \chip \chim \rightarrow $ &  
$ \nu_i \nu_j \ell_k   $  $ \nu_i \nu_j \ell_k   $ \\
\hline
\bf{(B)} 
4 leptons + $E_{T miss}$ & 
$ \chin \chin \rightarrow $ & 
$ \nu \ellp \ellm  $  $ \nu \ellp \ellm  $  \\
\hline
\bf{(C)} 
6 leptons + $E_{T miss}$ & 
$ \chip \chim \rightarrow $   
 W$^{(*)}$  W$^{(*)}$  $\chin$  $\chin$ $\ra$  &
$ \ellp \nu $  $ \ellm \nu $  $ \nu \ellp \ellm   $  $ \nu \ellp \ellm  $ \\
\hline
\bf{(D)} 
6 leptons & 
$ \chip \chim \rightarrow $   &
$ \ell_i \ell_j \ell_k $  $ \ell_i \ell_j \ell_k $  \\
\hline
\bf{(E)} 
leptons plus jets &
$ \chip \chim \rightarrow $   
W$^{(*)}$  W$^{(*)}$  $\chin$  $\chin$ $\ra$  &
W$^{(*)}$  W$^{(*)} $  $ \nu \ellp \ellm   $  $ \nu \ellp \ellm  $ \\
& 
&
W$^{(*)}$  W$^{(*)} $  $ \ell_i q q   $  $ \ell_i q q  $ \\
& 
&
W$^{(*)}$  W$^{(*)} $  $ \ell_i q q   $  $ \nu_i q q  $ \\
&
$ \chip \chim, \quad \chin \chin \rightarrow $   &
$ \ell_i q q  $  $ \ell_i q q   $ \\
&
&
$ \ell_i q q  $  $ \nu_i q q   $ \\
\hline
\bf{(F)} 
2 taus + $\ge$ 4 jets  &
$ \chip \chim, \quad \chin \chin \rightarrow $   &
$ \tau q q  $  $ \tau q q  $ \\
& $ \chip \chim \rightarrow $   
W$^{(*)}$  W$^{(*)}$  $\chin$  $\chin$ $\ra$  &
W$^{(*)}$  W$^{(*)} $  $ \tau q q   $  $ \tau q q  $ \\
\hline
\bf{(G)} 
4 jets + $E_{T miss}$  &
$ \chip \chim, \quad \chin \chin \rightarrow $   &
$ \nu q q  $  $ \nu q q  $ \\
\hline
\bf{(H)} 
$>$4 jets + $E_{T miss}$  &
$ \chip \chim \rightarrow $   
W$^{(*)}$  W$^{(*)}$  $\chin$  $\chin$ $\ra$  &
W$^{(*)}$  W$^{(*)} $  $ \nu_i q q   $  $ \nu_i q q  $ \\
\hline
\bf{(I)} 
$\ge$6 jets   &
$ \chip \chim \rightarrow $   
W$^{(*)}$  W$^{(*)}$  $\chin$  $\chin$ $\ra$  &
W$^{(*)}$  W$^{(*)} $  $ q q q   $  $ q q q  $ \\
&
$ \chip \chim \rightarrow $   
 &
$ q q q   $  $ q q q  $ \\
\hline
\end{tabular}
\end{center}
\caption{\it
List of the decay channels covered by the individual analyses,
as described in the text.
The leptons plus jets final states analysis includes 
all possible W$^{(*)}$ decay modes
and all lepton flavours, except for those cases covered by 
any of the other analyses.}
\label{tab:relation}
\end{table}

To be considered in the analyses, tracks in the central detector and clusters
in the electromagnetic calorimeter were required to satisfy the normal quality
criteria~\cite{ref:leptpairs}.  
It was also required that
the ratio of the number of tracks 
to the total number of reconstructed tracks be greater than 0.2
to reduce backgrounds from beam-gas and beam-wall events.
The visible energy, $\Evis$, the visible mass, $\Mvis$, 
and the total transverse momentum
of the event were calculated using the methods 
described in~\cite{ref:OPAL-Higgs} and
\cite{ref:OPAL-MT}.

\subsection{Multilepton Final States}
\label{sec:multileptons}

The event preselection and lepton identification is described 
in~\cite{ref:slept161}.
Multihadronic, cosmic and Bhabha 
scattering vetoes~\cite{ref:slept161} were applied
and the  number of tracks
was required to be at least two.

Only tracks with $|\cos \theta| <$ 0.95 were considered  
for lepton identification. A track was considered `isolated' 
if the total energy of other charged tracks within a cone 
of $10\degree$ half-opening angle centred
on this track
was less than 2~GeV.
A track was selected as an electron candidate if one of the three
algorithms was satisfied: {\it (i)} the output value of a 
neural net algorithm as described in
\cite{ref:NN} was larger than 0.8;
{\it (ii)} $0.5 < E/p < 2.0$,  where $p$ is 
the momentum of the electron candidate
and $E$ is the energy of the electromagnetic calorimeter cluster
associated with the track; {\it (iii)} a standard electron selection 
algorithm as described in \cite{ref:elecbarrel}
for the barrel region or in \cite{ref:elecendcap} for
the endcap region was satisfied. The electron algorithm {\it ii} complements 
algorithm {\it i} in the small polar angle region while algorithm {\it iii} was
used for redundancy since the electron identification was optimised for a 
high efficiency more than for a high purity. 
A track was selected as a muon candidate according to the 
criteria employed in the
analysis of \sm\ muon pairs~\cite{ref:leptpairs}.
That is, the track had associated activity in the muon chambers or hadron
calorimeter strips or it had a high momentum but was associated with only a
small energy deposit in the electromagnetic calorimeter.
Tau candidates were selected by requiring that there were at most three 
tracks within a cone of 35$^\circ$ 
half-opening angle centred on a track.
The invariant mass computed using all good tracks and EM clusters
within the above cone had to be less than 3~GeV.
For muon and electron candidates, the momentum was estimated 
from the charged track momentum measured in the central detector, while
for tau candidates the momentum was estimated from the vector sum of
the measured momenta of the charged tracks within the tau cone.

In each of the following multilepton final state analyses, 
tracks resulting from photon conversion
were also rejected using the algorithm described in
\cite{ref:conversion}. In the two- and six-lepton final states, 
the large background from two-photon
processes was reduced by requiring that  
the total energy deposited in each silicon tungsten calorimeter be
less than 5~GeV, be less than 5~GeV in each forward calorimeter, and
be less than 5~GeV in each side of the gamma-catcher.

In addition to the requirement that there be no 
unassociated electromagnetic cluster with an energy larger 
than 
25~GeV in the event, it was also required that there be no 
unassociated 
hadronic clusters with an energy larger than 10~GeV.

\subsubsection{Two-Lepton Final States with Missing Energy}

The analysis was optimised to retain good signal efficiency
while reducing the background, mainly due to two-photon processes and 
to $\ell \ell \nu \nu$ final states from $\WW$ production. 
The following cuts were applied.

\begin{figure}[htbp]
\centering
\epsfig{file=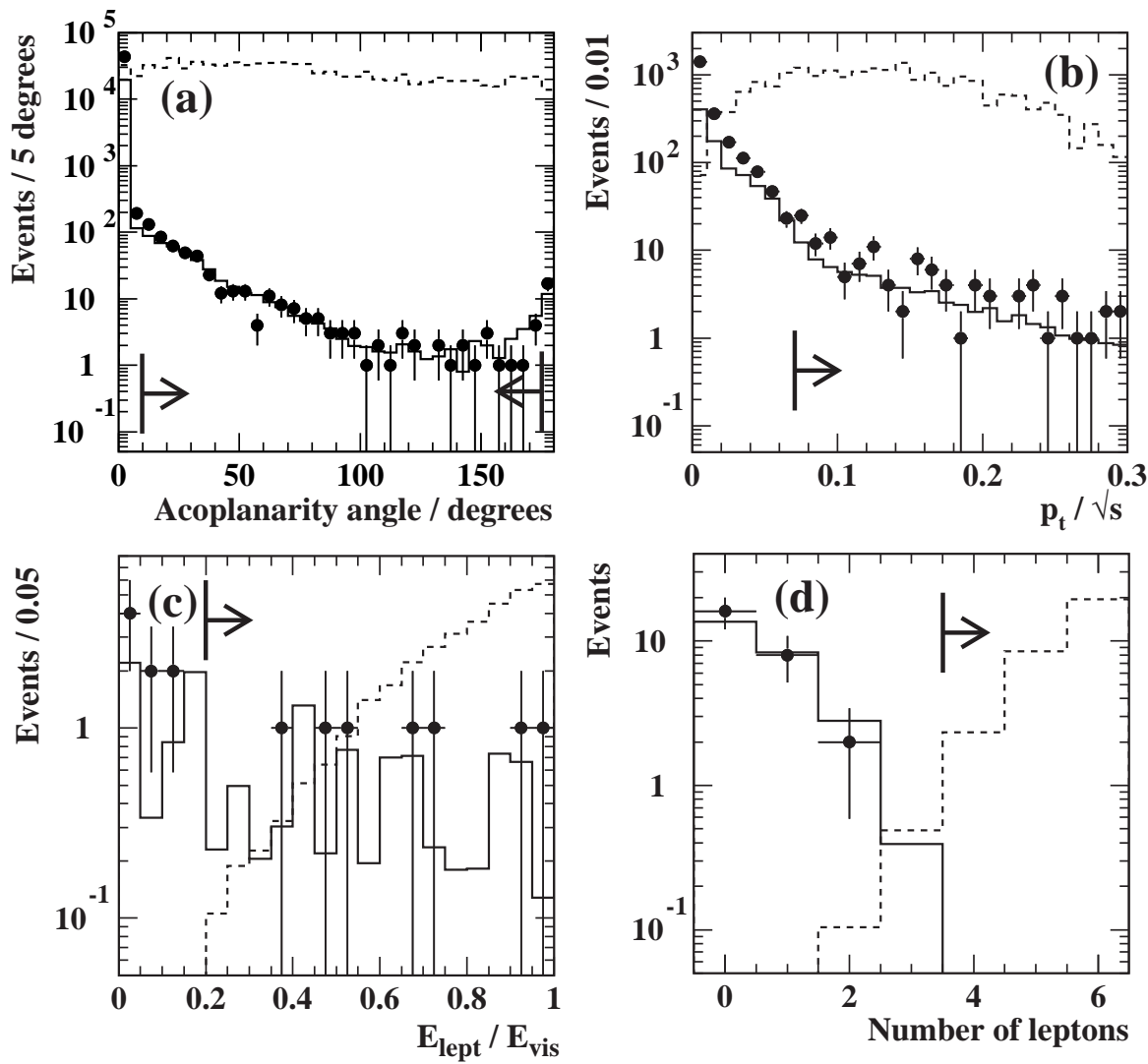,width=15.5cm} 
\caption[]{\sl
  (a) Chargino search (Analysis A): Distribution of the
  acoplanarity angle. 
  The dotted histogram 
  shows signal Monte Carlo events for direct decays of $\chargino$ with 
  $m_{\chargino} = 70$~GeV and for \lb$_{122}$.
  (b) Neutralino search (Analysis B): Distribution of the event 
  transverse momentum scaled by the centre-of-mass energy and
  calculated without the hadron calorimeter. 
  The dotted histogram 
  shows signal Monte Carlo events for direct decays of $\nt_1$ with 
  $m_{\nt_1} = 70$~GeV and for \lb$_{233}$.
  (c) Chargino search (Analysis C):  
  Distribution of the energy associated to the 
  identified leptons scaled by the total visible energy.
  The dotted histogram 
  shows signal Monte Carlo events for indirect decays of $\chargino$ with 
  $m_{\chargino} = 70$~GeV and for \lb$_{233}$. 
  (d) Chargino search (Analysis D): 
  Distribution of the number 
  of charged leptons, with a transverse 
  momentum with respect to the beam axis greater than 1.5~GeV.
  The dotted histogram 
  shows signal Monte Carlo events for direct decays of $\chargino$ with 
  $m_{\chargino} = 70$~GeV and for \lb$_{122}$. 
  Data are shown as points and the sum of all Monte Carlo background 
  processes is shown as the solid line. 
  The simulated signal events have arbitrary normalisation. 
  The arrows indicate the cut value.
 } 
\label{fig:multilepton1}
\end{figure}

\begin{description}

\item[(A1)]
Events had to contain 
exactly two identified 
and oppositely charged leptons, each with a transverse 
momentum with respect to the beam axis greater than 2~GeV.

\item[(A2)]
The background from two-photon processes and ``radiative return" events
($\ee \ra {\mathrm Z} \gamma$, where the $\gamma$ escapes
down the beam pipe) was reduced  by requiring
that the polar angle of the missing momentum, $\thmiss$,
satisfied \mbox{$\cosmiss < 0.9$}. 

\item[(A3)]
To reduce further the residual background from 
Standard Model lepton pair events,
it was required that $\Mvis /\sqrt{s} < 0.80$, 
where $\Mvis$ is the event visible mass.

\item[(A4)]
The acoplanarity angle\footnote
   {The acoplanarity angle, $\acop$,
    is defined as 180$\degree$ minus the angle
    between the two lepton momentum vectors  
    projected into the $x-y$ plane.}
($\acop$) between the two leptons was required to be
greater than 10$\degree$ in order to reject Standard Model leptonic events,
and smaller than 175$\degree$ in order to reduce the
background due to photon conversions.
The acoplanarity angle distribution is shown
in Figure~\ref{fig:multilepton1}~(a)
after cuts {\bf (A1)} to {\bf (A3)}.
The poor agreement between the data
and Monte Carlo expectation at this stage of the analysis 
is due partly to beam
related backgrounds and partly to incomplete modelling of two-photon
processes. 
The acollinearity angle\footnote
   {The acollinearity angle, $\acol$,
    is defined as 180$\degree$ minus the space-angle
    between the two lepton momentum vectors.}
($\acol$) was also required to be greater than 10$\degree$
and smaller than 175$\degree$.

\item[(A5)]
Cuts on $a_t^{\mathrm {miss}}$ 
and $p_t^{\mathrm {miss}}$ were applied, where
$a_t^{\mathrm {miss}}$ is the component
of the missing momentum vector perpendicular to the 
event thrust axis in the
plane transverse to the beam axis and $p_t^{\mathrm {miss}}$ 
is the missing transverse momentum~\cite{ref:slept161,ref:slep172}.

\item[(A6)]
The background 
was further reduced by requiring that the two 
identified leptons be of the same flavour.
Events were further selected by applying cuts on the momentum
of the two leptons as described in~\cite{ref:slept161}.

\end{description}

To maximise the detection efficiencies,  
the above selection was 
combined 
with the standard OPAL analysis to select $\WW$ pair 
events~\cite{ref:wwpaper}
where both W's decay leptonically. This combination
was performed after cut (A5) for events passing the
preselection 
criteria.
Events passing either set of criteria were accepted as 
candidates.

\begin{table}[htbp]
\begin{center}
\begin{tabular}{|r||c||c|c|c|}
\hline
Final State & Eff. (\%) & Selected Events & Tot. bkg MC  & 4-fermion \\
\hline
\hline
$ee     + E_{T \mathrm{miss}}$   & 44-74  & 11 & 13.8 & 13.5 \\
$\mu\mu + E_{T \mathrm{miss}}$   & 48-77  & 10 & 11.3 & 11.0 \\
$\tau\tau + E_{T \mathrm{miss}}$ & 20-45  & 10 & 15.5 & 12.5 \\
\hline
\end{tabular}
\end{center}
\caption{\it 
Detection efficiencies (in \%), events selected and background
predicted for the lepton-pair plus missing energy channels and for $\chargino$ 
masses between 45 and 90~GeV. 
}
\label{tab:2leptons}
\end{table}

Detection efficiencies are summarised in Table~\ref{tab:2leptons}
for the three lepton flavours considered.  
The efficiencies are quoted for $\chargino$ 
masses between 45 and 90~GeV. 
The 
expected background from all Standard Model
processes considered is normalised to the data luminosity of 56.5 pb$^{-1}$.
As can be seen in Table~\ref{tab:2leptons}, 
most of the background remaining comes from 4-fermion processes,
expected to be dominated by $\WW$ doubly-leptonic decays.
%
%
The second
most important contamination for the $\tau\tau$ final states arises
from two-photon processes leading to leptonic final states (up to 
1.9 events).

\subsubsection{Four-Lepton Final States}

The following cuts were applied to select a possible signal in the
four charged leptons and missing energy topology:

\begin{description}

\item[(B1)]
The background from two-photon processes and ``radiative return" events
($\ee \ra {\mathrm Z} \gamma$, where the $\gamma$ escapes
down the beam pipe) was reduced  by requiring
that the polar angle of the missing momentum direction, $\thmiss$,
satisfied \mbox{$\cosmiss < 0.9$}. 

\item[(B2)]
At least four tracks with a transverse 
momentum with respect to the beam axis greater than 1.0~GeV, 
were required.

\item[(B3)]
The event transverse momentum calculated without the hadron calorimeter
was required to be larger than 0.07 $\times \rs$.
This distribution 
is shown in Figure~\ref{fig:multilepton1}~(b) 
after cuts {\bf (B1)} and {\bf (B2)} have been applied. 
The poor agreement between the data
and Monte Carlo expectation at this stage of the analysis 
is due partly to beam
related backgrounds and partly to incomplete modelling of two-photon
processes. When the two-photon processes have been effectively reduced
after this cut, the agreement between data and Monte Carlo is good. 

\item[(B4)]
Events had to contain at least three well-identified 
isolated leptons, each with a transverse 
momentum with respect to the beam axis greater than 1.5~GeV.

\item[(B5)]
It was required that $\Evis /\sqrt{s} < 1.1$.

\item[(B6)]
The total leptonic energy, defined as the sum of the 
energy of all identified leptons,
was required to be greater than $0.5 \times \Evis$.

\item[(B7)]
To reduce further the total background from Standard Model di-lepton
production, it was required that the energy sum of the two most 
energetic leptons be smaller than 0.75~$\times \Evis$.

\end{description}

\begin{figure}[htbp]
\centering
\epsfig{file=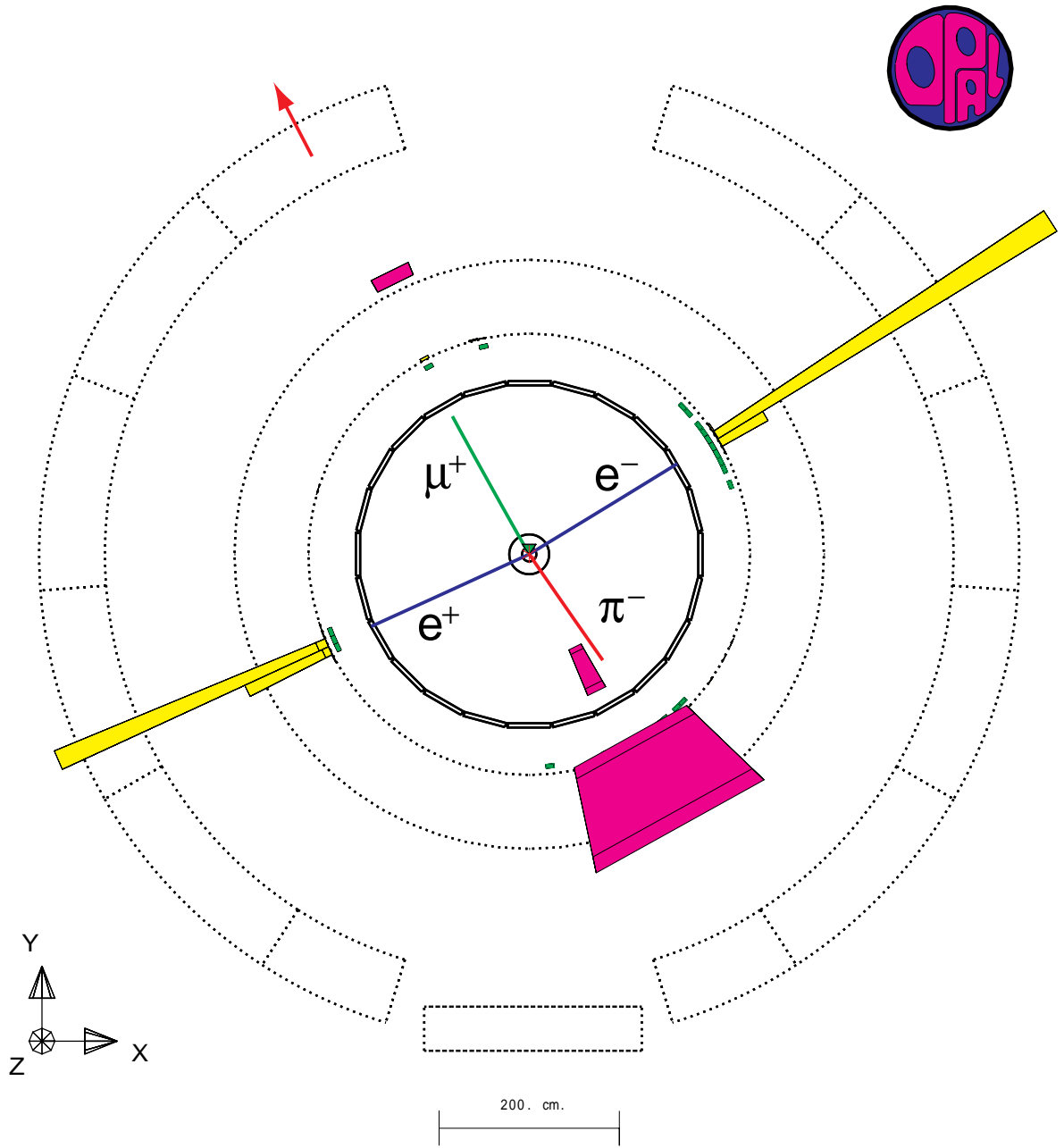,width=16.5cm}
\caption[]{\sl
Display of the event selected in the data by analysis B. This
candidate is compatible with the pair production of two on-shell Z
bosons, one decaying to an electron pair and the other to a
$\tau$ pair. One of the $\tau$'s decays to a pion and a neutrino and
the other $\tau$ decays to a muon plus two neutrinos.   
 } 
\label{fig:4lept_candi}
\end{figure}

To be independent of the types of decays, direct or indirect, 
which are searched for
and to maximise the detection efficiencies, no specific cut on the
lepton flavour present in the final state was applied. 
Detection efficiencies 
range from 16\% to 74\%
for neutralino masses between 30 and 90~GeV. 
The lower efficiency value arises from small neutralino masses (30 GeV) 
and decays with four taus in the final state.
The expected background is estimated to be 2.5 events. 
One candidate event has been selected from the data;
it is shown in Figure~\ref{fig:4lept_candi}.
This candidate is compatible with pair-production of two on-shell Z
bosons, one decaying to an electron pair and the other to a
$\tau$ pair. One of the $\tau$'s decays to a pion and a neutrino and
the other $\tau$ decays to a muon plus two neutrinos.

\subsubsection{Six-Lepton Final States}

The following cuts were applied:

\begin{description}

\item[(C1)]
To reduce the background from 
two-photon and di-lepton processes,
it was  required that $0.1 < \Evis /\sqrt{s} < 0.7$. 

\item[(C2)]
The event longitudinal momentum
was also required to be smaller than 0.9 $\times {p_{\mathrm{vis}}}$, 
where ${p_{\mathrm{vis}}}$ is the event total momentum.

\item[(C3)]
The event transverse momentum calculated without the hadron calorimeter
was required to be larger than 0.025 $\times \rs$.

\item[(C4)]
Events with 
fewer than five charged tracks (tracks from photon conversions
were not considered) with a transverse momentum with respect 
to the beam axis larger than 0.3~GeV were rejected.  

\item[(C5)]
Events had to contain at least three well-identified 
isolated leptons; at least two of them  must have a transverse 
momentum with respect to the beam axis greater than 1.5~GeV, and the
third one must have a transverse 
momentum with respect to the beam axis greater than 0.3~GeV.

\item[(C6)]
The total leptonic energy, defined as the sum of the energy of all 
identified leptons,
was required to be greater than $0.2 \times \Evis$. 
The distribution
of the total leptonic energy scaled by the visible energy, is shown
in Figure~\ref{fig:multilepton1}~(c), after cuts {\bf (C1)} to {\bf (C4)} 
have been applied.

\end{description}

In the case of final states without missing energy (chargino direct
decays without taus), the previous
selection cuts were replaced by the following ones:

\begin{description}

\item[(D1)]
To reduce further the residual background from 
two-photon processes,
it was required that $0.2 < E_{\mathrm{vis}} /\sqrt{s} < 1.2$. 

\item[(D2)]
Events with 
fewer than five charged tracks (tracks from photon conversions
were not considered) with a transverse momentum with respect 
to the beam axis larger than 1~GeV were rejected.  

\item[(D3)]
Events had to contain at least four well-identified 
isolated leptons, each of them with a transverse 
momentum with respect to the beam axis greater than 1.5~GeV.
The distribution of 
the number of charged leptons is shown in Figure~\ref{fig:multilepton1}~(d)
after cuts {\bf (D1)} and {\bf (D2)} have been applied.

\item[(D4)]
The total leptonic energy, defined as the sum of the energy of all 
identified leptons,
was required to be greater than $0.4 \times E_{\mathrm{vis}}$. 

\end{description}

Events passing either set of criteria 
were accepted. 
Detection efficiencies after combining the two analyses 
range from 22\% to 87\%
for chargino masses between 45 and 90~GeV.
The lower value of the selection efficiency arises from decays of a chargino
with a mass of 45 GeV leading to final states with 4 taus
and two leptons, while the higher value arises from decays of a chargino with a
mass of 90 GeV leading to final states with four muons and two electrons.
The background expectation 
is 1.7 events. There is one candidate event selected in the data.

\subsubsection{Inefficiencies and Systematic Errors}
\label{sec:syserr}

Variations in the efficiencies were estimated using events 
generated with $m_0$ = 48.4~GeV and
also events generated with  $\Delta m$ = 5~GeV, as described in 
Section~\ref{sec:MC}.
The inefficiencies due to variation of angular distributions 
were estimated for five different MSSM parameter sets, representing different 
neutralino field contents (gaugino/higgsino) and couplings,
and calculated separately for each analysis. The selection 
efficiencies varied by up to 10\%. 
In interpreting the results, 
a conservative approach was adopted by choosing the lowest 
efficiencies.

The inefficiency due to forward detector vetoes caused by
beam-related backgrounds or detector noise was estimated from a study
of randomly triggered beam crossings to be 
3.2\%. The quoted efficiencies 
are all scaled down to take this effect into account. 

The following systematic errors on the number of signal events expected 
have been considered~:
the statistical error on the determination of the efficiency from the 
Monte Carlo simulation;
the systematic error on the integrated luminosity, of 0.35\%; 
the systematic error due to the trigger efficiency was estimated to be
negligible because of the high lepton transverse momentum requirement;
the uncertainty due to the interpolation of the efficiencies was
estimated to be 4.0\% and the lepton identification uncertainty was 
estimated to be 2.4\% for muons, 3.9\% for electrons and
4.7\% for taus.
The total systematic error was calculated by summing in quadrature 
the individual errors. 
The total systematic error is incorporated into the limit calculation
using the method described in Reference~\cite{ref:cousins}.

\subsection{Jets plus Lepton Final States}
\label{sec:ivor}

The strategy to search for final states with jets and leptons is to look 
for signals with clear jets and well identified leptons. In the case of
neutrinos in the final state, background from two-photon processes can 
effectively be reduced by requiring some missing transverse momentum.
For most decays
the leptons will be isolated and therefore well distinguishable from the
background. The severest background in most analyses results from
W pair production.
However a kinematic fit on the invariant mass of 
jets and leptons gives a good mass resolution
and can therefore reduce most of this background.

This section describes the event selection for final states from the 
direct and indirect 
decay of gauginos via the couplings \lb\, and \lbp\ using an 
integrated luminosity\footnote{Detector status 
cuts different from the ones used in the 
analyses {\bf (A)} to {\bf (D)} have been used, resulting in a slightly  
different total integrated luminosity.}
of 55~pb$^{-1}$.
The selection cuts are organised as follows:

\begin{description}
\item[Preselection]
At least seven tracks, a minimum visible
energy of 0.3 $\cdot \sqrt{s}$, and at least one identified lepton 
with at least 3~GeV are required. 
\item [(E1)]
A cut on the visible energy scaled by the centre of mass energy 
with values depending on the expected number
of neutrinos, in the range between 0.4 and 1.2,  is applied.
In addition, the angle of the missing momentum 
with respect to the beam direction
has to fulfil \mbox{$\cosmiss < 0.95$},
if the final state contains neutrinos.
\item [(E2)]
The jets in the event have been reconstructed using the 
Durham~\cite{ref:durham} algorithm.
Cuts have been applied on the number 
of jets reconstructed with a cut parameter of 
0.005, and on the 
jet resolution $y_{i, i+1}$ 
at which the number of jets changes from $i$ to $i+1$ jets.
The value of $i$ depends on the expected number of jets in the final state,
and the cut takes into account  the high multiplicity of the signal events.
\item [(E3)]
To reduce the background from W pair production
for events with  missing momentum, 
a single constrained kinematic fit has been performed.
The inputs to the fit are the momenta of the lepton and the neutrino, 
taking the missing momentum to be the momentum of the neutrino, and the  
rest of the event reconstructed into 2 jets. The invariant mass is 
calculated (a) for the lepton and the  neutrino system and (b) for the two 
jet system, letting the masses of both systems be independent.
The reconstructed mass of at least one  system has to be outside a mass
window of  70~GeV$< m < 90$~GeV, or the probability for the fit has
to be less than 0.01.
\item [(E4)]
For the topologies with one 
charged lepton expected in 
the final state,
the background from W pair production is reduced further 
by a kinematic fit 
on the invariant mass of two pairs of jets, when reconstructing the 
whole event into 4 jets.
This kinematic fit assumes energy and momentum conservation and
the same mass for both jet pairs. 
The reconstructed mass has either to be outside a mass
window around the W mass, 
with a width varying between   8 and  20~GeV, 
depending on the signal to background ratio,  
or the probability for the fit has
to be less than 0.01.
\item [(E5)] For events with only one charged lepton expected from the
decay of the $\chipm$ or $\chin$, the momentum of the lepton has to be 
lower than 40~GeV to reduce the background from W pair production.
\item [(E6)] A certain number of identified leptons with a minimum energy is
required. 
For the indirect decay via the coupling \lb\ 
the requirement is at least two leptons with a
minimum energy of 10~GeV and 7~GeV for the most and second-most energetic,
respectively; 
for \lbp\ the number of leptons resulting from the 
$\chin$  decay, i.e. 0, 1, or 2.
In the direct decays via the  \lbp\ coupling 
also the number of charged leptons expected
(i.e. 1 or 2) is required.
\item [(E7)]
The identified leptons are required to be isolated.
The isolation criterion is that  
there be no charged track within a cone 
around the track of the lepton. 
If two leptons are required, both opening angles 
have to fulfil
$| \cos \theta | < $ 0.99;
if only one lepton is required, the opening angle 
has to fulfil
$| \cos \theta | < $ 0.98.
 
\end{description}

In the following it is described, which of the above cuts has been used 
for a given analysis. The number of observed and expected events, 
as well as the efficiencies for each analysis are listed in 
Table~\ref{tab:ivor_eff}.

\subsubsection{Indirect Decay via \lb\ }

Final states from the $\chipm\ $ decay via a W$^{(*)}$ can  have
jets and leptons in the final state. The signature of these final states
is at least four isolated leptons plus two or four jets, depending 
on whether one or both of the W$^{(*)}$'s decay hadronically.
If the mass difference 
between the $\chipm\ $ and the $\chin\ $ becomes small,
the jets might not be properly reconstructed.
Therefore the analysis is also sensitive to final states with at least 
two isolated leptons and some additional hadronic activity.

The same cuts are applied for any of the couplings \lb\ and also for 
the final states with two or four hadronic jets from the decays of the 
two W$^{(*)}$. The cuts {\bf (E3)}, {\bf (E4)}, and {\bf (E5)} have not 
been applied.

\subsubsection{Indirect Decay via \lbp\ }

In the indirect decay of $\chipm $  via \lbp\,
there are many different final state topologies possible.
The lepton from the \lbp\ decay of the $\chin $ can be either
charged or neutral. Therefore final states with 2, or 1 
charged leptons are analysed separately. 
The different decay modes of the W$^{(*)}$ are all covered with the same cuts.
Especially in the region of a small mass difference between the 
$\chipm$ and the $\chin$, all different topologies due
to different  decay modes of the W$^{(*)}$  look similar.

In the final states with two charged leptons from the $\chin$ decay, the cuts 
for electrons and muons are very similar.
The cuts {\bf (E3)}, and {\bf (E4)} have not 
been applied.   

In the selection for final states with one charged lepton, 
the cuts are the same for all three lepton flavours.



\subsubsection{Direct Decay via \lbp\ }

In the direct decay of $\chipm $ and $\chin $ via \lbp\ the final states 
consist of 2 leptons of the same flavour 
(neutral or charged) plus 4 hadronic jets.
The experimental signature is the same for $\chipm$ and for $\chin$.
Separate analyses have been performed for
final states with 1 or 2 charged leptons (electrons or muons).
In the case of 2 charged leptons, the cuts 
{\bf (E3)}, {\bf (E4)}, and {\bf (E5)} have not 
been applied.


\begin{figure}[p]
\begin{center}
\epsfig{file=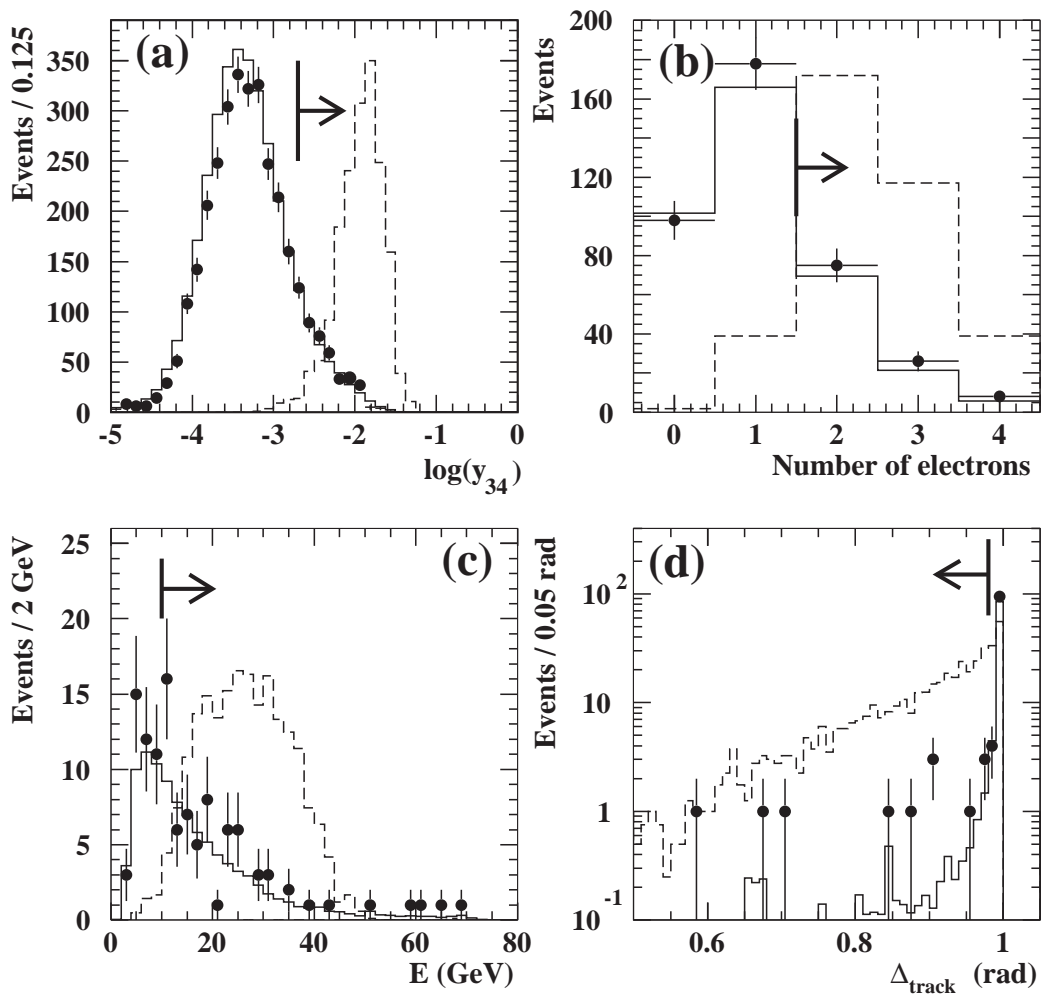,width=15.5cm} 
\end{center}
\caption{ \it This figure shows distributions 
of event variables from the analyses for final states with jets 
plus leptons 
for data (points) and MC 
(full histogram). Also indicated is a signal MC (dashed histogram) for
a $\chipm$ of mass 90~GeV, decaying via a \lbp\ coupling into final states 
with  two electrons and a hadronic decay of both W$^{(*)}$.
The scale of the  signal MC is arbitrary. 
The arrows point into the direction accepted by the cuts applied.
(a) The logarithm of the jet resolution, $y_{45}$, 
at which the number of reconstructed jets changes between 4 and 5, is shown,
after the cut on the visible energy  has been applied;
(b) the number of electrons after the  cut on the jet resolution, as 
indicated in Figure(a), 
has been applied. 
(c) The energy of the most energetic electron and
(d) the distance between the track of the most energetic electron 
and the nearest track, $\Delta_{\mathrm{track}}$, requiring at least two
identified electrons.
  }
\label{fig:ivor_mix}
\end{figure}

For the decay mode of a $\chipm$ via the coupling \lbp\ into a 
final state with two electrons the event distributions 
and several of the cuts used are shown in 
Figure~\ref{fig:ivor_mix}. The signal is shown for a $\chipm$ of 
90~GeV and the hadronic decay mode of the W$^{(*)}$'s, but the
cuts are the same for all decay modes of the W$^{(*)}$'s.
Therefore the data distributions do not depend on it.  

Figure~\ref{fig:ivor_mix}(a) shows the event 
distributions for the visible  energy, 
scaled by the centre of mass energy after the preselection cuts. 
The lower cut is chosen so that also the leptonic decay mode
of the W$^{(*)}$'s is selected.
Figure~\ref{fig:ivor_mix}(b) shows the jet 
resolution, $y_{4}$, where the number of jets
changes from 4 to 5 jets after the cut on the visible energy has been 
applied.
For Figures~\ref{fig:ivor_mix}(c) and (d) the cuts 
up to (E5) have been applied and show the
number of electrons and the distance of the most energetic electron 
with respect to the closest track, respectively.

\subsubsection{Efficiencies and Backgrounds}

The efficiencies resulting from the analyses above lie between
30 and 80\% for final states without at least one electron or muon
from the $\chin$ decay
for $\chipm$ masses around 90~GeV. For final states with taus the
efficiencies lie in the range between 10\% and 30\%.
The efficiencies are best for the topologies resulting from the decays
via couplings \lb\/, as many well isolated leptons are present.
Also the topologies with two electrons or two muons in the indirect decays
via \lbp\ have efficiencies above 50\% for $\chipm$ masses of 90~GeV.

\begin{table}[htb]
\begin{center}
\begin{tabular}{|l||c||c|c|}
\hline
Final State & Eff. (\%) & Selected Events & Tot. bkg MC  \\
\hline
\hline
 \lb\ indirect & 12 -- 82  & 1 & 3.3  \\
\hline
\hline
\hline
\multicolumn{4}{|l|}{\lbp\ indirect}  \\
\hline
$ee      + \ge$ 4 jets  & 23 -- 55  & 1 & 1.4  \\
$e \nu      + \ge$ 4 jets  & 3 -- 30  & 0 & 1.0  \\
$\mu \mu   + \ge$ 4 jets  & 27 -- 60  & 0 & 1.5  \\
$\mu \nu   + \ge$ 4 jets  & 3 -- 31  & 2 & 0.5  \\
$\tau \nu   + \ge$ 4 jets  & 1 -- 11  & 6 & 4.9  \\
\hline
\hline
\hline
\multicolumn{4}{|l|}{\lbp\ direct} \\
\hline
$ee     $ + 4 jets  & 24 -- 57  & 1 & 0.9  \\
$e \nu     $ + 4 jets   & 7 -- 39  & 1 & 2.0  \\
$\mu \mu     $ + 4 jets   & 30 -- 60  & 0 & 1.0  \\
$\mu \nu     $ + 4 jets   & 8 -- 46  & 4 & 1.4  \\
$\tau \nu     $ + 4 jets   & 2 -- 14  & 1 & 4.1  \\
\hline
\end{tabular}
\end{center}
\caption{\it 
Detection efficiencies (in \%)
for the final states considered for $\chargino$ 
masses varying between 45 and 90~GeV and for 
$\chin$ masses between 30 and 90~GeV. 
The number of events remaining after the selection cuts and the 
expected backgrounds from all Standard Model
processes considered are quoted. The main contribution to the total
background estimate derives 
from $\WW$ leptonic decays (4-fermion processes), while other processes
contribute less than 10\%.}
\label{tab:ivor_eff}
\end{table}

The background from two-photon processes is negligible, same as the 
background from multihadronic final states for most analyses. 
W pair production is the
major background. The number of expected background events is 
estimated to be between 0.5 and 2.0 for events with at least one  
electron or muon from the $\chin$ decay.  For events with taus from the
$\chin$ decay, the expected background lies in between 1.6 and 4.9 events, and
2.9 events are expected in the case of no charged lepton in the decay of
the $\chin$.

The numbers of events observed in the data 
show no significant discrepancy over the expected number of events.

\subsubsection{Systematic Errors}

For the lepton identification a systematic error of 4\% was estimated for the 
electrons, 3\% for the muons and 6\% for the taus. 
The systematic error on the measured luminosity is 0.35\%.
The systematic error due to the uncertainty in the trigger efficiency was 
estimated to be negligible, because of the requirement of at least seven good
tracks.
The statistical error on the determination of the efficiency from the MC
samples has also been treated as a systematic error.
To check the dependence on the quark flavour of the jets, samples with 
different quark flavours have been produced. 
The standard samples used to determine the efficiencies always 
resulted in the lowest 
efficiency. Therefore no additional error has been assigned due to the 
quark flavour.

\subsection{Jets Plus Two $\tau-$Lepton Final States}
\label{sec:susan}

Final states containing at least two $\tau$-leptons and between four and
eight jets
can be produced via the processes 
e$^+$e$^-\rightarrow\chipm\chipm$ or $\chin\chin$,
with $\chi\rightarrow
\tau + qq$.
In the case of direct decay, and $\chi^\pm\rightarrow\chi^0
f\bar{f^\prime}\rightarrow\tau +qq+ f\bar{f^\prime}$ in the case of
indirect decay, leading to the following four possible signal topologies:
\begin{itemize}
     \item Direct decay:
        \begin{description}
           \item[(a)] 
               four jets and two $\tau$-leptons  
        \end{description}
     \item Indirect decay:
        \begin{description}
           \item[(b)] 
               four jets, two $\tau$-leptons plus two additional
leptons (of any flavour) and their associated neutrinos 
           \item[(c)] 
               six jets, two $\tau$-leptons plus one additional
lepton (of any flavour) and its associated neutrino
           \item[(d)] 
               eight jets and two $\tau$-leptons  
        \end{description}
\end{itemize}

The backgrounds come predominantly from 
($Z\gamma)^*\rightarrow q\bar{q}(\gamma)$ and four-fermion processes.

In each of the four cases, the selection begins with the 
identification of $\tau$-lepton candidates~\cite{ref:smpaper}, using 
three algorithms 
designed to identify electronic,muonic and hadronic $\tau$-lepton decays.
The original $\tau$-lepton direction is approximated
by that of the visible decay products.
The following preselection,
was made:

\begin{description}
\item[(F1)]  
Events are required to contain at least nine charged tracks, and must have at 
least two $\tau$-lepton
candidates, each with electric charge $|q|=1$ and whose charges sum to zero.

\item[(F2)]
Events must have no more than a total of 20 GeV of energy
deposited in the forward detector, gamma-catcher, and 
silicon-tungsten luminosity monitor; a missing momentum vector
satisfying $|\cos\theta_{\rm miss}| < 0.97$,
total vector transverse momentum of at least 2\% of $\sqrt{s}$, and a scalar sum
of all track and cluster transverse momenta larger than 40~GeV.
  
\item[(F3)]
Events must contain at least
three jets, reconstructed using
the cone algorithm as in~\cite{ref:smpaper}\footnote{Here, single 
electrons and muons from $\tau$-lepton decays are allowed to be recognised 
as low-multiplicity ``jets''.}, and no energetic isolated photons\footnote{
An energetic 
isolated
photon is defined as an electromagnetic cluster
with energy larger than 15~GeV and no track within a cone
of $30^\circ$ half-angle.}.

\item[(F4)]
Events must contain no track or cluster with 
energy exceeding $0.3\sqrt{s}$. 
\end{description}

In order to select a final $\tau$-candidate pair for each
event, and to further suppress the remaining background,
a likelihood method similar to that described in~\cite{ref:mssmpaper}
is applied to those events passing the above preselection. 
For each $\tau$-candidate pair and its associated hadronic ``rest of the
event'' (RoE), composed of those tracks and clusters not having been
identified as belonging to the pair,
a joint discriminating variable, ${\cal L}$, is constructed 
using normalised reference distributions generated from Monte Carlo samples
of events belonging to the following four classes:

\begin{enumerate}
   \item Signal events where the selected pair is composed of two real 
$\tau$-leptons 
   \item Signal events where the selected pair contains at most one real 
$\tau$-lepton 
   \item SM four-fermion events where the selected pair is composed of up to 
two real $\tau$-leptons
   \item Events from the process 
$Z\gamma^*\rightarrow q\bar{q}(\gamma)$ containing no real $\tau$-leptons
\end{enumerate}

For classes 1 and 2, different signal reference distributions are generated
for the four topologies a)-d).
The variable ${\cal L}$ is related to the probability that the selected
$\tau$-candidate pair and RoE belong to class 1. The set of input variables
to the reference distributions includes those which characterise each of the
two $\tau$-lepton candidates individually, those which describe their behaviour
as a pair and those which characterise the RoE. For those variables
describing the $\tau$-candidates individually, 
separate reference distributions are generated for 
leptonic (electron or muon), 
hadronic 1-prong and hadronic 3-prong $\tau$-candidates, in order to exploit
the differences between the three categories,
as follows:
\begin{itemize}
 \item Used for all three categories:
     \begin{itemize}
          \item $|\cos\alpha_i|$, where $\alpha_{i}$ is the
angle between the direction of the $i$-th $\tau$ candidate and that of the
nearest track not associated with it.
         \item $|\vec{p}_i|$, the momentum of this nearest non-associated track
         \item  $R_{em}^{11/30}\cdot R_{cd}^{11/30}$, where 
$R_{em(cd)}^{11/30}$ is the ratio of the electromagnetic cluster
energies (charged track momenta) within a cone of $11^\circ$ half-angle
centred on the $\tau$ candidate axis
to that within a $30^\circ$ half-angle cone.
         \item $|\vec{p}_\tau|$, the magnitude of the momentum of the 
$\tau$-candidate.
         \item The type of $\tau$ candidate itself, i.e. lepton,
1-prong, 3-prong. 
         \end{itemize}
  \item Used for 1- and 3-prong categories: 
       \begin{itemize}
           \item $\sum_{i}|p_T|_{i,\tau}/E_{\tau}$, the sum of the
transverse momenta (with respect to the $\tau$-candidate axis)
of tracks $i$ in the $\tau$-candidate
            \item $m_{\tau}$, the invariant mass of the $\tau$ candidate
          \end{itemize}
    \item Used for lepton and 1-prong categories:
            \begin{itemize}
        \item The magnitude of the impact parameter, in three dimensions,
of the $\tau$-candidate
                 \end{itemize}
     \item Used for 1-prong category only:
         \begin{itemize}
             \item $N_{hd}/N_{em}$, the ratio of hadronic calorimeter
to electromagnetic calorimeter clusters associated to the $\tau$ candidate.
         \end{itemize}  
     \item Used for 3-prong category only:
         \begin{itemize}
                \item The vertex significance in three dimensions of the
$\tau$-candidate.
         \end{itemize}
\end{itemize}

\begin{figure}[htbp]
\centerline{\hfill
\epsfig{file=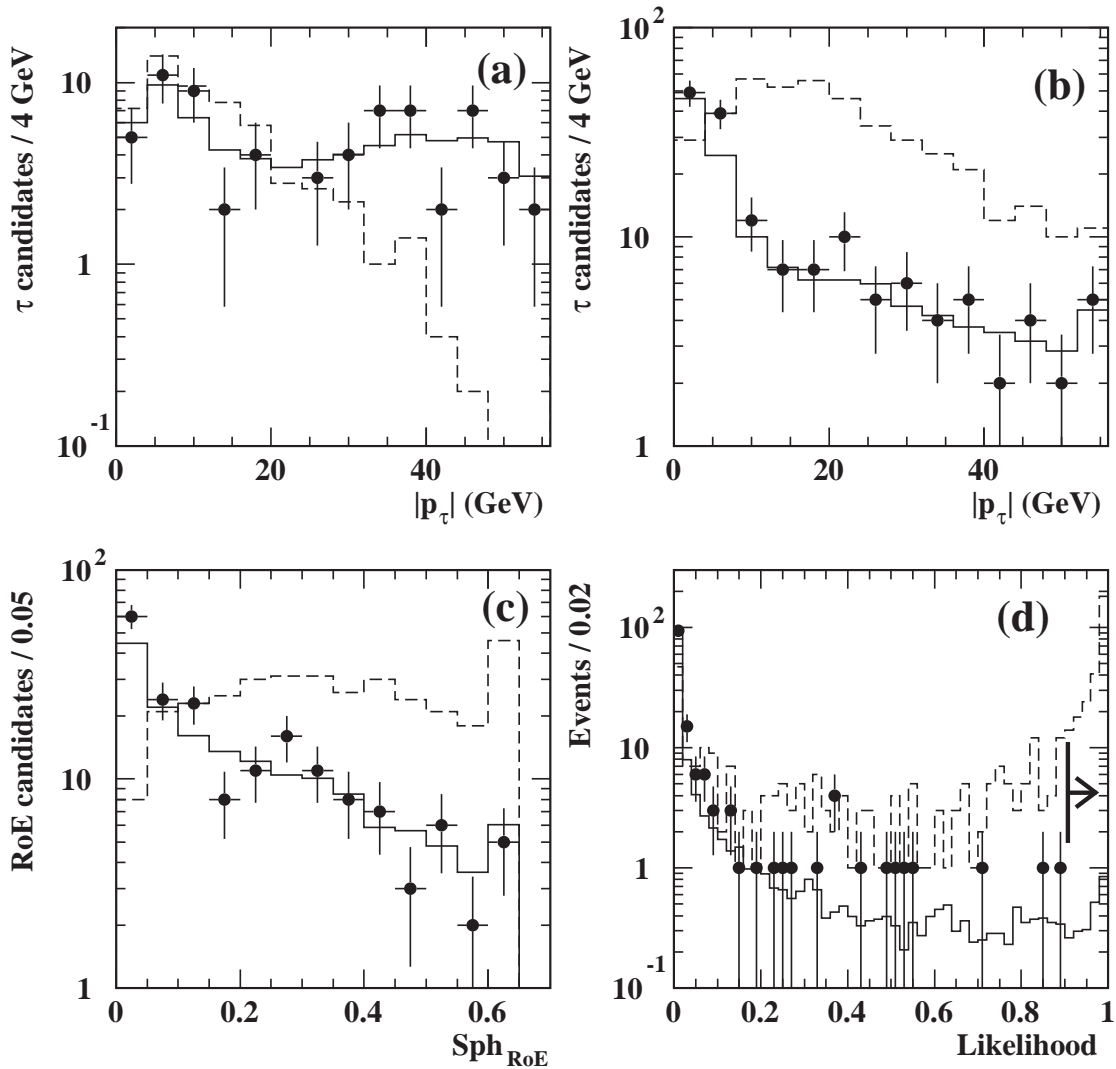,height=15.5cm}\hfill
}
\caption[]{\label{fig:twotaufourjet}\sl
         Jets plus at least two $\tau$-leptons, topology a): Distributions 
of relevant
quantities for data(points), estimated SM background (full histogram)
normalised to the integrated luminosity of the data, and a simulated signal
(dashed histogram, arbitrary normalisation) corresponding to 
$m_{\chi^{\pm}}=70$~GeV (direct decay). Figures (a),(b) and (c) show some 
of the variables 
input
to the likelihood: (a) The momentum of leptonic $\tau$ candidates;
(b)The momentum of 1-prong hadronic $\tau$ candidates; (c) The sphericity
of the hadronic RoE. The likelihood distribution  ${\cal L}$ is shown in (d);
events to the
right of the arrow indicating the cut position are accepted. All 
distributions are after the imposition
of cut (F4).      
}
\end{figure}

The following five input variables are used to 
characterise the $\tau$-candidates
as a pair or the RoE associated 
with the pair:
\begin{itemize}
  \item  The angle between the two $\tau$-candidates 
  \item The sphericity of the RoE
  \item  The sum of the number of charged tracks and 
electromagnetic calorimeter clusters
in the RoE
   \item The number of hadronic calorimeter clusters in the RoE
    \item The jet resolution parameter $y_{23}^{RoE}$, at which the 
number of jets in the RoE changes from 2 to 3.
\end{itemize}

Distributions of some of the input variables as well as that of 
${\cal L}$ are shown in 
Fig.~\ref{fig:twotaufourjet}, for the case of topology a).

The $\tau$-candidate pair having the highest value of ${\cal L}$
is chosen in each event.
Then, for topology b) only, the following requirement is imposed reflecting
the expectation of two additional leptons other than the two $\tau$-leptons:
\begin{description}
\item[(F5)] The sum of the number of $\tau$-lepton candidates plus the number
of identified electrons 
(as in~\cite{ref:smpaper}) 
and muons not tagged as 
$\tau$-lepton 
candidates must be at least 4. 
\end{description}
Finally, the following
requirements are made on the values of ${\cal L}$:

\begin{description}
\item[(F6)] ${\cal L}>0.9, 0.6, 0.65$ and $0.75$ respectively for signal
topologies a), b), c), d).
\end{description}

For topologies a)-d) respectively, zero, two, one and one 
events survive the selection while the background is 
estimated to be
2.27, 2.31, 3.19 and 1.93 events for an integrated luminosity of 
55.8 pb$^{-1}$.
The detection efficiencies for 
chargino masses between 70 and 90 GeV range from approximately 24 to 28\%,
21 to 32\%, 22 to 24\%, and 14 to 19\%,
while those for neutralinos in the same mass range lie between 
17 and 18\%. For charginos with masses of 45 GeV and below,
the detection efficiency
falls to approximately 5\%, 9\%, 5\% and 1\%, and for neutralinos
to (8\%).

These efficiencies are affected by the following
uncertainties:
Monte Carlo statistics, typically 5.0\%;
uncertainty in the tau-lepton preselection efficiency, 1.2\%;
uncertainty in the modelling of the other preselection variables, 2.0\%;
uncertainties in the modelling of the likelihood input variables, 10.0\%;
uncertainties in the modelling of fragmentation and hadronisation, 6.0\%;
and uncertainty on the integrated luminosity, 0.5\%~\cite{ref:lumino}.
Taking these uncertainties as independent and adding them in quadrature
results in a total systematic uncertainty  of 12.9\%  (relative errors).
The uncertainty in the number of expected background events was 
estimated to be 18\%.

\subsection{Four Jets plus Missing Energy }
\label{sec:gabi}

Direct decays of charginos and neutralinos via $\lambda^\prime$ coupling
can lead to final states with four jets and missing energy
due to the two undetected neutrinos. The dominant backgrounds come from
four-fermion processes and radiative or mismeasured two-fermion events.
The selection procedure is described below:

\begin{description}
\item[(G0)] The event has to be classified as multi-hadron final-state as 
described in \cite{LEP2MH}.
\item[(G1)] The visible energy of the event is required to be less than 
0.85$\sqrt{s}$.   
\item[(G2)] To reject two-photon and radiative two-fermion events
the transverse momentum should be larger than 10 GeV, 
the total energy measured in the forward calorimeter, gamma-catcher 
and silicon tungsten calorimeter should be less than 20 GeV, 
and the missing momentum should not point to the beam direction,
$| \cos \theta_{\mathrm{miss}}| <$ 0.96.   
\item[(G3)] The events are forced into four jets using the Durham 
jet-finding algorithm, and rejected if the jet resolution parameter 
$y_{34}$ is less than 0.001. 
\item[(G4)] An additional cut is applied 
against semi-leptonic four-fermion events, vetoing on  
isolated leptons being present in the event.
The lepton identification is
based on an Artificial Neural Network routine~\cite{tauID},
which was 
originally written to identify
tau leptons but 
is 
efficient for electrons and muons, as well. 
The ANN output is required to be larger than
0.97 for lepton candidates.
\item[(G5)] Finally, a likelihood selection
is employed to classify the remaining events
as two-fermion, four-fermion or qqqq$\nu\nu$ processes.
The method is described in \cite{ref:mssmpaper}.
The information of the following variables are combined:
\begin{itemize}
\item the effective centre-of-mass energy~\cite{sprime} of the event;
\item the transverse momentum of the event;
\item the cosine of the polar angle of the missing momentum vector;
\item the D parameter~\cite{dpar} of the event;
\item the logarithm of the $y_{34}$ parameter;
\item the minimum number of charged tracks in a jet;
\item the minimum number of electromagnetic clusters in a jet;
\item the highest track momentum;
\item the highest electromagnetic cluster energy;
\item the number of leptons in the event, using loose selection criteria 
for the lepton candidates (the ANN output is larger than 0.5);
\item the mass of the event excluding the best lepton candidate (if any)
after a kinematic fit using the W$^+$W$^-$ $\rightarrow$ qq$l\nu$ hypothesis;
\item the cosine of the smallest jet opening angle, defined by the half-
angle of the smallest cone containing 68\% of the jet energy; 
\end{itemize}
The event is rejected if its likelihood output is less than 0.95.
\end{description}

\begin{figure}
\begin{center}
\epsfig{file=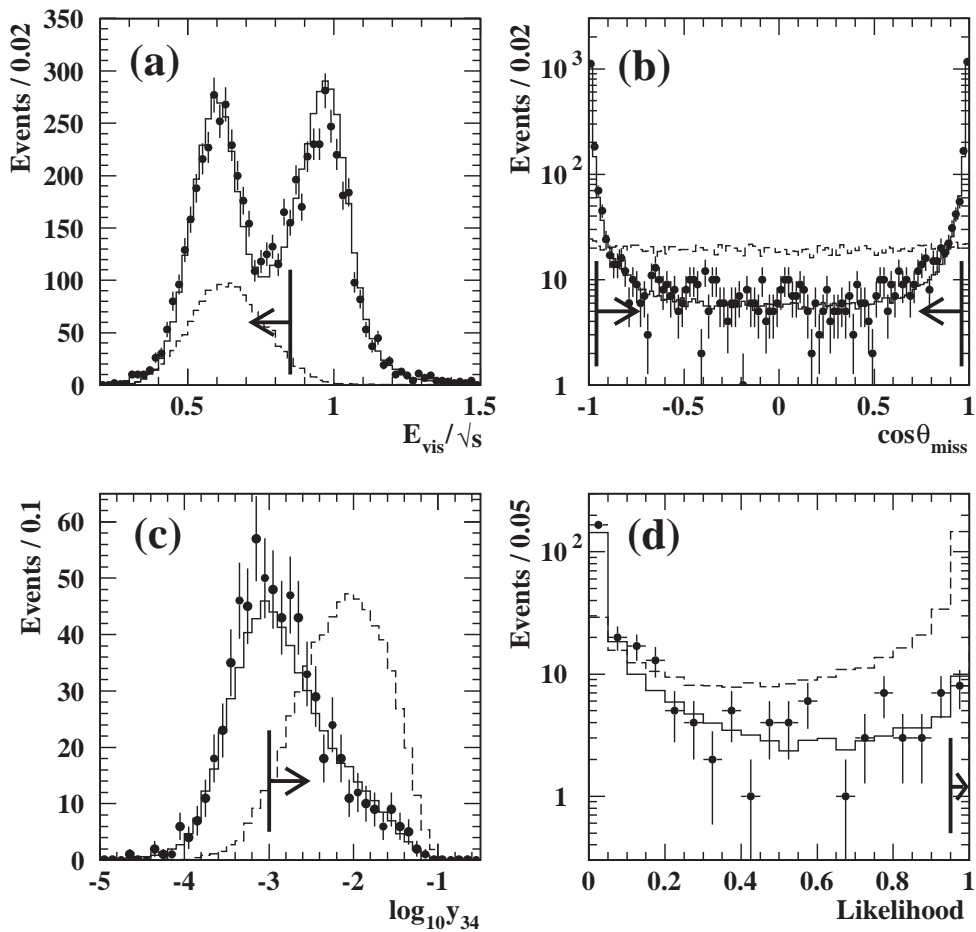,width=14.5cm} 
\end{center}
\caption{ \it Four jets plus missing energy: 
Distributions for data (points), for the estimated SM background  
(full histogram) and  for simulated signal (dotted histogram).
Figure (a) shows the visible energy, 
E$_{vis}$, scaled by the centre of mass energy, $\sqrt{s}$, for multihadron
events after cut (G0). 
In Figure (b) the cosine of the polar angle of the missing momentum
vector is plotted after cut (G1). 
In Figure (c) the logarithm of the jet resolution, $y_{34}$, 
at which the number of reconstructed jets changes between 3 and 4, is shown
after cut (G2) have been applied.
Figure (d) shows the final selection using the likelihood output.
The arrows indicate the accepted regions in each plot.
The SM background is normalised to the integrated luminosity of the
data, while the normalisation of the  signal distribution is arbitrary. 
}
\label{fig:gabi_fig}
\end{figure}

Figure~\ref{fig:gabi_fig} shows experimental plots of the selection
variables for the data, the estimated background and simulated signal
events. The distributions are well described by the Monte Carlo simulation.

After all cuts, 8 events are selected in the data sample, while 
9.47$ \pm$ 0.33 (stat) $\pm$ 2.07 (syst) events 
are expected from Standard Model processes, of which 72\% originates 
from four-fermion processes. The signal detection  efficiency varies 
between 7\% and 60\% for gaugino masses of 45 -- 90 GeV
for $\lambda^\prime_{121}$ and $\lambda^\prime_{123}$ couplings.

The small efficiency for light gaugino masses is the result of 
initial-state radiation and the larger boost  of the particles, which make the
event similar to the QCD two-fermion background.

The background expectation is subject to the following systematic errors
and inefficiencies:
inefficiency due to the forward energy veto (1.8\%);
the statistical error due to the limited number of Monte Carlo events
(3.4\%); 
the statistical and systematic error on the luminosity measurement
(0.45\% in total); 
error on the lepton veto (1\%);
uncertainty on modelling the SM background processes by comparing
different event generators (3.3\%) 
and the modelling of kinematic variables used in the analysis (21\%,
dominated by the error on the visible energy). 

The signal detection efficiency is affected by the following systematics:
inefficiency due to the variation of m$_0$ (0 -- 5\%) and due to the
forward energy veto (1.8\%); the statistical error due to the limited
number of Monte Carlo events (2 -- 12\%);
error on the lepton veto (1\%);
uncertainty on modelling the kinematic variables used in the
analysis (6\%).

\subsection{More than Four Jets plus Missing Energy }
\label{sec:mar1}

This analysis applies to chargino indirect decays via \lbp, where both
neutralinos decay into quarks and neutrinos.
The selected events must have clear reconstructed jets, missing energy and
missing transverse momentum. To
account for the possibility of leptonic decays of the W$^{(*)}$, the presence of
charged leptons in the events has to be allowed; an upper bound on the
lepton momentum is imposed, in order to reduce the background from semi-leptonic
W-pair events.

The total integrated luminosity amounts to 56.5~pb$^{-1}$. The selection cuts 
are
described below. An event is retained as a candidate if it satisfies the {\bf
Preselection} and any of the requirements {\bf (H1)}, {\bf (H2)} or
{\bf (H3)}.
\begin{description}
\item[Preselection] 
    Events have to be classified as multi-hadron final states as 
    described in \cite{LEP2MH}.
    The visible energy $E_{\mathrm{vis}}$, scaled by the centre-of-mass energy 
    must be in the range $0.4<E_{\mathrm{vis}}/\sqrt{s}<0.9$.
    The most energetic identified lepton (e or $\mu$) must have a momentum
    lower than 25~GeV.
    The number of tracks plus the number of EM clusters must exceed 60.
\item[(H1)]
    The jet resolution $y_{34}$, at which the event switches from 3 to 4 jets,
    must be $y_{34}>0.02$. The missing momentum $p_t$ of the event, scaled by
    the visible energy, must satisfy $p_t/E_{\mathrm{vis}}>0.15$.
\item[(H2)]
    To increase the efficiency for small $\Delta m$ values, where the missing 
    energy is
    larger but some jets are softer, $y_{34}>0.015$ and 
    $p_t/E_{\mathrm{vis}}>0.2$ are
    required.
\item[(H3)]
    To improve the efficiency for $m_{\chipm}=45$~GeV, where many events are 
    affected
    by initial state radiation and have therefore smaller visible energy and
    softer jets, $E_{\mathrm{vis}}/\sqrt{s}<0.7$, $y_{34}>0.01$ and 
    $p_t/E_{\mathrm{vis}}>0.13$
    are required.
\end{description}

\begin{figure}[p]
\begin{center}
\epsfig{file=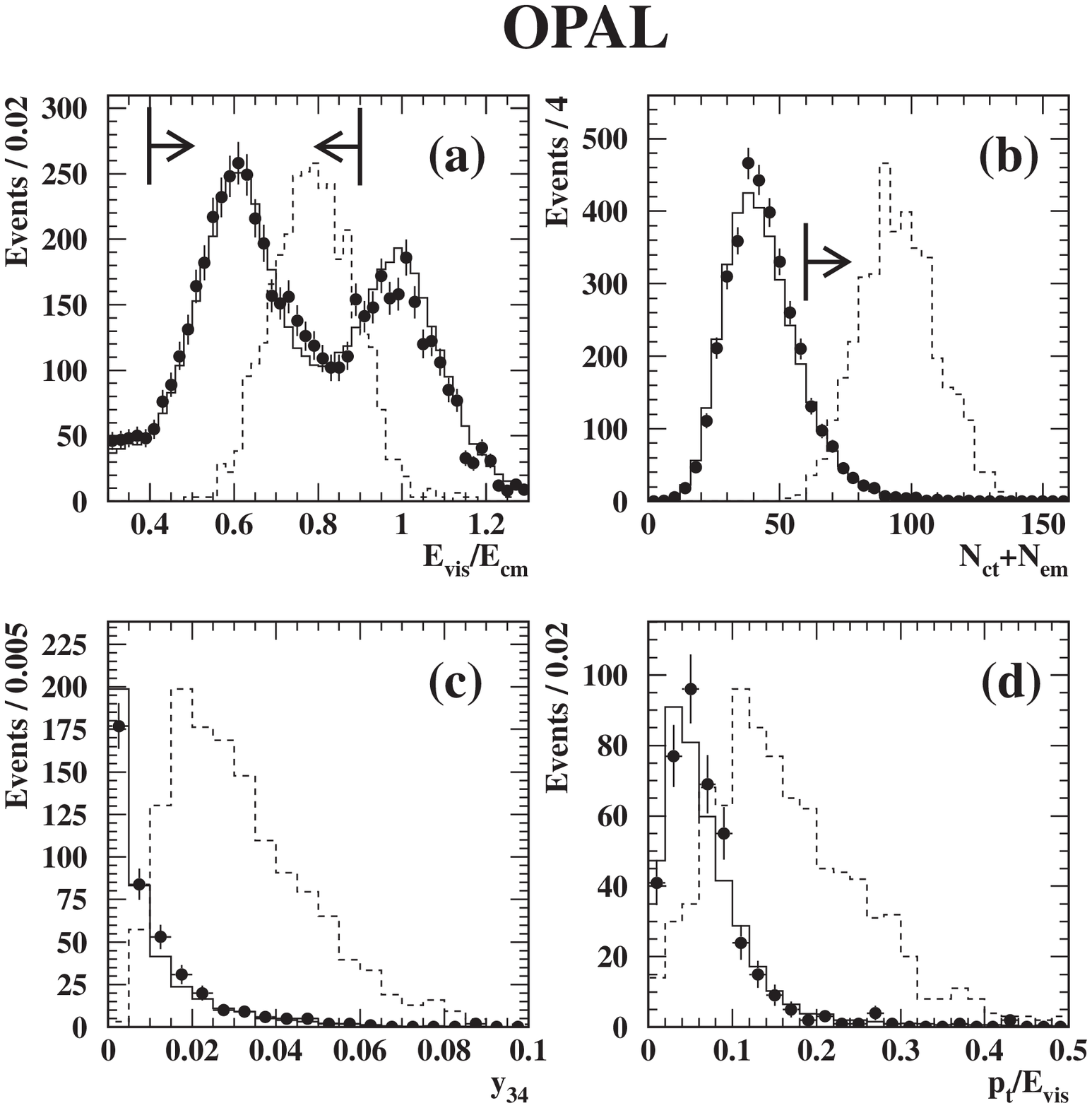,width=15.5cm} 
\end{center}
\caption{ \it Distributions 
of event variables for final states with 
more than four jets and missing energy,
for data (points) and MC 
(full histogram). Also indicated is a signal MC (dashed histogram) for
$\chipm$ pairs of mass 90~GeV and $\Delta m=45$~GeV. Both charginos are assumed
to decay indirectly via a \lbp\ coupling into $\nu$qq, and both W$^{(*)}$ into
hadrons, leading to 
qq~qq~$\nu$qq~$\nu$qq final states.
The scale of the  signal MC is arbitrary. 
(a) Visible energy scaled by the centre-of-mass energy, 
$E_{\mathrm{vis}}/\sqrt{s}$,
for events selected as multi-hadron final states. The lower cut rejects 
$\gamma\gamma$ events. 
(b) The number of charged tracks $N_{ct}$ plus the number of EM clusters
$N_{em}$, after all the other cuts of the preselection.
(c) The jet resolution parameter $y_{34}$ at which the number of reconstructed
jets switches between 3 and 4, plotted after the preselection cuts.
(d) The event transverse momentum $p_t$ scaled by the visible energy 
$E_{\mathrm{vis}}$,
plotted after the preselection cuts.
In Figures~(a) and (b) the arrows point to the region accepted by the 
applied cuts.
  }
\label{fig:marcello}
\end{figure}

The distributions of the selection variables are shown in 
Figure~\ref{fig:marcello} for experimental data, Standard Model background 
and signal.
After the full selection 7 events survive, where the expected background
from Standard Model is 10.9 events. The efficiency is in the range 6\%~--~33\%.

The systematic error due to the Monte~Carlo statistics is less than 1.5\%; 
the systematic error on the collected luminosity is 0.35\%; the systematic error
due to the trigger efficiency is assumed to be negligible, due to the large
track and cluster multiplicity required.
The total systematic error due to the applied cuts is 2.3\%, where the most 
relevant components arise from the cuts on $y_{34}$ (2.2\%) and 
$p_t/E_{\mathrm{vis}}$ (0.6\%).

The sensitivity of the selection to quark flavours has been studied, showing 
that final states containing heavy quarks yield larger efficiencies; 
conservatively, the quark flavours yielding the lowest efficiencies have been 
considered to evaluate limits.
The variation in the efficiency due to variations in $m_0$ and $\Delta m$
has been studied and the lowest efficiency has been used.

\subsection{More than Four Jets and No Missing Energy}
\label{sec:mar2}

This analysis applies to chargino direct and indirect decays 
via \lbpp. Events are expected to have at least six quarks in the final 
state. The event thrust and 
sphericity are also used to reduce the background.

\begin{figure}[p]
\begin{center}
\epsfig{file=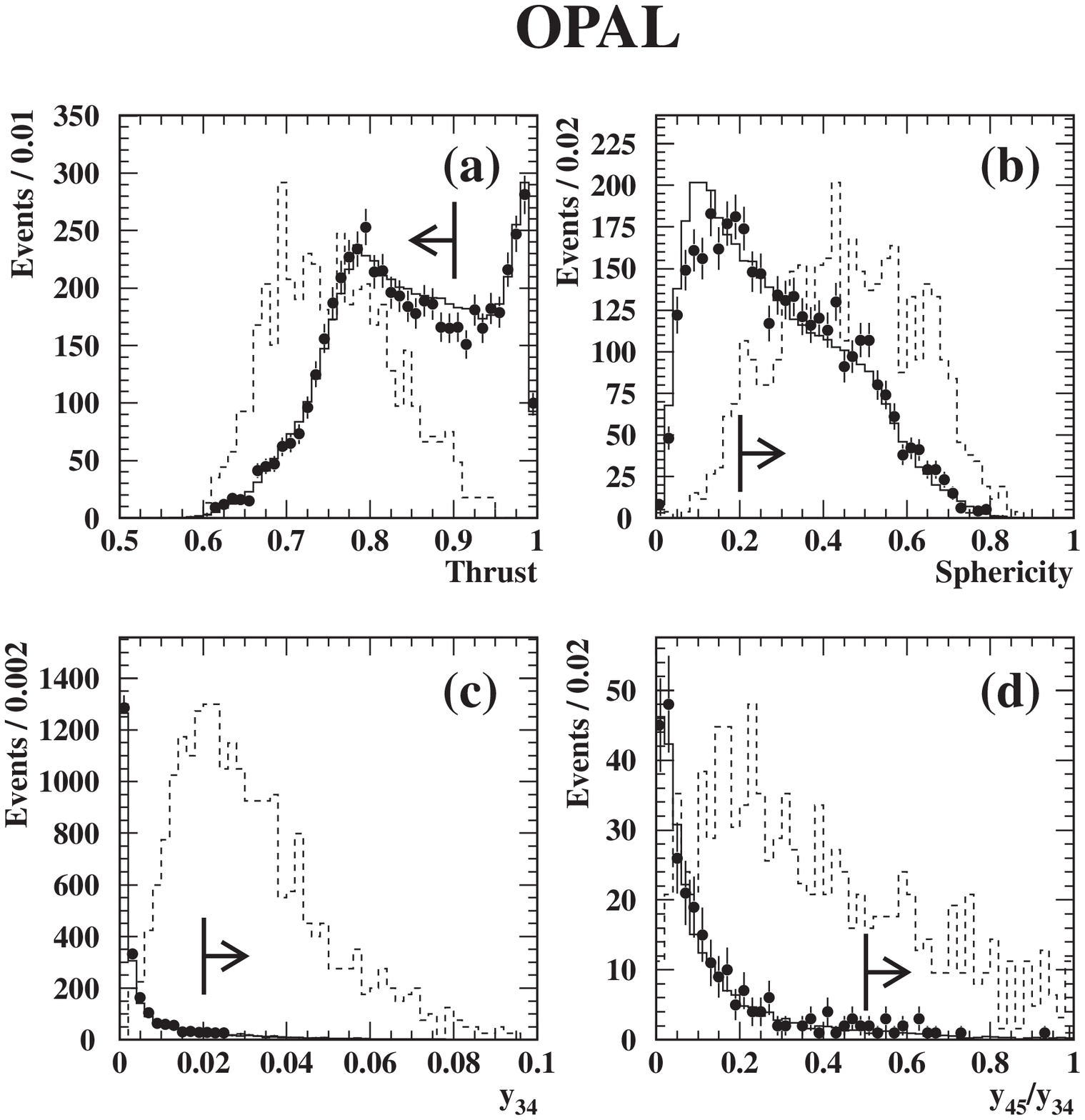,width=15.5cm} 
\end{center}
\caption{ \it Distributions 
of event variables for final states with 
more than four jets,
for data (points) and MC 
(full histogram). Also indicated is a signal MC (dashed histogram) for
$\chipm$ pairs of mass 90~GeV, decaying directly via \lbpp\ coupling into 
$\mathrm{qqq~qqq}$ final states.
The scale of the  signal MC is arbitrary. 
(a) Thrust distribution, plotted after the preselection cuts.
(b) Sphericity distribution, plotted after the cut on the thrust.
(c) Distribution of the jet resolution parameter $y_{34}$ 
at which the number of reconstructed
jets switches between 3 and 4, plotted after the cut on the sphericity.
(d) Distribution of $y_{45}/y_{34}$, plotted after the cut on $y_{34}$.
In each figure, the arrows point to the region accepted by the 
applied cut.
  }
\label{fig:marcello_2}
\end{figure}

The total collected luminosity amounts to 56.5~pb$^{-1}$. 
The selection cuts are
described below. 
An event is retained as a candidate if it satisfies the {\bf
Preselection} and any of the two requirements {\bf (I1)} or {\bf (I2)}.
\begin{description}
    \item[Preselection] 
        Events have to be classified as multi-hadron final states as 
        described in \cite{LEP2MH}.
        The visible energy $E_{\mathrm{vis}}$, scaled by the centre-of-mass 
        energy, 
        must be in the range $E_{\mathrm{vis}}/\sqrt{s}>0.4$.
    \item[(I1)] To reduce the $\qq$ background, an event 
                $\mathrm{thrust}$ less than 0.9
                and a $\mathrm{sphericity}$ larger than 0.2 are required. 
		To reduce $\qq$,
                $\qq\ell\nu$ events, $y_{34}>0.02$ is required. 
                To reduce $\qq\qq$ events,
                $y_{45}/y_{34}>0.5$ is required.
    \item[(I2)] To reduce 4-fermion events and part of two-jet $\qq$
                events, $0.8<\mathrm{thrust}<0.9$ is required. To reduce $\qq$
                events, $y_{56}>0.003$ is required.
\end{description}
        The condition {\bf (I1)} is optimised for large chargino masses, but
        becomes inefficient for smaller masses, where 
        the chargino decay products are very boosted, and jets cannot 
        be easily resolved. The condition {\bf (I2)} recovers efficiency in this
        latter case, exploiting the sharper thrust distribution, due to the
        large boots.

\begin{figure}[tb]
\begin{center}
\epsfig{file=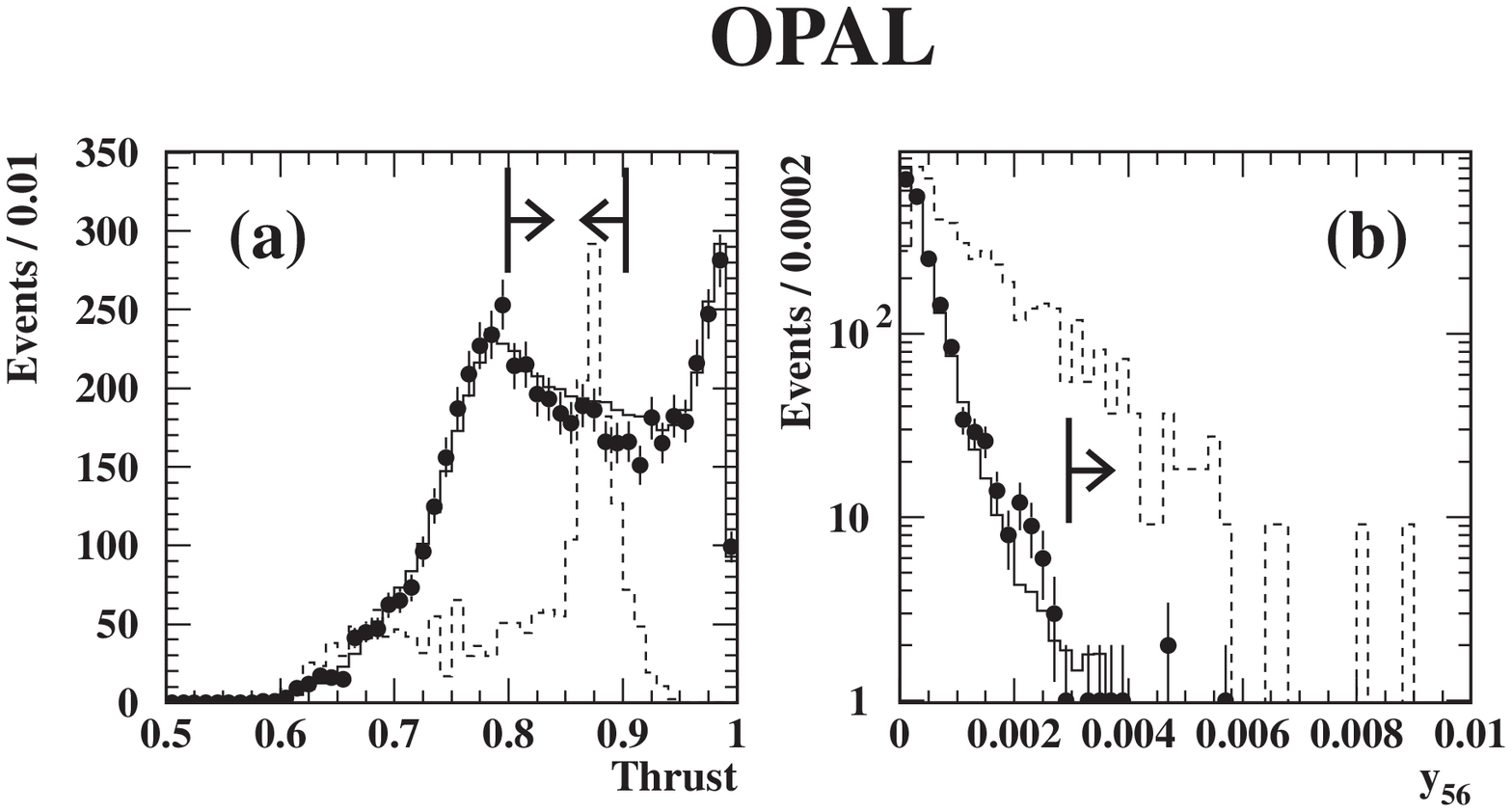,width=15.5cm} 
\end{center}
\caption{ \it Distributions 
of event variables for final states with 
more than four jets,
for data (points) and MC 
(full histogram). Also indicated is a signal MC (dashed histogram) for
$\chipm$ pairs of mass 45~GeV, decaying directly via \lbpp coupling into 
$\mathrm{qqq~qqq}$ final states.
The scale of the  signal MC is arbitrary. 
(a) Thrust distribution, plotted after the preselection cuts.
(b) Distribution of the jet resolution parameter $y_{56}$ 
at which the number of reconstructed
jets switches between 5 and 6, plotted after the cut on the thrust.
In each figure, the arrows point to the region accepted by the 
applied cut.
  }
\label{fig:marcello_3}
\end{figure}

The distributions of the selection variables are shown in 
Figures~\ref{fig:marcello_2} and \ref{fig:marcello_3}, 
for experimental data, Standard Model background 
and signal.
After the selection, 24 events survive, where the expected background
from Standard Model is 22.4 events. The efficiency is in the range 
9\%~--~23\%.

The systematic error due to the Monte~Carlo statistics is less than 1.4\%; 
the systematic error on the integrated luminosity is 0.35\%; 
the systematic error
due to the trigger efficiency is assumed to be negligible, due to the large
track and cluster multiplicity required.

The variation in the efficiency due to variations in $m_0$ and $\Delta m$ 
has been studied and the lowest efficiency has been used.
The sensitivity of the selection to quark flavours has been studied, showing 
that final states containing heavy quarks yield larger efficiencies; 
conservatively, the quark flavours yielding the lowest efficiencies have been 
considered to evaluate limits.

The total systematic error due to the applied cuts is 4.0\%, where the 
most relevant components arise from the cuts on $y_{34}$ (2.2\%) and 
$y_{45}/y_{34}$ (3.3\%). The systematic error due to the incorrect 
simulation of the parton shower for quark triplets,
such as those originating from gaugino decays, is estimated to be 1.2\%.

\section{Limits on Topological Cross-Sections}

In this chapter the results from the individual topological analyses
presented in the previous chapter are given.
As for each topological search the observation is in good agreement with the 
\sm\ expectations, 
there is no claim for a signal, and 
95\% confidence level (CL)
cross-section upper limits 
are presented for final state topologies expected from \Rparity\
violating $\chipm$ and $\chin$ decays.

\subsection{Multilepton Final States}

\begin{figure}[p]
\begin{center}
\epsfig{file=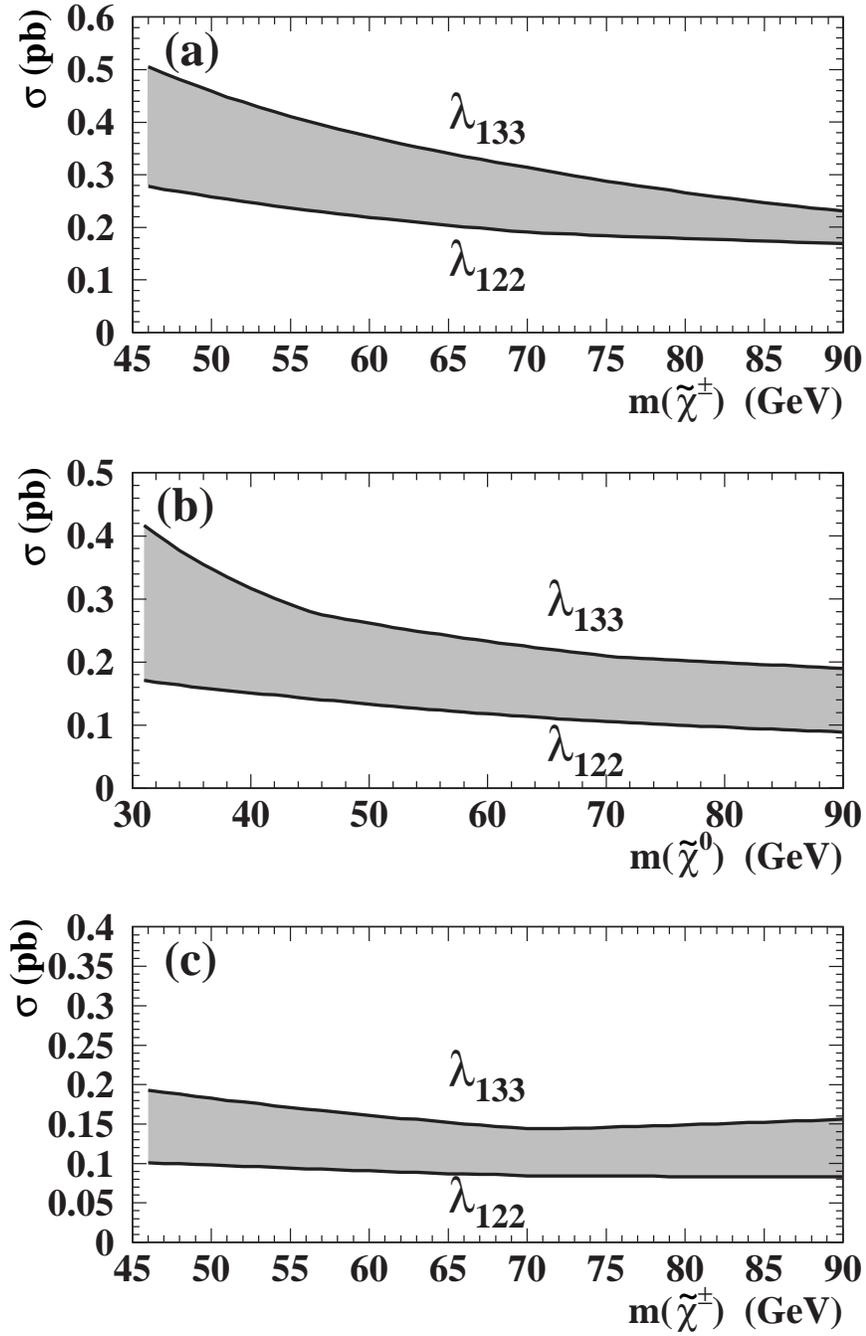,width=12.5cm} 
\caption{ \it Upper limits at 95\% CL on the cross-sections for final states
with  (a) 2 leptons, (b) 4 leptons, and (c) 
6 leptons. 
For each curve the coupling \lb\ that was assumed 
to be different from zero is given.
Limits arising from any other coupling \lb\ different from zero 
lie in the band between the two curves.
}
\label{fig:cross_leptons}
\end{center}
\end{figure}

Figure~\ref{fig:cross_leptons} 
shows the upper limits obtained
for the cross-section times branching ratio
of leptonic final states from the selection described in 
Section~\ref{sec:multileptons}.
In each of the 
Figures~\ref{fig:cross_leptons}(a), (b), (c), 
corresponding to two-, four-, and six-lepton final states,
the two curves represent the cross-section limits for lepton
flavour mixtures yielding the most and the least stringent limits;
the latter usually corresponds to several taus in the final state.
The limits corresponding to any other lepton flavour mixture are located
within the band between the two curves. The specific couplings 
$\lambda_{ijk}$ which lead to the limiting curves are indicated in each case.

\subsection{Final States with Leptons plus Jets}

\begin{figure}[p]
\begin{center}
\epsfig{file=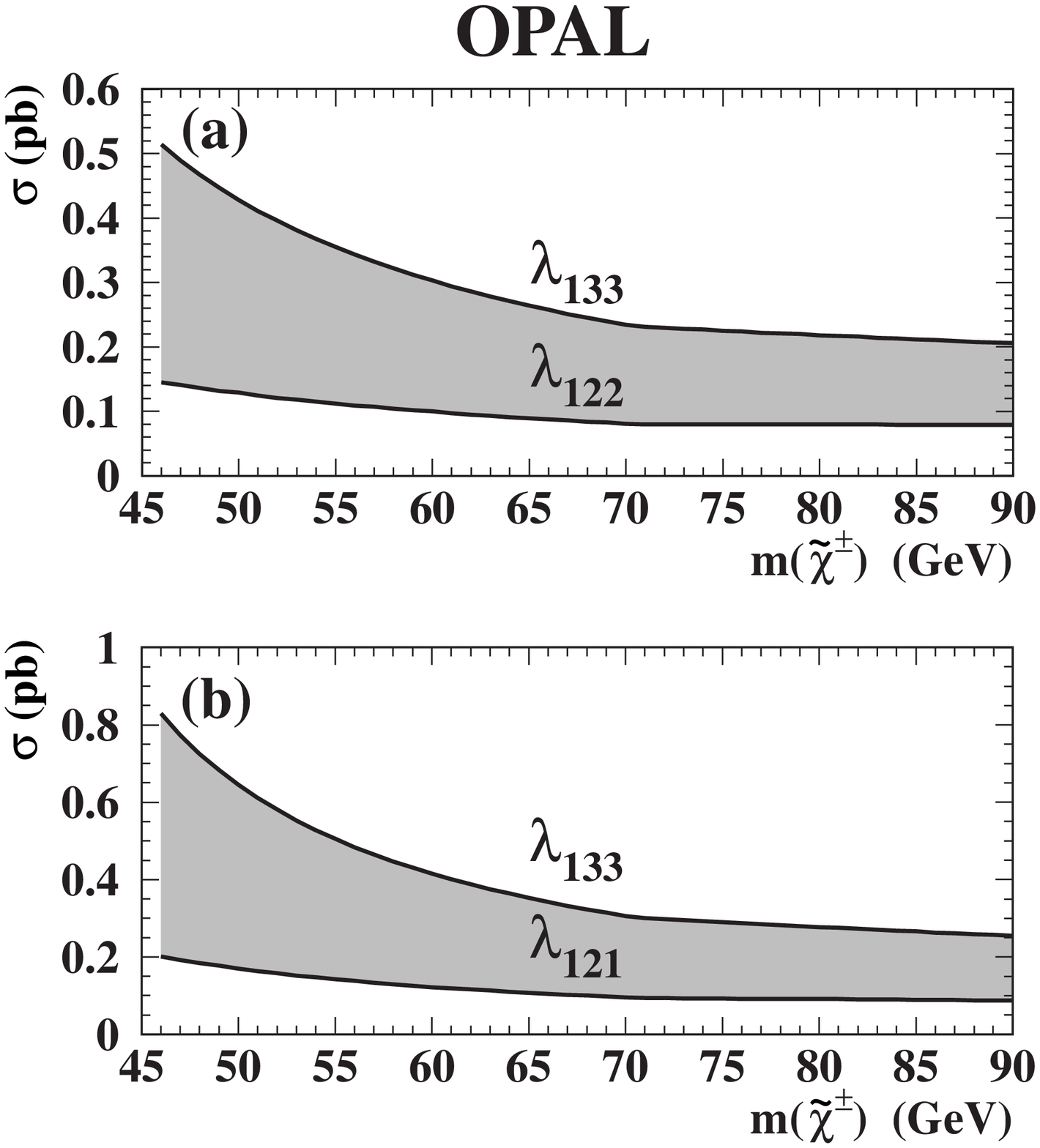,width=12.5cm} 
\caption{ \it Upper limits at 95\% CL on the cross-sections for final states 
with jets 
and leptons resulting from the  indirect decay via a coupling \lb\/. 
In the upper plot one \Wstar\ is decaying leptonically and the other 
hadronically, while in the lower plot both decay hadronically.
For each curve, the coupling \lb\ that was assumed 
to be different from zero is indicated.
Limits arising from any other coupling \lb\ different from zero 
lie in the gray zone between the two curves shown.
}
\label{fig:cross_mix1}
\end{center}
\end{figure}

\begin{figure}[p]
\begin{center}
\epsfig{file=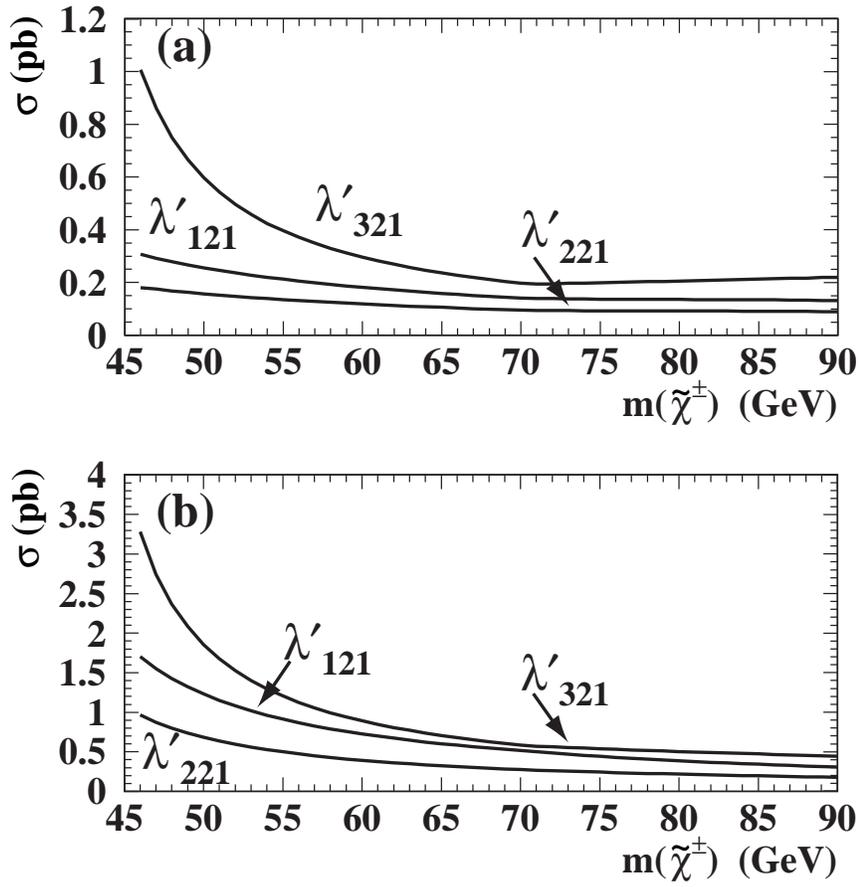,width=12.5cm} 
\caption{ \it Upper limits at 95\% CL
on the cross-sections for final states with jets 
and leptons resulting from the direct decay via a coupling \lbp\
for (a) two lepton plus four jets final states and 
(b) for one lepton plus four jets final states.
For each curve the coupling \lbp\ that was assumed 
to be different from zero is given.
Limits arising from any other  coupling \lbp\ lie below the limit 
shown for the coupling with the same first index.
}
\label{fig:cross_mix2}
\end{center}
\end{figure}

\begin{figure}[p]
\begin{center}
\epsfig{file=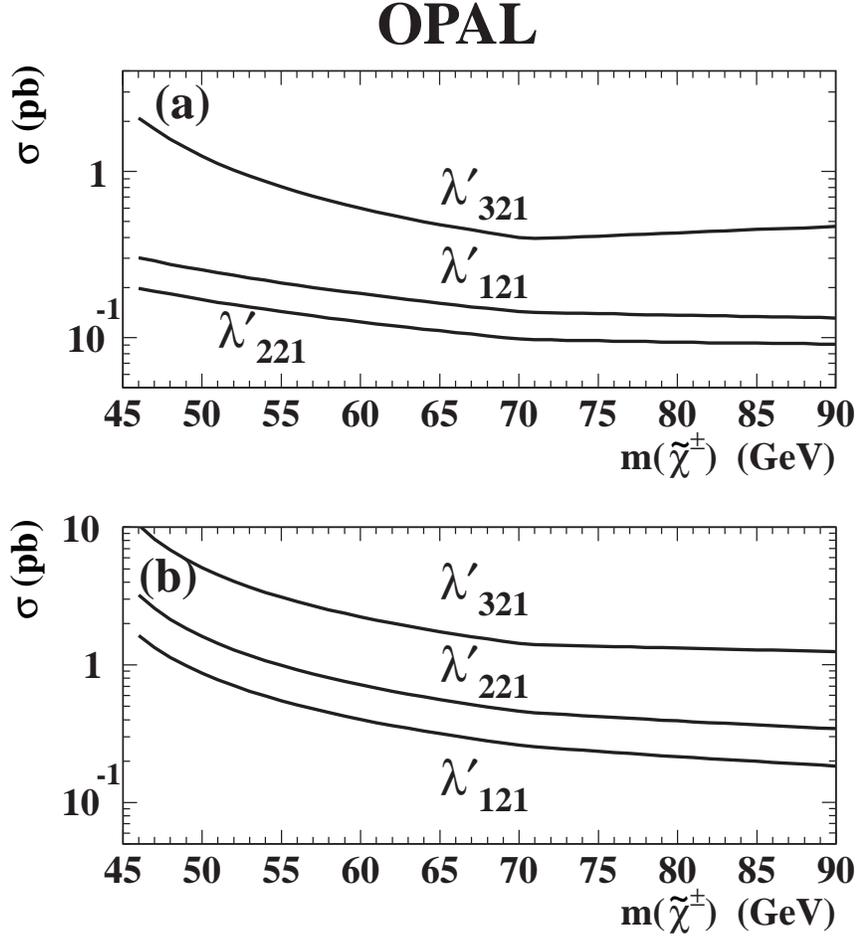,width=12.5cm} 
\caption{ \it Upper limits at 95\% CL
on the cross-sections for final states with jets 
and leptons resulting from the  indirect decay via a coupling \lbp\/. 
In Figure (a) final states with two charged leptons of the same flavour 
are shown. Figure (b) shows the final states with one charged lepton. 
Each Figure contains all decay modes of the two \Wstar.
For each curve the coupling \lbp\ that was assumed 
to be different from zero is given.
Limits arising from any other coupling \lbp\ lie below the limit 
shown for the coupling with the same first index.
}
\label{fig:cross_mix3}
\end{center}
\end{figure}

Figures~\ref{fig:cross_mix1}, 
\ref{fig:cross_mix2}, 
and \ref{fig:cross_mix3} show the cross-section limits for final
states with leptons plus jets from the selections described in 
Sections~\ref{sec:ivor} and \ref{sec:susan}.
Figure~\ref{fig:cross_mix1} corresponds to final states with 
(a) five charged leptons plus two jets and 
(b) four charged leptons plus four jets.
Again, the limits obtained for various mixtures of lepton flavours lie 
in the band between the two limiting curves.
Figure~\ref{fig:cross_mix2} corresponds to final states with
(a) two charged leptons of the same flavour plus four jets and 
(b) one charged lepton plus four jets. 
Here the limits are given separately for each of the lepton 
flavours, fixed by the first index of the coupling $\lambda_{ijk}^{'}$.
Limits arising from any other coupling \lbp\ lie below the limit 
shown for the coupling with the same first index.
The final states shown in Figure~\ref{fig:cross_mix3}
are like the ones in Figure~\ref{fig:cross_mix2} plus the decay
products of two \Wstar. Again,
the limits are given separately for each of the lepton 
flavours, fixed by the first index of the coupling $\lambda_{ijk}^{'}$ and
limits arising from any other coupling \lbp\ lie below the limit 
shown for the coupling with the same first index.

\subsection{Multi-jet Final States}

\begin{figure}[p]
\begin{center}
\epsfig{file=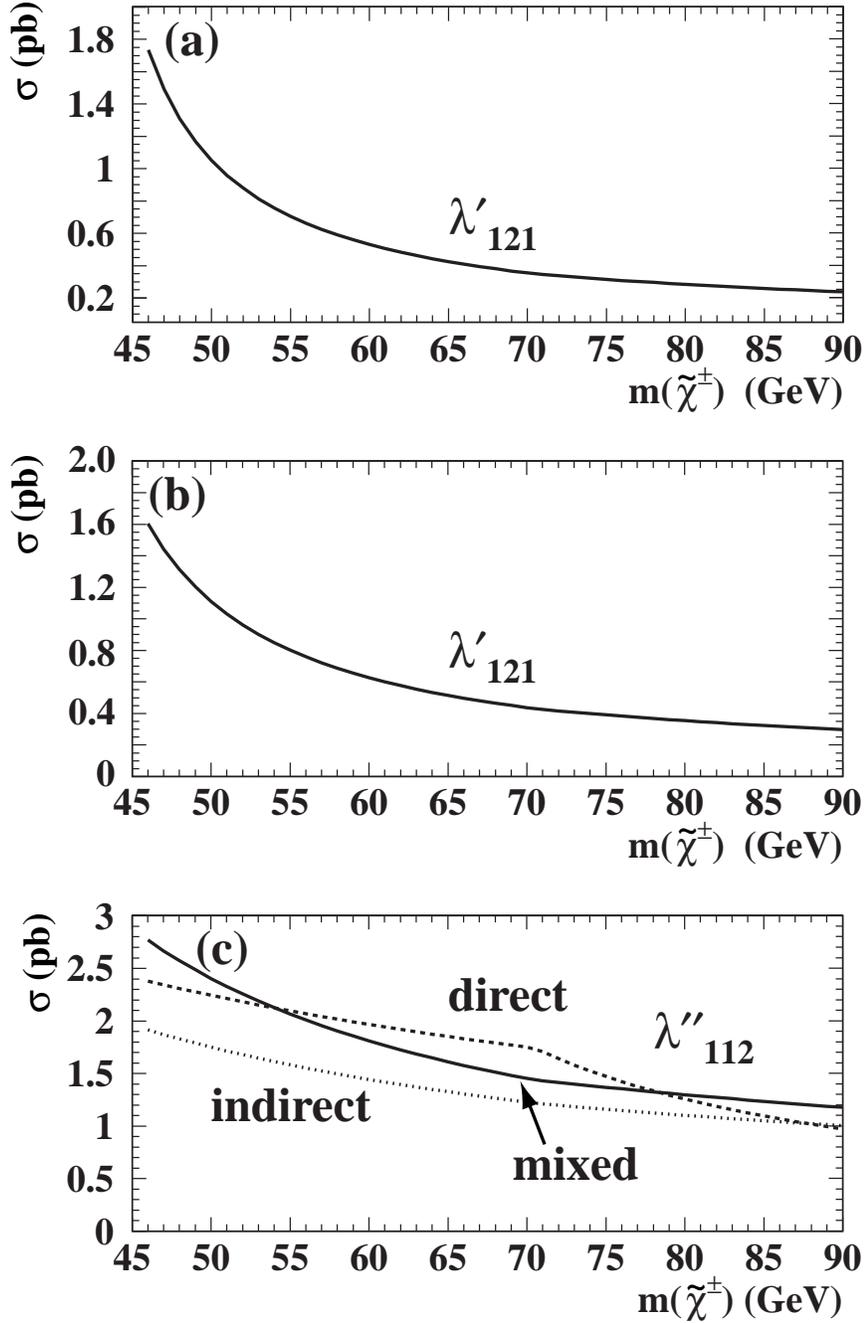,width=12.5cm} 
\caption{ \it Upper limits at 95\% CL
on the cross-sections for final states with 
(a) four jets and missing energy,
(b) four jets and missing energy plus two \Wstar, and
(c) at least six jets.
For each curve the  coupling \lbp\ or \lbpp\ that was assumed 
to be different from zero is given.
Limits arising from any other  coupling \lbp\ or \lbpp\ lie below the limits
shown.
}
\label{fig:cross_mix4}
\end{center}
\end{figure}

Figure~\ref{fig:cross_mix4} 
shows the cross-section limits for final
states with at least four jets
from the selections described in 
Sections~\ref{sec:gabi}, \ref{sec:mar1}, and \ref{sec:mar2}.
Figure~\ref{fig:cross_mix4} (a) corresponds to final states with
four jets plus missing energy and 
(b) to four jets plus missing energy plus two \Wstar.
The limits shown are independent on the first index of the 
coupling $\lambda_{ijk}^{'}$, corresponding to the neutrino flavour.
Limits arising from any other coupling \lbp\ lie below the limit 
shown.
In Figure~\ref{fig:cross_mix4} (c) 
limits for final states with at least six jets are shown.
The three limits labelled direct, mixed, and indirect correspond
to final states with  six jets plus 0, 1, and 2 \Wstar, respectively. 
Limits arising from any other coupling \lbpp\ lie below the limits 
shown.

\section{Limits on the Gaugino Production Cross-Sections}
\label{sec:interpretation}

We now proceed to derive upper bounds on the cross-section of the 
processes \mbox{e$^+$e$^- \ra \chip \chim $} and $\chin \chin$. 
These are presented separately for the direct and indirect modes
for decays being mediated by the couplings \lb, \lbp, and \lbpp.
In addition a limit independent of a decay via the direct or indirect 
mode is given.
The limits are valid for values of the  couplings \lb, \lbp,
and \lbpp\ greater than 10$^{-5}$, assuming a prompt decay at the 
interaction vertex. 
Further requirements are that the  mass difference 
$\Delta m = m(\chipm) - m(\chin) $ be larger than 5~GeV, and that
only one \lb-coupling  be different from zero.
For the indirect decay of the $\chipm$, we assume the decay via a W$^{(*)}$,
and combine the results according to the leptonic and hadronic branching
ratio of the W boson. 
Decays via a sfermion are not considered in this analysis.

The  likelihood ratio method~\cite{likelihood}  has been used to
combine the results from several analyses
to determine the excluded cross-sections.
This method assigns a greater weight to analyses 
with a higher expected sensitivity, taking into account the expected 
background.
All upper bounds on the cross-sections are given at the 95\% CL.

The same method has been used in the determination of limits which are 
independent of whether the decay is direct or indirect.
The branching fractions into these two decay modes varies with
the parameters of the MSSM. To achieve a limit independent of the 
branching ratio and thus of the MSSM parameters, the branching 
fractions 
into the direct and indirect decays have been varied simultaneously 
between 0 and 1.
The efficiency for the process of one chargino decaying via the direct mode 
and the other via the indirect mode has been set to zero, except for the
case of \lbpp. 
A limit is then calculated for all branching ratios at each mass value
using the  likelihood ratio method.
This results in a cross-section limit as a function of the 
branching ratio and the chargino mass. By taking the 
worst limit at each  mass value, a limit independent of the branching ratio
is determined.

\subsection{Chargino Decays via \lb }

\begin{figure}
\begin{center}
\epsfig{file=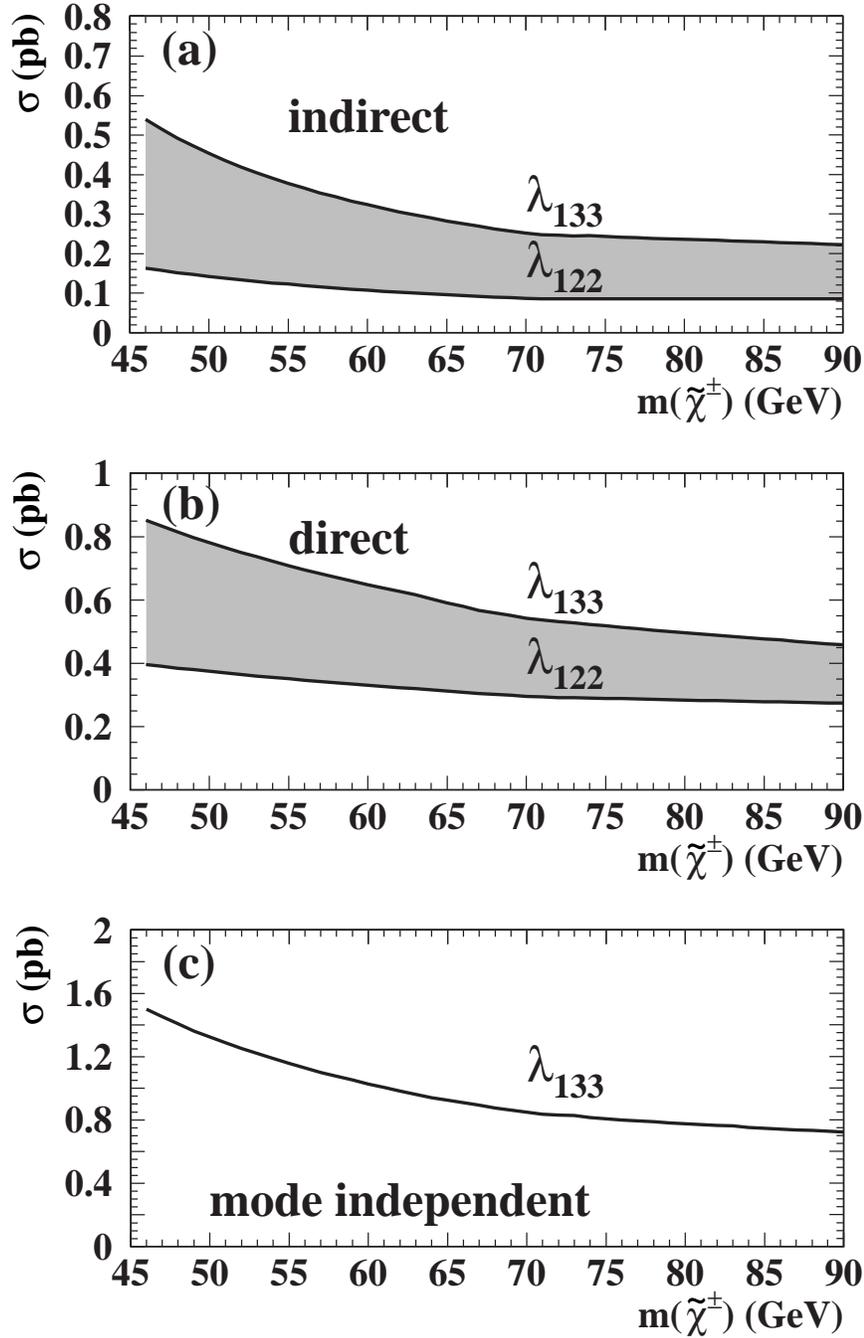,width=12.5cm} 
\caption{ \it 95\% CL upper limits on cross-sections from the decay of a 
$\chipm$ via a coupling \lb\/, assuming a mass difference 
$\Delta m = m(\chipm) - m(\chin) > 5$~GeV. 
The cross-section limits are shown
(a) for the indirect decay mode 
(b) for the direct decay  mode and
(c) independent of the decay mode.
The cross-section limits for any coupling \lb\ lie below the 
limit for a decay via $\lambda_{133}$.
For the (a) direct decay mode  and the (b) indirect decay mode also the 
best cross-section limit, corresponding to $\lambda_{122}$, is shown.}
\label{fig:cross_chipm_lb}
\end{center}
\end{figure}

Figure~\ref{fig:cross_chipm_lb} shows 
the upper limits on the 
cross-sections for decays via a
coupling \lb\  
for (a) the indirect decay 
(b) the direct decay and
(c) independent of the decay mode.
The upper limits on the cross-section 
for any coupling \lb\ lie between the two curves shown
in (a) and (b) and below the one shown in (c). 
The results vary between 0.1 and 1.0~pb for masses above 
70~GeV.
The limits for the direct decay are 
not as strong as those  for the indirect decay, because
the direct decay with one $\chipm$ decaying into three charged leptons and 
the other decaying into one charged lepton has not been generated.
The analysis for four leptons plus missing transverse momentum should be
sensitive to these final states, but we have conservatively set the 
efficiency to zero.
The same applies for the limit independent of the decay mode, which is also 
worse than any of the direct or indirect limits, because of setting
the efficiency for one $\chipm$ decaying 
directly and the other indirectly to zero. 

The mass limits derived from the cross-section limits
are only slightly degraded
by the small lack in Monte Carlo
samples, as the expected cross-section for chargino pair-production is 
large for the kinematical allowed region.

\subsection{Chargino Decays via \lbp}

In the decay of a gaugino via a \lbp\ coupling, the branching ratio
of the gaugino into a final state with a charged or a neutral lepton
is dependent on the mass of the sneutrinos, the mass of the 
sleptons and on the gaugino composition. 
To avoid a dependence of the bounds of the excluded cross-section
on the MSSM parameters in this decay mode, 
the branching ratio of both modes has been varied  between
0 and 1, using the  likelihood ratio method~\cite{likelihood}  
to determine the excluded cross-section, like in the determination of the
limit independent of the direct and indirect decay mode.

\begin{figure}
\begin{center}
\epsfig{file=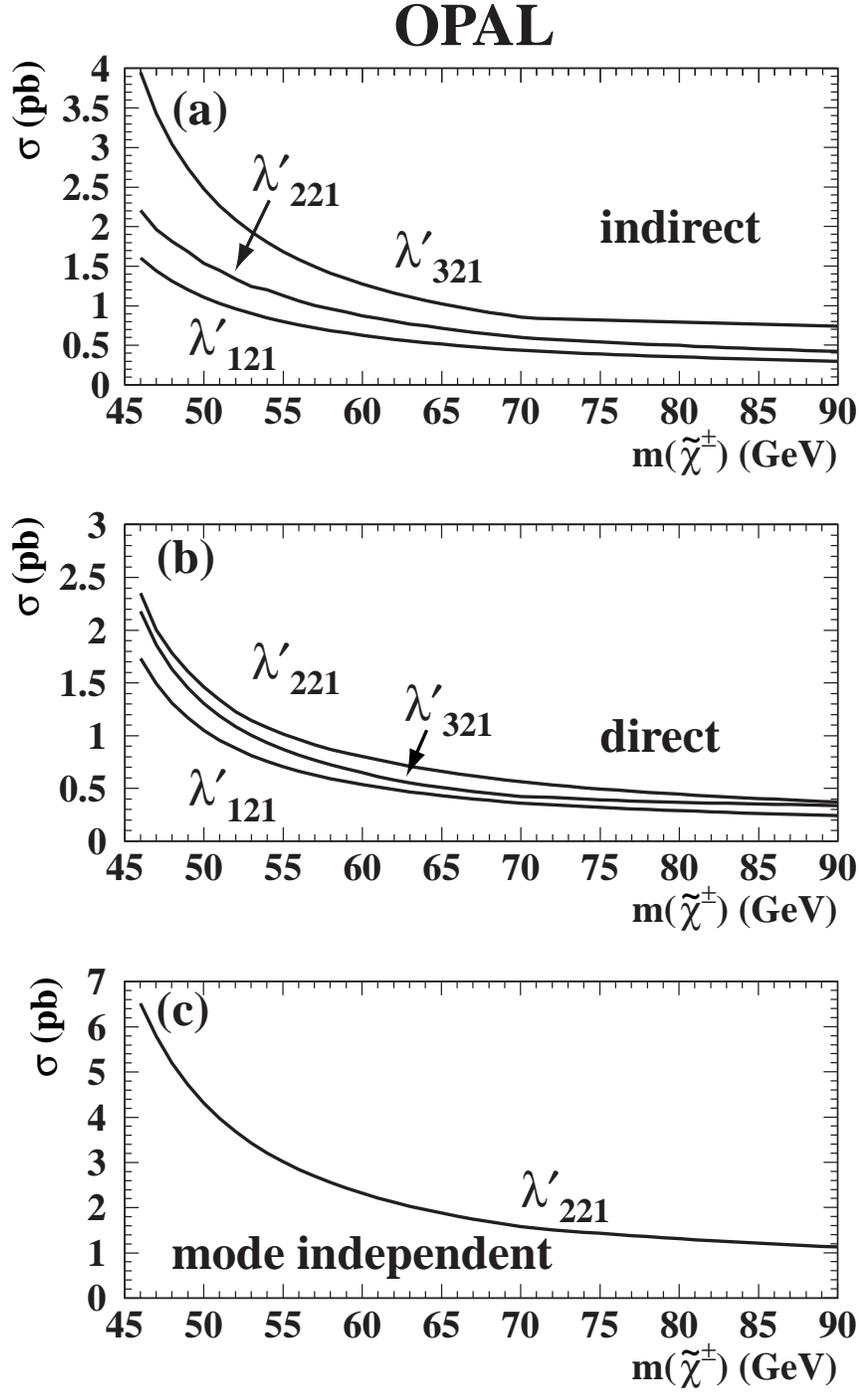,width=12.5cm} 
\caption{ \it 95\% CL upper limits on cross-sections from 
(a) the indirect and 
(b) the direct decay and 
(c) independent of the decay mode
of a $\chipm$ via a coupling \lbp\/. 
In (a) and (b) the excluded cross-sections for any 
coupling \lbp\ lie between the upper and lower curves.
Also indicated are the couplings for which the best and the worst upper 
limits have been achieved.
In (c) the upper limit is given for any coupling $\lambda^{'}_{ijk}$ with
$i = 1,2$. 
}
\label{fig:cross_chipm_lbp}
\end{center}
\end{figure}

Figure~\ref{fig:cross_chipm_lbp} shows the upper limits on the cross-sections 
for the decay of a $\chipm$ via \lbp\ for (a) the indirect decay and (b) 
the direct decay mode for any of the 27 \lbp\ couplings.
Limits arising from any other coupling \lbp\ lie below the limit 
shown for the coupling with the same first index.
Figure~\ref{fig:cross_chipm_lbp}(c) shows the upper limits on 
the cross-sections independent of the decay mode.
Limits arising from any other coupling 
$\lambda^{'}_{ijk}$, with $ i=1,2; j,k = 1,2,3$
lie below the limit shown.
The limits are not as good as  for decays via 
couplings \lb\/, since the signal looks more like the Standard Model processes,
and lie between 0.3 and 1.8~pb for masses above 70~GeV.

\subsection{Chargino Decays via \lbpp}

\begin{figure}
\begin{center}
\epsfig{file=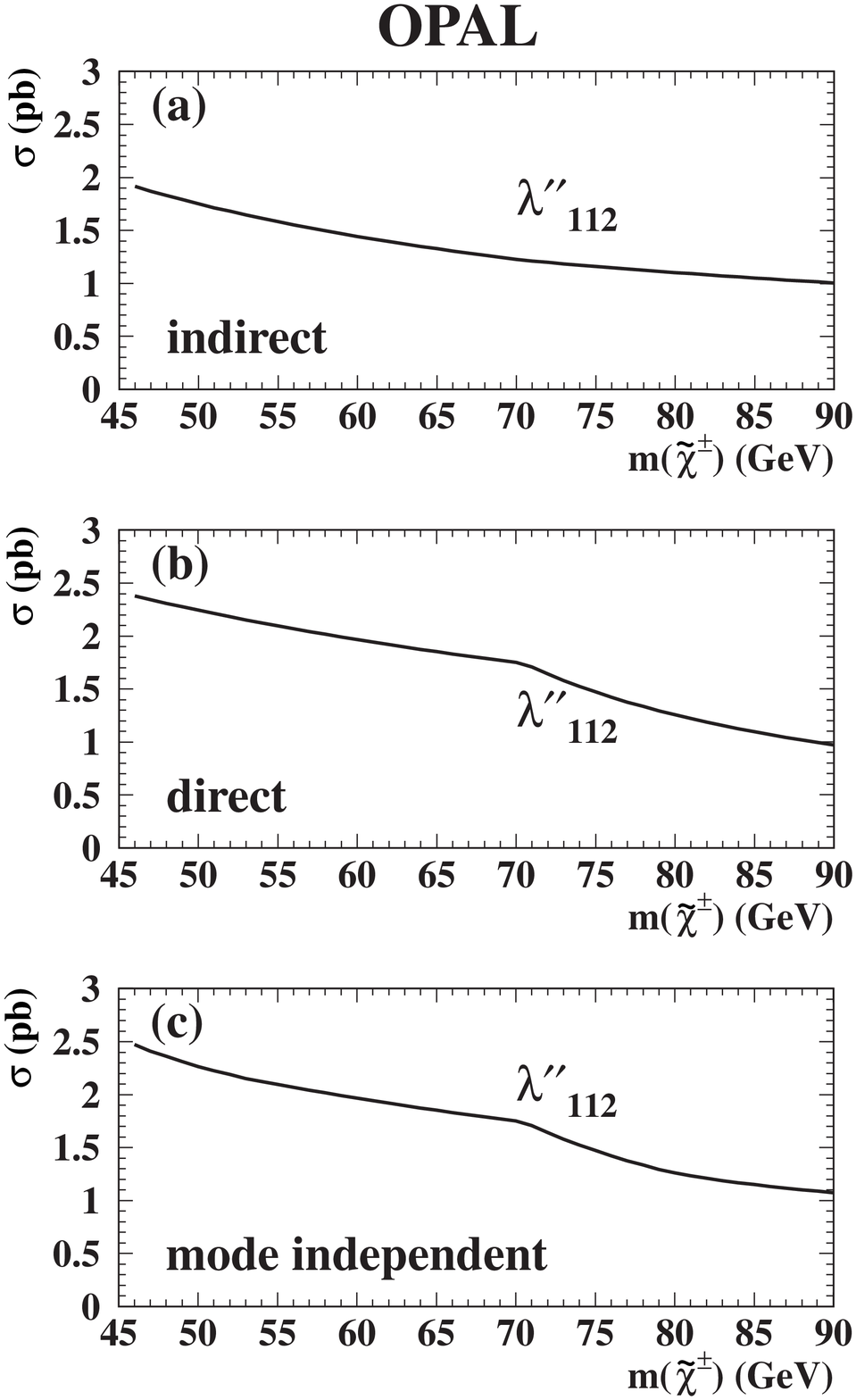,width=12.5cm} 
\caption{ \it 95\% CL upper limits on cross-sections from the decay of a 
$\chipm$ via a coupling \lbpp\/, assuming a mass difference 
$\Delta m = m(\chipm) - m(\chin) > 5$~GeV. 
The cross-section limits are shown
(a) for the indirect decay mode 
(b) for the direct decay  mode and
(c) independent of the decay mode.
The cross-section limits for any coupling \lbpp\ lie below the 
limit for a decay via $\lambda^{''}_{122}$.}
\label{fig:cross_chipm_lbpp}
\end{center}
\end{figure}

Figure~\ref{fig:cross_chipm_lbpp} shows the upper limits on the cross-sections 
for the decay of a $\chipm$ via \lbpp\ for (a) the indirect decay, (b) 
the direct decay mode, and (c) independent of the decay mode. 
Limits arising from any other coupling \lbpp\ lie below the limit 
shown.
As for this 
case the decay of one chargino decaying directly and the other indirectly 
has been simulated, the mode independent limit is similar to 
the direct and  indirect one.
The limits lie  between 1.0 and 1.8 pb for masses above 70~GeV.  

\subsection{Neutralino Decays}

\begin{figure}
\begin{center}
\epsfig{file=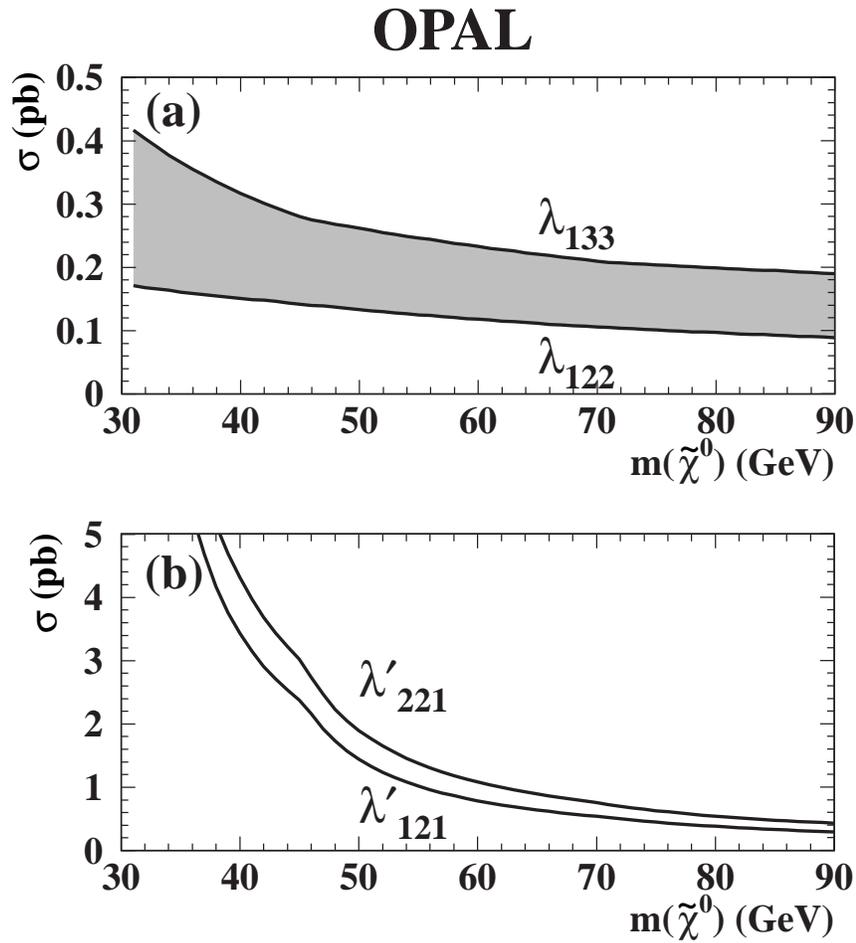,width=12.5cm} 
\caption{ \it 95\% CL upper limits on cross-sections from the 
direct decay of a 
$\chin$ via (a) a coupling \lb\ and (b) a coupling \lbp\/. 
The cross-section limits for any 
coupling \lb\ lie in the gray zone between the two limits shown.
The limit for $\lambda^{'}_{321}$ lies on top of that for 
$\lambda^{'}_{121}$.
Limits arising from any other coupling \lbp\ lie below the limit 
shown for the coupling with the same first index.
}
\label{fig:cross_chin}
\end{center}
\end{figure}

Figure~\ref{fig:cross_chin} shows the upper limits on the cross-sections 
for the decay of a $\chin$.
Limits arising from a decay via a coupling \lb\ lie between the two limits
shown in (a). In (b) the upper limits for a decay via \lbp\ are shown.
Limits arising from any other coupling \lbp\ lie below the limit 
shown for the coupling with the same first index.

\section{Interpretation in the MSSM}

\begin{figure}
\begin{center}
\epsfig{file=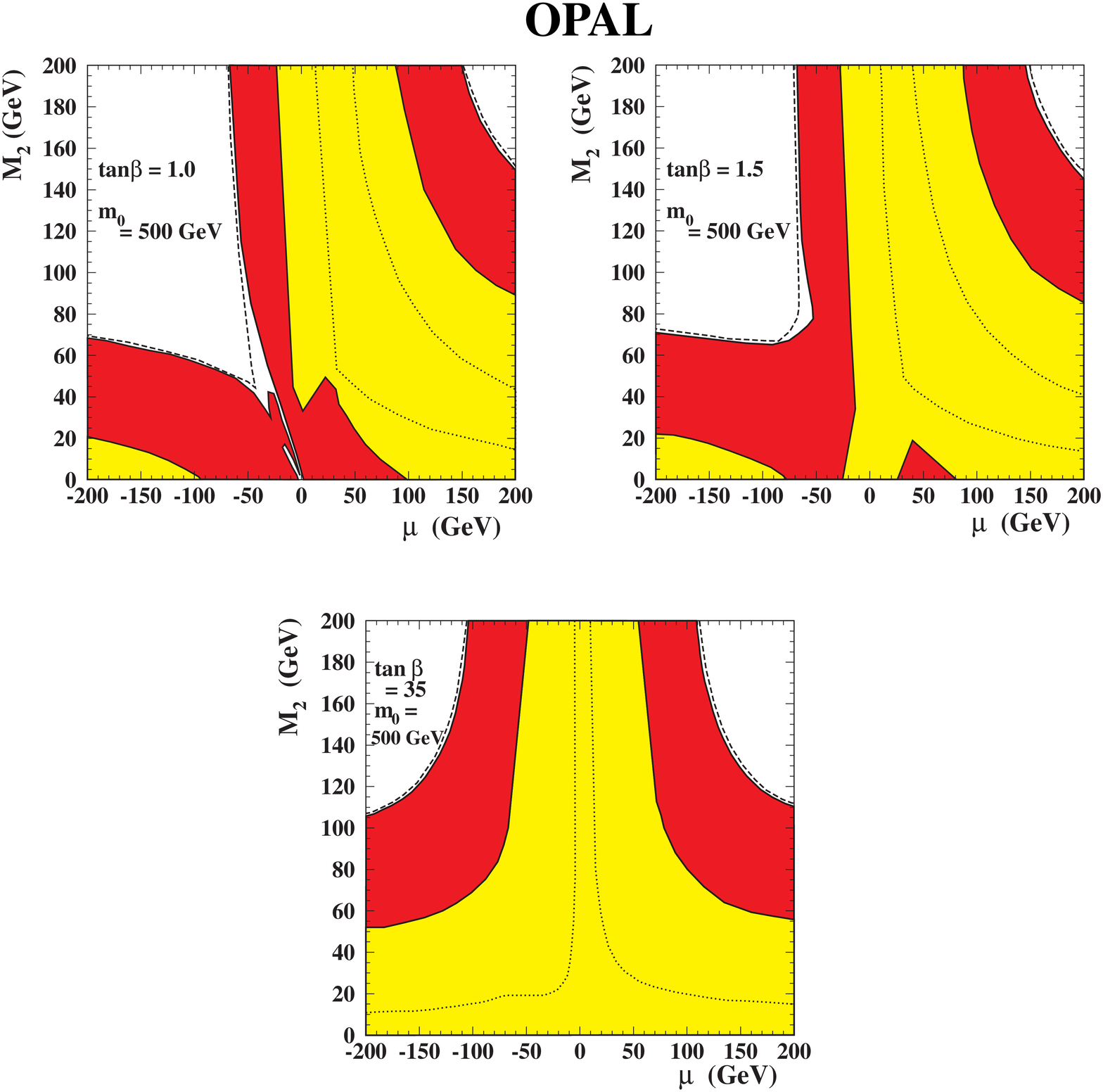,width=16.0cm} 
\caption{\it Excluded region in the M$_2$ -- $\mu$ plane from the 
decay
of $\chipm$, for \lb~$\neq 0$, \lbp=\lbpp=0, for m$_0 = 500$~GeV.
The dark area shows the points excluded by the LEP2 searches and the 
light area
the points excluded from the Z$^0$ width. 
The dashed line shows the kinematic limit for 
$\sqrt{s} = 183$~GeV, and the 
dotted line shows the area with $\Delta m < 5$~GeV.}
\label{fig:m2mu_indirect_lb_m500}
\end{center}
\end{figure}

\begin{figure}
\begin{center}
\epsfig{file=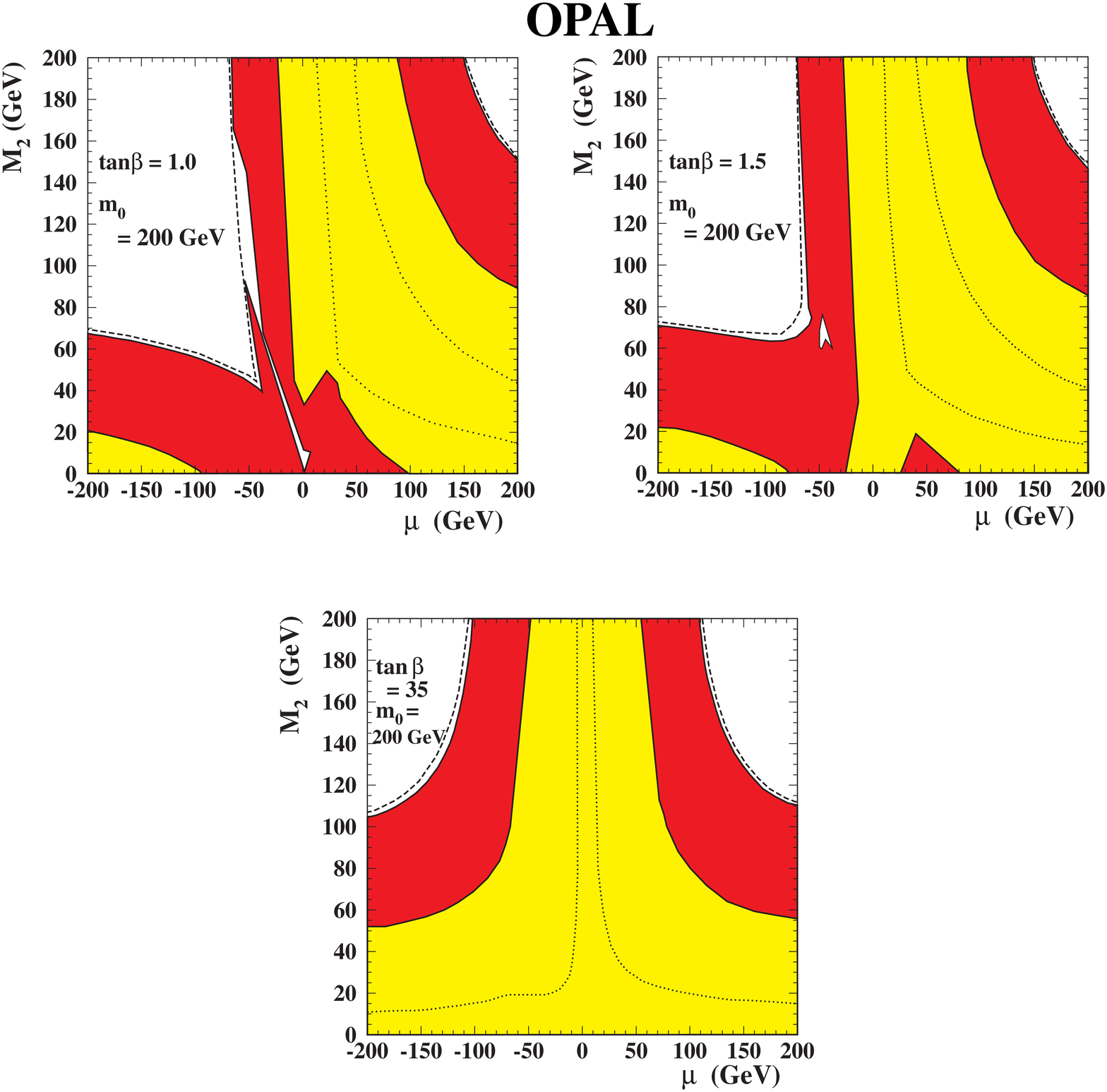,width=16.0cm} 
\caption{\it Excluded region in the M$_2$ -- $\mu$ plane from the 
decay
of $\chipm$ for \lb~$\neq 0$, \lbp=\lbpp=0, for m$_0 = 200$~GeV. 
The dark area shows the points excluded by the LEP2 searches and the 
light area
the points excluded from the Z$^0$ width. 
The dashed line shows the kinematic limit for 
$\sqrt{s}  = 183$~GeV, and the 
dotted line shows the area with $\Delta m < 5$~GeV.}
\label{fig:m2mu_indirect_lb_m200}
\end{center}
\end{figure}

\begin{figure}
\begin{center}
\epsfig{file=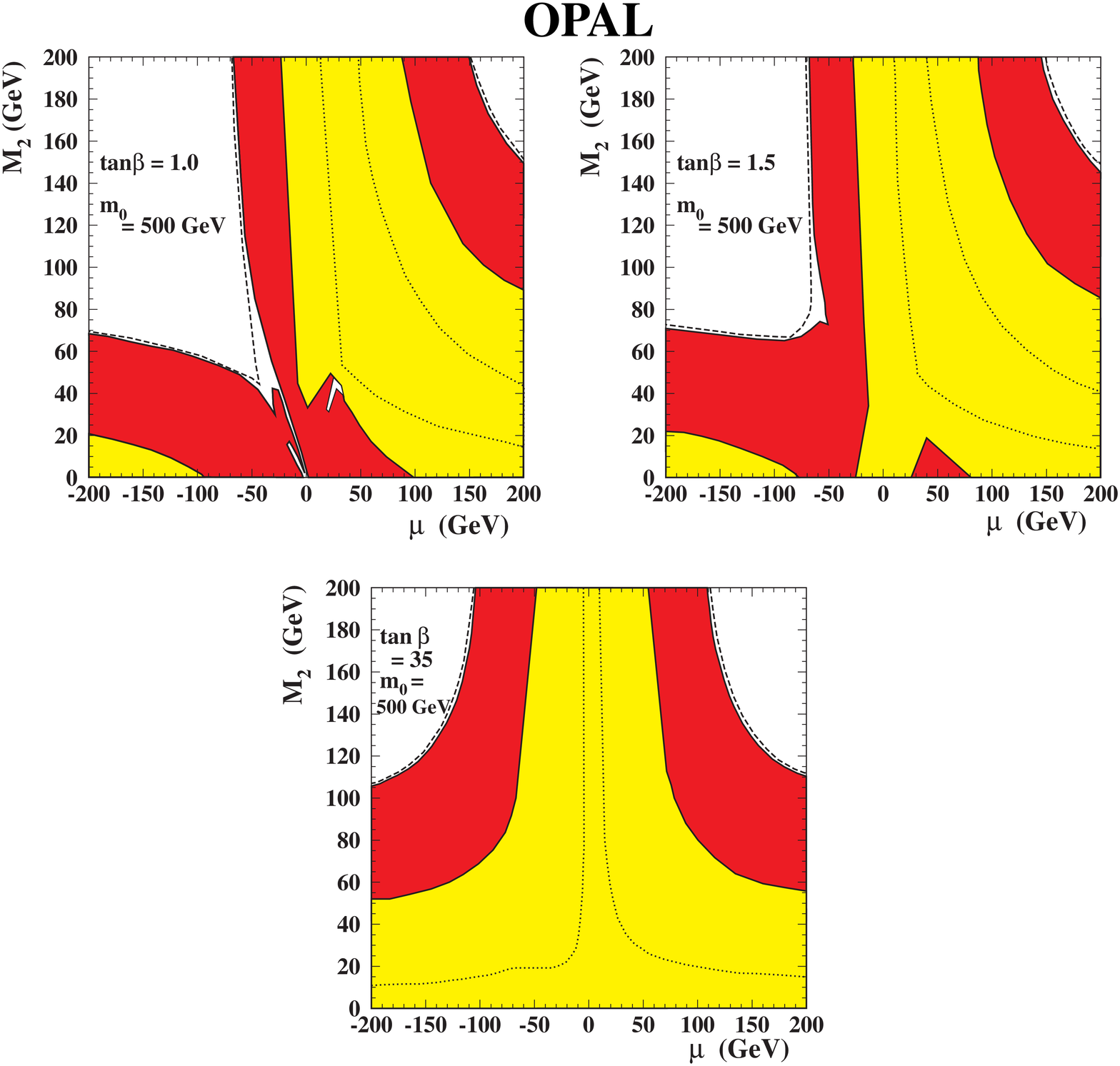,width=16.0cm} 
\caption{\it Excluded region in the M$_2$ -- $\mu$ plane from the 
indirect decay
of $\chipm$ for \lbp~$\neq 0$, \lb=\lbpp=0, for m$_0 = 500$~GeV. 
The dark area shows the points excluded by the LEP2 searches and the 
light area
the points excluded from the Z$^0$ width. The dashed line shows the kinematic
limit for 
$\sqrt{s}  = 183$~GeV, and the 
dotted line shows the area with $\Delta$~m $<$ 5~GeV}
\label{fig:m2mu_indirect_lbp_m500}
\end{center}
\end{figure}

\begin{figure}
\begin{center}
\epsfig{file=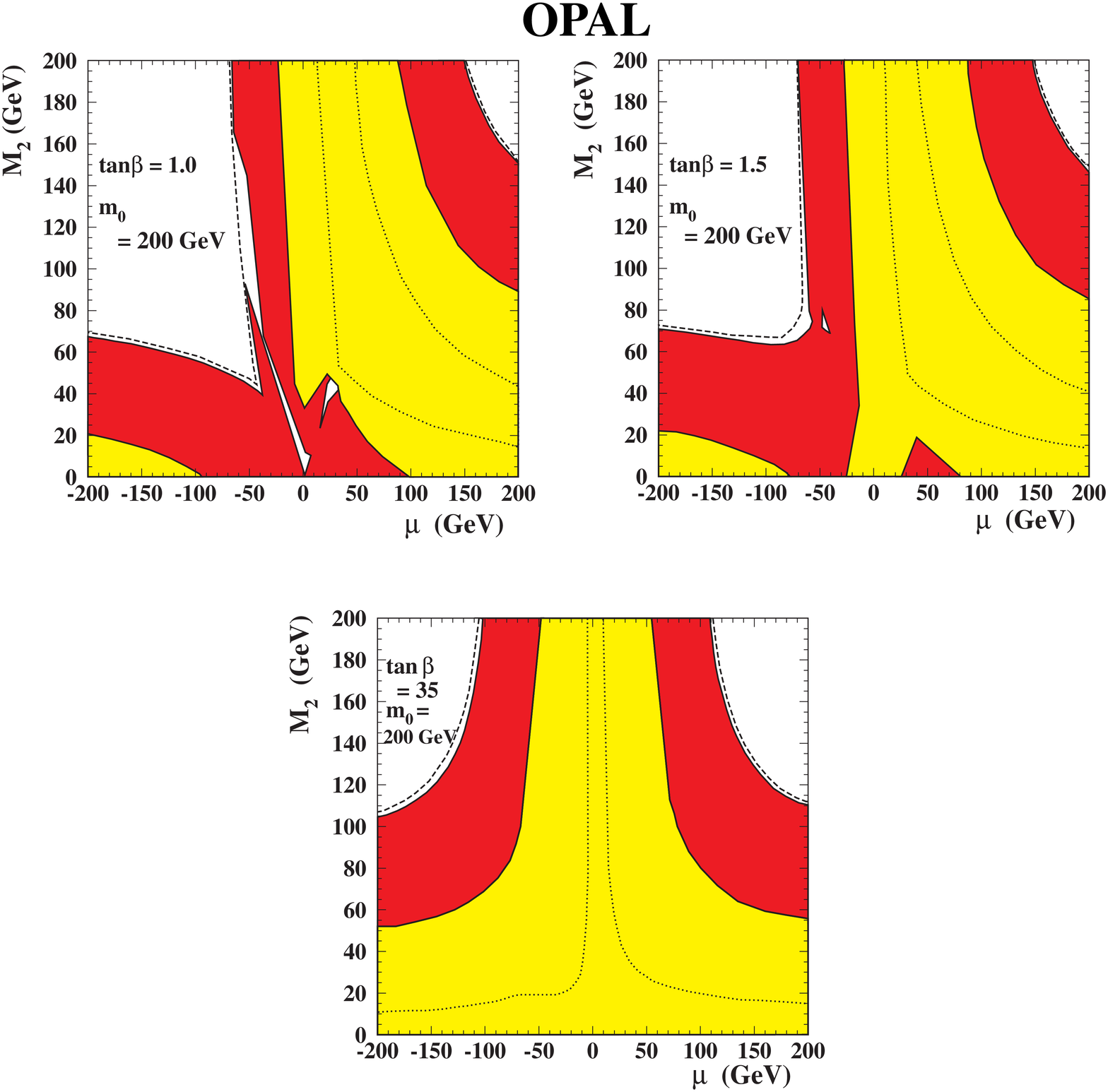,width=16.0cm}
\caption{\it Excluded region in the M$_2$ -- $\mu$ plane from the 
indirect decay
of $\chipm$ for \lbp~$\neq 0$, \lb=\lbpp=0, for m$_0 = 200$~GeV. 
The dark area shows the points excluded by the LEP2 searches and the 
light area
the points excluded from the Z$^0$ width. 
The dashed line shows the kinematic limit for 
$\sqrt{s} = 183$~GeV, and the 
dotted line shows the area with $\Delta m < 5$~GeV.}
\label{fig:m2mu_indirect_lbp_m200}
\end{center}
\end{figure}

\begin{figure}
\begin{center}
\epsfig{file=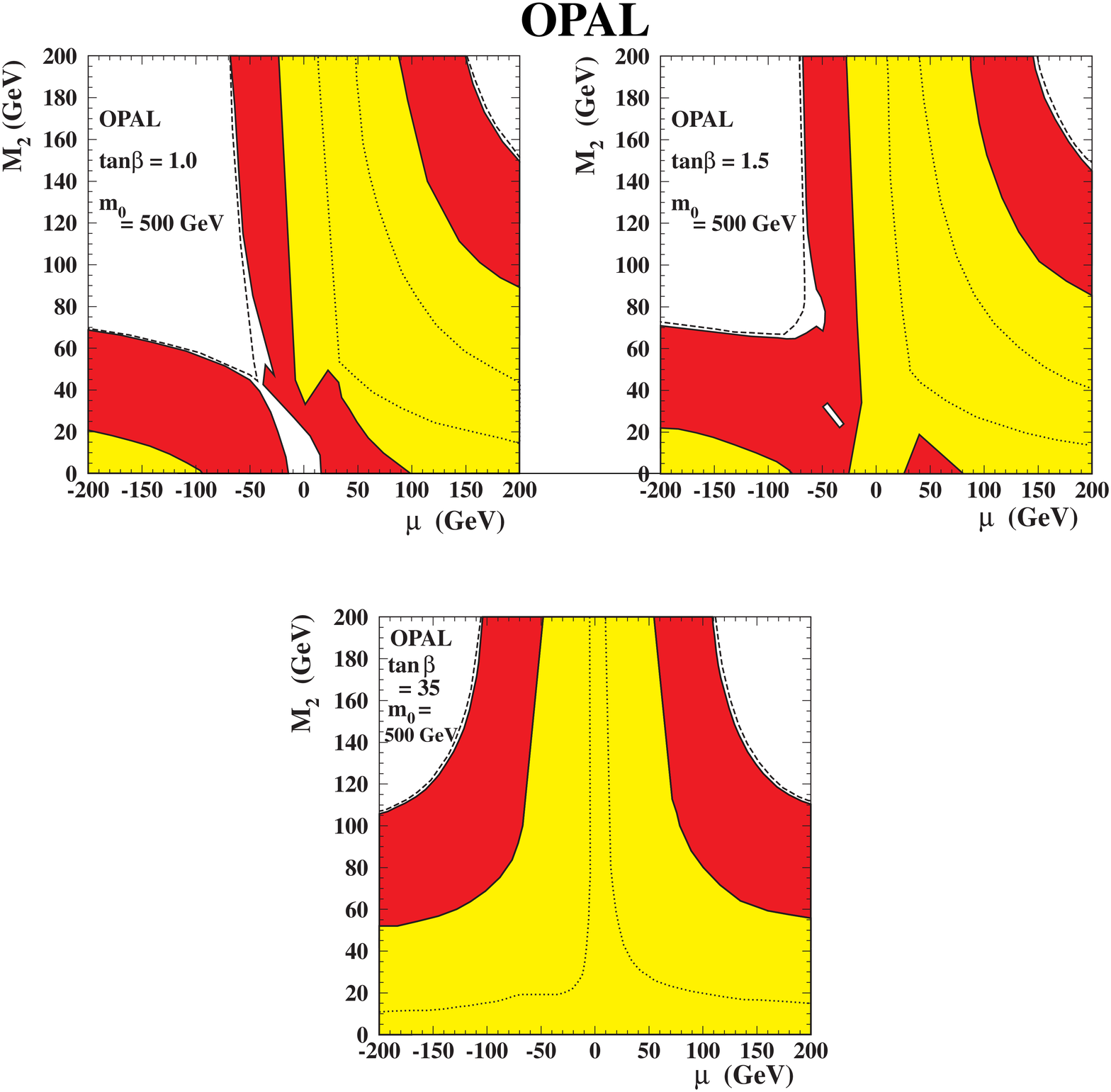,width=16.0cm} 
\caption{\it Excluded region in the M$_2$ -- $\mu$ plane from the 
decay
of $\chipm$ for \lbpp~$\neq 0$, \lb=\lbp=0, for m$_0 = 500$~GeV.
The dark area shows the points excluded by the LEP2 searches and the 
light area
the points excluded from the Z$^0$ width. The dashed line shows the kinematic
limit for 
$\sqrt{s} = 183$~GeV, and the 
dotted line shows the area with $\Delta m < 5$~GeV.}
\label{fig:m2mu_indirect_lbpp_m500}
\end{center}
\end{figure}

\begin{figure}
\begin{center}
\epsfig{file=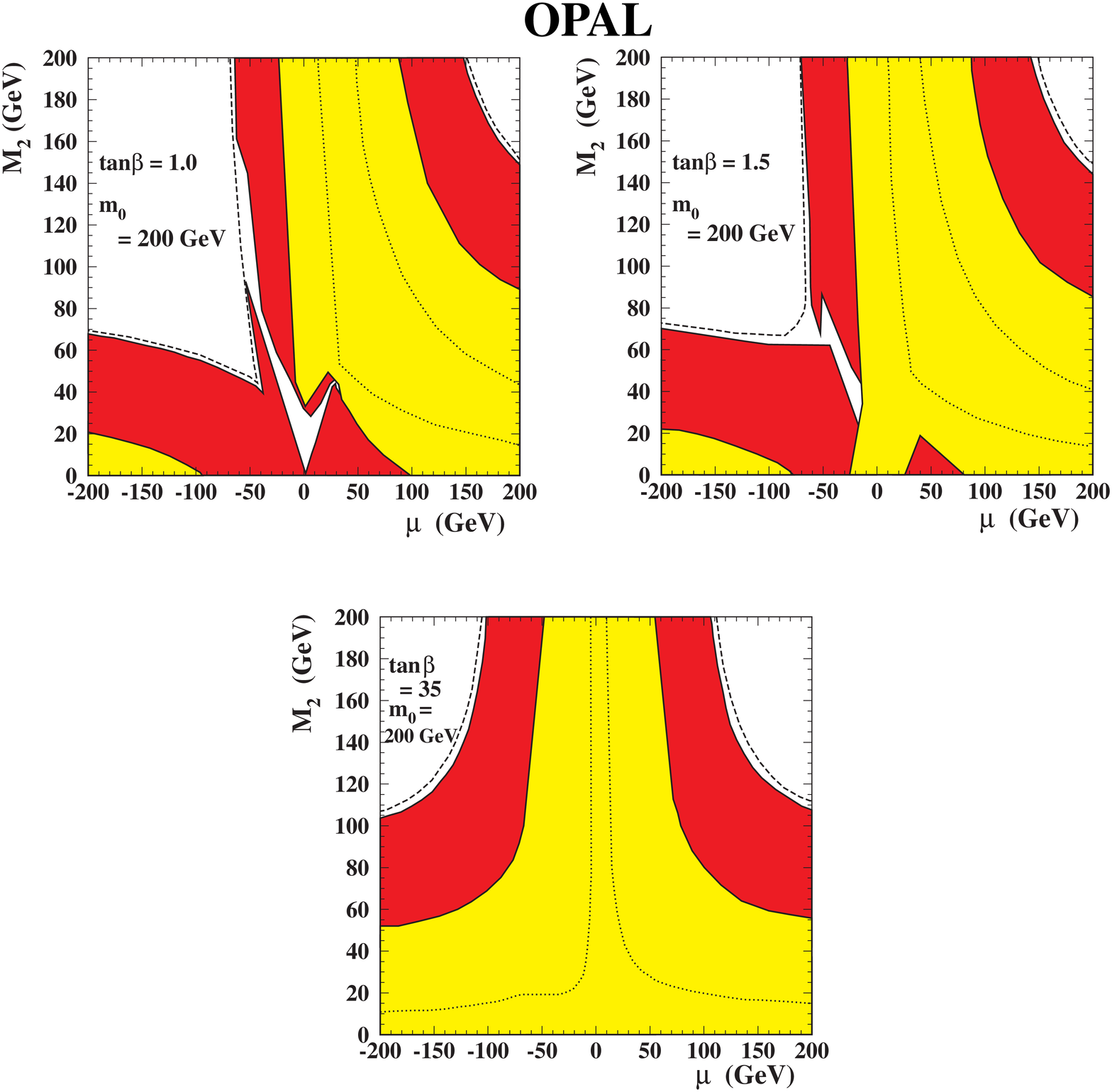,width=16.0cm} 
\caption{\it Excluded region in the M$_2$ -- $\mu$ plane from the 
decay
of $\chipm$ for \lbpp~$\neq 0$, \lb=\lbp=0, for m$_0 = 200$~GeV.
The dark area shows the points excluded by the LEP2 searches and the 
light area
the points excluded from the Z$^0$ width. The dashed line shows the kinematic
limit for 
$\sqrt{s} = 183$~GeV, and the 
dotted line shows the area with $\Delta m < 5$~GeV.}
\label{fig:m2mu_indirect_lbpp_m200}
\end{center}
\end{figure}

From the cross-section upper limits presented in 
Chapter~\ref{sec:interpretation}, regions in the 
MSSM parameter space can be excluded.
These regions are shown in Figures~\ref{fig:m2mu_indirect_lb_m500}
to ~\ref{fig:m2mu_indirect_lbpp_m200} 
 in the ($M_2$, $\mu$) plane
for $m_0$ = 500~GeV and  200~GeV
and for $\tan \beta$ = 1.0, 1.5, and 35.0.
The results are presented separately for decays with the
\lb, \lbp, and \lbpp couplings being different from zero.

The exclusion limits are determined by combining the following:
the excluded cross-sections 
from the excess Z$^0$ 
width~\cite{pdg},
by comparing the measured and predicted width of the Z$^0$
(light grey area); 
the cross-section upper limits 
from the pair production of $\chipm$ and their decay via a \lb-coupling
(black area),
using the worst limit and consequently making 
the excluded region  independent of decay type. 
The production cross-section for $\chin$ pairs is so small that 
limits from the $\chipm$  analyses 
are more precise than the $\chin$ limits everywhere.
The regions excluded for $m_0$ = 500~GeV are also valid for 
$m_0 >$ 500~GeV.

Although the Monte Carlo events have only been generated for 
$\Delta m = m(\chipm) - m(\chin) > $5~GeV, the regions with 
$\Delta m < $5~GeV are excluded from the total Z$^0$ width in the regions
shown in the figures. Only for  large values of $M_2$ ( $ > $800~GeV)  
regions with $\Delta m < $5~GeV cannot be excluded.
The area with $\Delta m < $5~GeV lies inside the two dotted lines in
the figures.

For values of $\tan \beta$ smaller than 2, the branching ratio of
$\chipm \ra W^{(*)} \neutrala $ 
with the subsequent decay of $\neutrala \ra \chin \gamma$
can become as large as one for
small $M_2$ and for $\mu$
negative and small.
We have checked that our analyses are sensitive to these decays.
For $\tan \beta \le 2.0$ regions with the above decay exist, that cannot be 
excluded with the present data with the mode independent 
cross-section upper limit.

The regions excluded in the MSSM parameter space 
for exactly one coupling \lb\ not equal to zero are shown in 
Figures~\ref{fig:m2mu_indirect_lb_m500} and
\ref{fig:m2mu_indirect_lb_m200} independent of direct or 
indirect decay mode.
The figures show that 
the excluded area is very close to the 
kinematic limit for $\chipm$ pair production.

The regions excluded in the MSSM parameter space 
for exactly one coupling \lbp\ not equal to zero are shown in 
Figures~\ref{fig:m2mu_indirect_lbp_m500} and
\ref{fig:m2mu_indirect_lbp_m200} for the indirect decay.
For large values of $\tan \beta$ the excluded region goes up to the kinematic 
limit, but  for smaller values, unexcluded regions exist even for chargino 
masses as small as 45~GeV.

For one coupling \lbpp\ not equal to zero 
the regions excluded in the MSSM parameter space are shown in  
Figures~\ref{fig:m2mu_indirect_lbpp_m500} and
\ref{fig:m2mu_indirect_lbpp_m200} independent of the decay mode.
Also here unexcluded regions exist for small values of $\tan \beta$.

\begin{figure}
\begin{center}
\epsfig{file=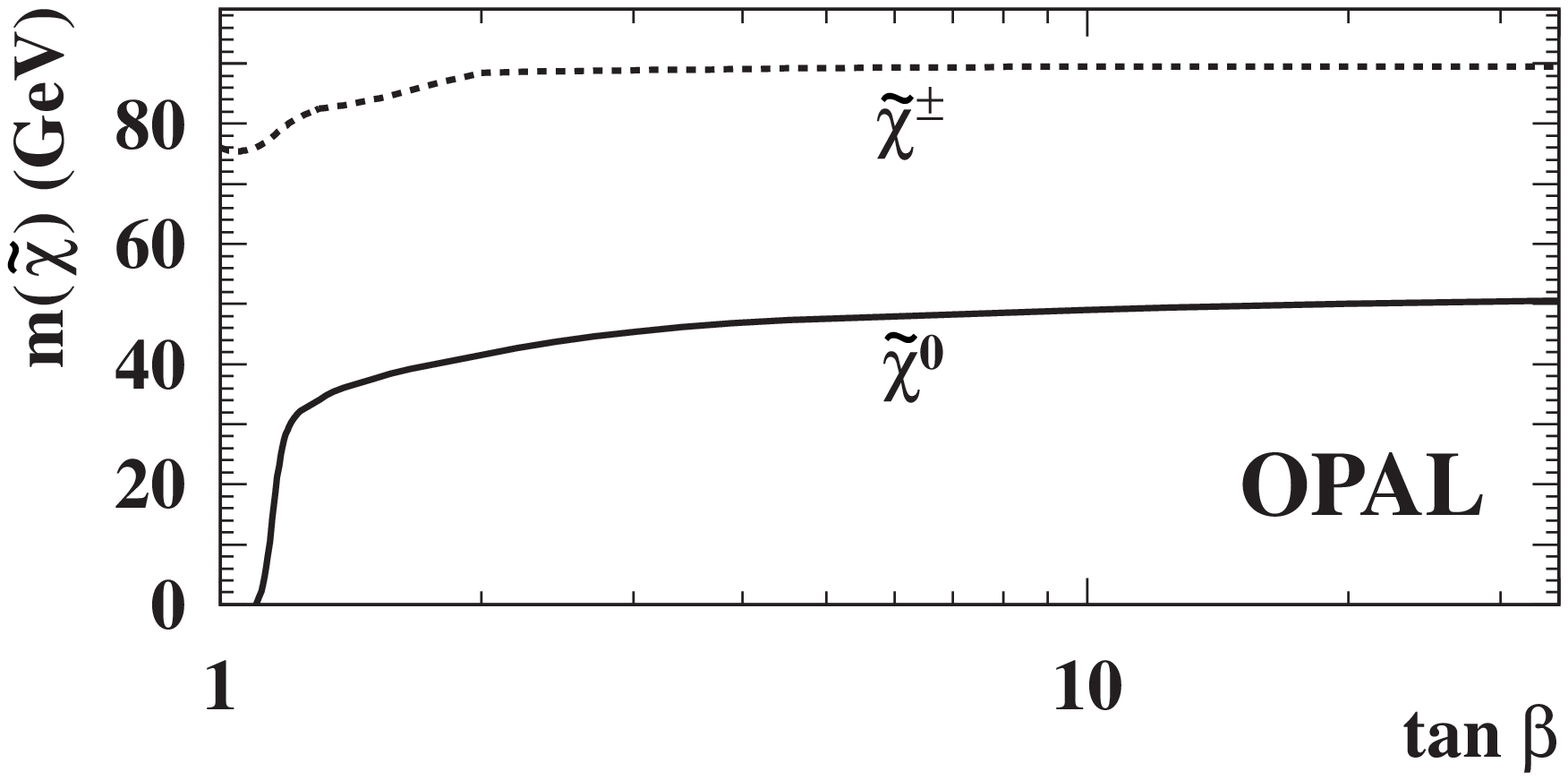,width=12.5cm} 
\caption{\it Excluded masses for $\chin$ (solid line) and $\chipm$ 
(dashed line) as a function of $\tan \beta$ 
for m$_0 = 500$~GeV for any coupling \lb\/. 
The exclusion limits for the $\chin$ do not result  from the direct search 
for $\chin$ decays but from the excluded CMSSM parameter space from
the $\chipm$ searches. 
}
\label{fig:tanbeta} 
\end{center}
\end{figure}

Each point in the MSSM parameter space corresponds to a $\chipm$ and a 
$\chin$ mass pair. By excluding regions of this parameter space one 
can therefore
also limit the allowed mass domains for these particles. 
The excluded masses for $\chin$ for a given $m_0$ depend on the 
value of $\tan \beta$ and are shown in Figure~\ref{fig:tanbeta}
for any coupling \lb\ greater than zero.
Due to the small unexcluded regions in the MSSM parameter space for
values of $\tan \beta \sim $1.0, no mass limit can be given for that region.
For $\tan \beta > $1.2 the lower limit on the $\chin$ mass is 29~GeV
for $m_0$ = 500~GeV, increasing up to 50~GeV for  $\tan \beta > $20.

The limits for the $\chipm$ depend much less on
$\tan \beta$ than those of the $\chin$. 
For the $\chipm$ a mass up to 76~GeV  is excluded for any coupling
\lb\  for any point in the MSSM parameter space,
with $m_0$ = 500~GeV,
under the assumption that it decays via a W$^{(*)}$.

\clearpage

\clearpage

\clearpage

\clearpage

\section{Conclusions}
\label{sec:conclusions}

We have been searching for \Rparity\ violating decays of charginos 
and neutralinos via the 
Yukawa couplings \lb, \lbp, and \lbpp. 
Analyses have been presented for a large number of final states arising from
these 
decays, varying from two leptons up to
more than four jets. We have not observed any excess of events. Limits on 
the cross 
section times branching ratio, ranging from  0.08~pb up to several pb,
have been presented separately for all these topologies. From these, 
upper limits on the pair production cross-section of $\chipm$ and 
$\chin$ are obtained and presented separately for \lb, \lbp,  and \lbpp 
for direct, indirect and mixed decay modes.
For \lb\ and \lbpp\, limits have also been presented which are valid
independently of whether the decays are direct or indirect.

Finally, the limits are interpreted in the framework of the MSSM.  
Most of the kinematically accessible regions 
in the $M_2-\mu$ plane for $m_0 \ge$ 500~GeV
have been excluded.
Lower mass limits, of 76~GeV for $\chipm$ at $m_0$~=~500~GeV and 
$\tan \beta \ge $1.0,
and 29~GeV for
 $\chin$ at  $m_0$~=~500~GeV and $\tan \beta \ge $1.2, have been 
obtained independently of which of the couplings \lb, \lbp, and \lbpp\, 
is assumed to be different from zero.
All the limits are valid at the 95\% confidence level, for couplings larger 
than 10$^{-5}$.

\bigskip\bigskip\bigskip
\appendix
\par
{\Large\bf Acknowledgements}
\par
We particularly wish to thank the SL Division for the efficient operation
of the LEP accelerator at all energies
 and for their continuing close cooperation with
our experimental group.  We thank our colleagues from CEA, DAPNIA/SPP,
CE-Saclay for their efforts over the years on the time-of-flight and trigger
systems which we continue to use.  In addition to the support staff at our own
institutions we are pleased to acknowledge the  \\
Department of Energy, USA, \\
National Science Foundation, USA, \\
Particle Physics and Astronomy Research Council, UK, \\
Natural Sciences and Engineering Research Council, Canada, \\
Israel Science Foundation, administered by the Israel
Academy of Science and Humanities, \\
Minerva Gesellschaft, \\
Benoziyo Center for High Energy Physics,\\
Japanese Ministry of Education, Science and Culture (the
Monbusho) and a grant under the Monbusho International
Science Research Program,\\
Japanese Society for the Promotion of Science (JSPS),\\
German Israeli Bi-national Science Foundation (GIF), \\
Bundesministerium f\"ur Bildung, Wissenschaft,
Forschung und Technologie, Germany, \\
National Research Council of Canada, \\
Research Corporation, USA,\\
Hungarian Foundation for Scientific Research, OTKA T-016660,
T023793 and OTKA F-023259.\\

\end{document}